\documentclass[3p]{elsarticle}
\usepackage{graphicx}
\usepackage{epstopdf}
\usepackage[percent]{overpic}
\usepackage{psfrag}
\usepackage{fancyhdr}
\usepackage{amsmath}
\usepackage{amssymb}
\usepackage{booktabs}
\usepackage{color}
\usepackage[bookmarks=false]{hyperref}
\usepackage{afterpage}

\usepackage{SIunits}
\newcommand{\fprompt}{F$_{\mathrm{prompt}}$}
\newcommand{\fpromptmath}{\mathrm{F}_{\mathrm{prompt}}}
\newcommand{\pleak}{P$_{\mathrm{leak}}$}
\newcommand{\pleakmath}{\mathrm{P}_{\mathrm{leak}}}
\newcommand{\zfit}{Z$_{\mathrm{fit}}$}
\newcommand{\kevee}{keV$_{\mathrm{ee}}$}
\newcommand{\kevr}{keV$_{\mathrm{r}}$}

\definecolor{light-gray}{gray}{0.6}

\newcommand{\mk}{black}

\journal{Astroparticle Physics}

\begin{document}
\begin{frontmatter}
\title{Measurement of the scintillation time spectra and pulse-shape discrimination of low-energy $\boldmath{\beta}$ and nuclear recoils in liquid argon with DEAP-1}

\author[TRIUMF]{P.-A.~Amaudruz}
\author[LU]{M.~Batygov\fnref{carleton}}\fntext[carleton]{Current address: Carleton University, Ottawa, ON, K1S 5B6, Canada.}
\author[UofA]{B.~Beltran}
\author[QU]{J.~Bonatt}
\author[Carleton]{K.~Boudjemline}
\author[Carleton,QU]{M.\,G.~Boulay\corref{cor1}}
\cortext[cor1]{Corresponding authors.}
\ead{MarkBoulay@cunet.carleton.ca, mark.boulay@queensu.ca}
\author[QU]{B.~Broerman}
\author[UofA]{J.\,F.~Bueno}
\author[RHUL]{A.~Butcher}
\author[QU]{B.~Cai}
\author[UPenn]{T.~Caldwell}
\author[QU]{M.~Chen}
\author[UofA]{R.~Chouinard}
\author[LU,SNOLAB]{B.\,T.~Cleveland}
\author[QU]{D.~Cranshaw}
\author[QU]{K.~Dering}
\author[LU,SNOLAB]{F.~Duncan}
\author[RHUL]{N.~Fatemighomi}
\author[LU,SNOLAB]{R.~Ford}
\author[QU]{R.~Gagnon}
\author[QU]{P.~Giampa}
\author[UNM]{F.~Giuliani}
\author[UNM]{M.~Gold} 
\author[QU]{V.\,V.~Golovko\fnref{aecl}}\fntext[aecl]{Current address: CNL, Chalk River, ON, K0J 1J0, Canada.}
\author[UofA]{P.~Gorel}
\author[RHUL]{E.~Grace}
\author[Carleton]{K.~Graham}
\author[UofA]{D.\,R.~Grant}
\author[UofA]{R.~Hakobyan}
\author[UofA]{A.\,L.~Hallin}
\author[Carleton]{M.~Hamstra}
\author[QU]{P.~Harvey}
\author[QU]{C.~Hearns}
\author[LU]{J.~Hofgartner\fnref{cornell}}\fntext[cornell]{Current address: Cornell University, Ithaca, NY 14850, USA.}
\author[LU,SNOLAB]{C.\,J.~Jillings}
\author[QU]{M.~Ku\'zniak\corref{cor1}\fnref{carleton}}\ead{kuzniak@owl.phy.queensu.ca}
\author[LU,SNOLAB]{I.~Lawson}
\author[RHUL]{F.~La~Zia}
\author[SNOLAB]{O.~Li}
\author[QU]{J.\,J.~Lidgard}
\author[SNOLAB]{P.~Liimatainen}
\author[YU]{W.\,H.~Lippincott\fnref{fermi}}\fntext[fermi]{Current address: Fermilab, Batavia, IL 60510, USA.}
\author[QU]{R.~Mathew}
\author[QU]{A.\,B.~McDonald}
\author[UofA]{T.~McElroy}
\author[SNOLAB]{K.~McFarlane}
\author[YU]{D.\,N.~McKinsey}
\author[Carleton]{R.~Mehdiyev}
\author[RHUL]{J.~Monroe}
\author[TRIUMF]{A.~Muir}
\author[QU]{C.~Nantais}
\author[QU]{K.~Nicolics}
\author[YU]{J.~Nikkel}
\author[QU]{A.\,J.~Noble}
\author[QU]{E.~O{'}Dwyer}
\author[UofA]{K.~Olsen}
\author[Carleton]{C.~Ouellet\corref{cor1}}\ead{couellet@physics.carleton.ca}
\author[QU]{P.~Pasuthip}
\author[Sussex]{S.\,J.\,M.~Peeters}
\author[QU,SNOLAB]{T.~Pollmann\corref{cor1}}\ead{tina.pollmann@snolab.ca}
\author[QU]{W.~Rau}
\author[TRIUMF]{F.~Reti\`ere}
\author[UNC]{M.~Ronquest\fnref{lanl}}\fntext[lanl]{Current address: LANL, P.O. Box 1663, Los Alamos, NM 87545, USA.}
\author[RHUL]{N.~Seeburn}
\author[QU]{P.~Skensved}
\author[TRIUMF]{B.~Smith}
\author[QU]{T.~Sonley}
\author[UofA]{J.~Tang}
\author[SNOLAB]{E.~V\'azquez-J\'auregui\fnref{unam}}\fntext[unam]{Current address: Instituto de F\'isica, UNAM, P.O. Box 20-364, 01000 M\'exico, D.F., M\'exico.}
\author[QU]{L.~Veloce}
\author[RHUL]{J.~Walding}
\author[QU]{M.~Ward}

\address[UofA]{Department of Physics, University of Alberta, Edmonton, Alberta, T6G 2R3, Canada}
\address[Carleton]{Department of Physics, Carleton University, Ottawa, Ontario, K1S 5B6, Canada}
\address[LU]{Department of Physics and Astronomy, Laurentian University, Sudbury, Ontario, P3E 2C6, Canada}
\address[UNM]{Department of Physics, University of New Mexico, Albuquerque, NM 87131, United States}
\address[UNC]{University of North Carolina, Chapel Hill, NC 27517, United States}
\address[UPenn]{Department of Physics, University of Pennsylvania, Philadelphia, PA 19104, Unites States}
\address[QU]{Department of Physics, Engineering Physics, and Astronomy, Queen's University, Kingston, Ontario, K7L 3N6, Canada}
\address[RHUL]{Royal Holloway, University of London, Egham Hill, Egham, Surrey TW20 0EX, United Kingdom}
\address[SNOLAB]{SNOLAB, Lively, Ontario, P3Y 1M3, Canada}
\address[Sussex]{University of Sussex, Sussex House, Brighton, East Sussex BN1 9RH, United Kingdom}
\address[TRIUMF]{TRIUMF, Vancouver, British Columbia, V6T 2A3, Canada}
\address[YU]{Department of Physics, Yale University, New Haven, CT 06520, United States}

\begin{abstract}
The DEAP-1 low-background liquid argon detector was used to measure scintillation pulse shapes of electron and nuclear recoil events and to demonstrate the feasibility of pulse-shape discrimination down to an electron-equivalent energy of 20~\kevee.

In the surface dataset using a triple-coincidence tag we found the fraction of $\beta$ events that are misidentified as nuclear recoils to be $<1.4\times 10^{-7}$~(90\% C.L.) for energies between 43--86~\kevee\ and for a nuclear recoil acceptance of at least 90\%, with 4\% systematic uncertainty on the absolute energy scale. The discrimination measurement on surface was limited by nuclear recoils induced by cosmic-ray generated neutrons. This was improved by moving the detector to the SNOLAB underground laboratory, where the reduced background rate allowed the same measurement to be done with only a double-coincidence tag.

The combined data set contains $1.23\times10^8$ events. One of those, in the underground data set, is in the nuclear-recoil region of interest. Taking into account the expected background of 0.48 events coming from random pileup, the resulting upper limit on the level of electronic recoil contamination is~$<2.7\times10^{-8}$~(90\% C.L.) between 44--89~\kevee\ and for a nuclear recoil acceptance of at least 90\%, with 6\% systematic uncertainty on the absolute energy scale.

We developed a general mathematical framework to describe pulse-shape-discrimination parameter distributions and used it to build an analytical model of the distributions observed in DEAP-1. Using this model, we project a misidentification fraction of approximately $10^{-10}$ for an electron-equivalent energy threshold of 15~\kevee\ for a detector with 8~PE/\kevee\ light yield. This reduction enables a search for spin-independent scattering of WIMPs from 1000~kg of liquid argon with a WIMP-nucleon cross-section sensitivity of $10^{-46}$~cm$^2$, assuming negligible contribution from nuclear recoil backgrounds.

\end{abstract}
\end{frontmatter}
\twocolumn
\section{Introduction}
The ability to separate electron-recoil ($\beta$-$\gamma$) interactions from nuclear-recoil interactions is critical for many nuclear and particle astrophysics experiments, including direct searches for dark matter particles.  Liquid argon provides very sensitive pulse-shape discrimination based on scintillation timing~\cite{boulay_astro}, and it is a favourable target for dark matter particle searches since it can be used to construct a very large target mass detector.  It is the target of choice in ArDM~\cite{ardm}, MiniCLEAN~\cite{miniclean}, DarkSide~\cite{darkside}, and WArP~\cite{warp} detectors.  In this paper we present results on the pulse-shape discrimination of $\beta$-$\gamma$ events from nuclear recoils with the DEAP-1 liquid argon detector, substantially extending the initial analysis~\cite{d1arxiv}.

Argon has many desirable properties as a scintillator, among them a high light yield of approximately 40 photons per keV~\cite{Miyajima:1974zz} and ease of purification, so that it can meet the radio-purity requirements of a rare-event search experiment. However, argon that is condensed from the atmosphere is known to contain cosmogenically-produced $^{39}$Ar, which undergoes $\beta$-decay at a rate of approximately 1~Bq per kg~\cite{benetti,Looslipaper}. The scintillation properties of liquid argon provide a method for discriminating these $\beta$-decays from WIMP interactions in the detector~\cite{boulay_astro}.

Scintillation in argon is a result of the formation of excited dimers after exposure to ionizing radiation~\cite{Mul70}. These occur in singlet and triplet states.  On decaying to the ground state, they emit light at a peak wavelength of 128~nm, lower in energy than the lowest excited atomic state~\cite{Ged72,hitachi}. The scintillation light can thus pass through pure argon without being absorbed.

The scintillation yield of nuclear recoils in liquid argon is quenched to about 0.25(2)~\cite{gastleretal} of the yield for electron recoils\footnote{We are aware of more recent results of approximately 0.29\cite{Cao2014,Creus2015}, which is slightly higher. Changed quenching factor does not affect our conclusions significantly.}. When referring to energies of nuclear recoils, units of either \kevee\ (``electron equivalent'') or \kevr\ are used, with the latter being the full energy of the recoil, and [\kevr]$=$0.25$\cdot$[\kevee].

The two argon dimer states have vastly different lifetimes, about 6~ns for the singlet and approximately 1.5~\micro s for the triplet state~\cite{hitachi}. Moreover, the relative population of singlet and triplet states is determined by the linear energy transfer (LET), such that fewer triplet excimers are produced at higher LET, and by the track structure of the exciting radiation~\cite{hitachi, doke}. With the large difference in lifetimes, the percentage of light signal in the first few tens of nanoseconds is a good estimate for the relative population of the singlet state, allowing for an effective way to discriminate between particles of different LETs, such as low-energy electrons and nuclei. The exact value of the triplet state lifetime is debated in the literature and values as low as 1110~ns~\cite{suemoto} and as high as 1590~ns~\cite{hitachi} have been reported (for a review of recent results see Ref.~\cite{Acciarri:2008kv}). 

Measurements of the pulse-shape discrimination of $\beta$-$\gamma$ events from nuclear recoils in liquid argon have also been reported in \cite{Lippincott:2008ad} and \cite{darkside}.  In this work the upper limit on the  $\beta$-$\gamma$ event misidentification probability is improved by a factor of $\sim$5, due to higher statistics.  A new improved analytic model for the pulse-shape discrimination parameter distribution, presented in Section~\ref{section:analytic}, is consistent with the data and provides a more general framework than the previously used ratio-of-Gaussians model from Ref.~\cite{d1arxiv}. \textcolor{\mk}{It has been applied to the case of a much larger detector.}

\section{Experimental apparatus}
\begin{center}
\begin{figure}[htb]
\includegraphics[width=3.5in]{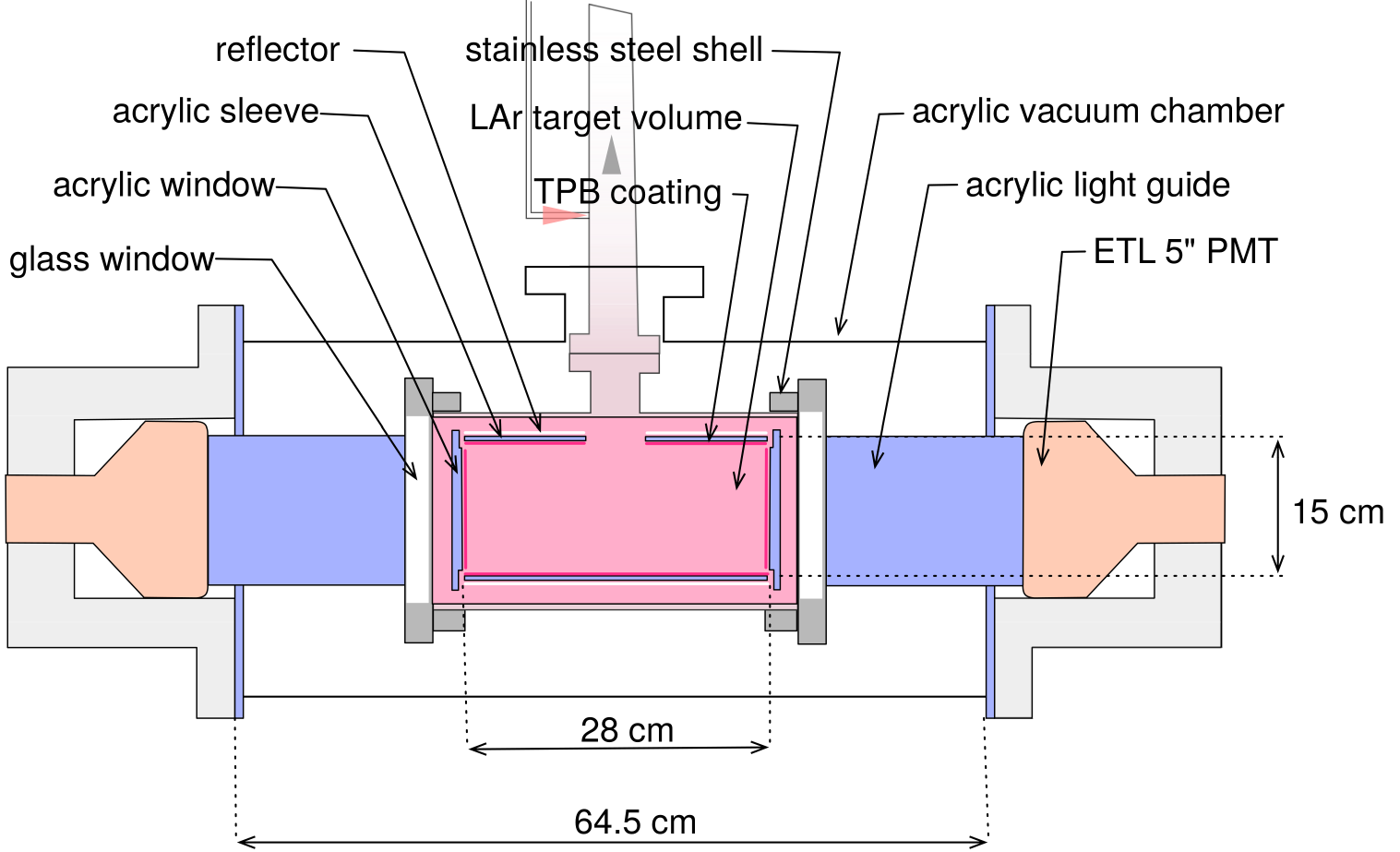}
\caption{DEAP-1 detector configuration.  Liquid argon is contained in a stainless steel target chamber, which includes an inner acrylic cylinder painted with or wrapped in diffuse reflector and coated on the inside with TPB.  Flow directions of liquid argon and boil-off vapor are indicated with arrows.  Scintillation light from the liquid argon is wavelength-shifted by the TPB and transmitted to PMTs operating at room temperature through PMMA acrylic lightguides. The PMTs (5-inch ETL 9390B) are held in place by spring-loaded polyethylene supports, which maintain contact between the PMTs and lightguides.}
\label{fig:deap1_detector}
\end{figure}
\end{center}

The target volume of the DEAP-1 detector (shown in
Fig.~\ref{fig:deap1_detector}) is a cylinder 28~cm in length and 15~cm
in diameter, containing 5.1~L (7~kg) of liquid argon at about 87~K. It
is defined by a 1/4-inch thick polymethyl-methacrylate (PMMA) sleeve
and two PMMA windows, which were coated on the inside, using standard
vacuum deposition techniques, with a roughly 0.1~mg/cm$^{2}$ thick
layer of the wavelength-shifter 1,1,4,4-tetraphenyl-1,3-butadiene
(TPB)~\cite{burtonpowell}. The TPB shifts the 128~nm liquid argon
scintillation light to the visible range so that it may pass the glass
and acrylic windows. In order to increase scintillation light
collection efficiency, the outside of the acrylic sleeve is coated in
TiO$_{2}$ paint (Bicron BC-620) for the first two datasets presented
here and wrapped in Gore\textregistered\ Diffuse Reflector for the
third dataset.

The target volume is contained inside a cylindrical stainless steel ultra-high vacuum shell with a 6-inch diameter glass window on each end. An 8-inch long cylindrical PMMA light guide rests against each glass window. The stainless steel shell and part of each light guide are located inside a 12~inch diameter PMMA acrylic vacuum chamber, and further thermally insulated by multi-layer super-insulation. The light guides are o-ring sealed to PMMA flanges on the insulating acrylic vacuum chamber, and the outer light guide face is at laboratory atmosphere and room temperature. A ETL 5" 9390B photo multiplier tube (PMT) is coupled to each lightguide using Bicron BC-630 optical gel and operated at room temperature.

The light guides allow the PMTs to be operated at room-temperature by thermally insulating them from the liquid argon target volume (the measured heat load in this configuration is approximately 7 watts per light guide) while at the same time transporting visible light from the target volume. The lightguides also moderate neutrons emitted from the PMT glass so that the background rate in the target volume is reduced. They are constructed from UV-absorbing (UVA) acrylic to minimize potential backgrounds from \v{C}erenkov radiation generated by cosmic-ray muons.

All detector components were selected to minimize radioactivity in the target volume, in particular neutrons from ($\alpha$,n) interactions in detector materials.

The argon is cooled by passing it through a 50-foot-long coil immersed
in a liquid-nitrogen tower. The nitrogen is maintained at an absolute
pressure of 32.5 to 33.5 psi. There is an auto-fill system connected to
an external 230~L Dewar to maintain nitrogen level. The argon passes through a
1/4-inch tube to the neck of the argon chamber. Argon which boils from
the top of the chamber flows in gas phase through an argon return line
to the top of the cooling tower, located above and slightly to the
side of the target chamber. During operation the gas-phase flow rate
of argon is 2.5 standard litres per minute through the return line.

The absolute pressure of the argon chamber was typically 13~psia which
corresponds to a temperature of 86.5~K. The temperature is higher in
the argon chamber than in the nitrogen tower because of heat loads
from conduction through the light guides, heating from the residual
gas in the insulating vacuum and heating by conduction through metal
components of the argon chamber. These processes are described
in~\cite{Lidgard}.

At the top of the tower there is an argon inlet port which is normally
valved off. During fill ultra-high purity (grade 5.0) argon is passed
through a SAES getter~\cite{SAES} at rates between 5 and 12
standard litres per minute. (The return line is valved off for fill
forcing the gas to pass through the cooling coil.) The argon at the
outlet of the purification system contains less than 1~ppb of
impurities. For the underground dataset, an additional charcoal
trap~\cite{Eoin}, maintained at -110~C, was used to remove
the trace amounts of radon in the argon.

The target chamber is surrounded by a neutron and $\gamma$-ray shield consisting of a minimum of two layers of ultra-pure water in 12-inch cubical polyethylene containers, for a total water shielding  thickness of 60~cm. A Geant4~\cite{geant4,geant4b} Monte-Carlo simulation showed that the water shield is sufficient to reduce nuclear recoil events in the liquid argon from external-source ($\alpha$,n) neutrons to less than one per year.

The data presented here come from three datasets: 67 days of operation
in the surface laboratory at Queen's University without any overburden
for cosmic-ray shielding; and $\simeq$ 1 month and $\sim9$ months of
measurements underground at SNOLAB~\cite{snolab} at a depth of 6000
meters water equivalent.  At that depth the cosmic-ray muon flux
passing through the detector is of the order of one per year, reduced
from approximately 10 per second at Queen's University. The first
dataset at SNOLAB was a period of testing of the data acquisition and
optimization of triggering. Between the first and second data sets at
SNOLAB the acrylic sleeve was replaced and the bicron paint was
replaced with Gore Tex reflector. 

Details on data analysis (see Sec.~\ref{sec:baseline}), data-quality
cuts (see Sec.~\ref{sec:cuts}) and detector stability (see
Sec.~\ref{sec:stability}) are included.  The surface and two
underground datasets had similar light yield and comparable stability,
which justifies extracting the upper limit on pulse-shape
discrimination from the combined dataset, presented in
Sec.~\ref{sec:snolab}.  For the sake of comparison with the analytic
model, discussed in Sec.~\ref{section:analytic}, surface and long
underground datasets are treated individually, in order to better
account for small differences in light yield, resolution, noise
parameters and associated systematic uncertainties. The short
underground dataset is not used in detailed fitting to the model.
We refer to the longer underground dataset as ``V1720 data'' for
reasons that will be apparent in the next sections.

\section{Data and analysis}
\subsection{Electronics and Trigger Configurations}

Signals from the PMTs are sent to Phillips 778 amplifiers. One output
of each amplifier is sent to a digitizer,
either LeCroy digitizing oscilloscope (1~sample per ns) or CAEN V1720 board (1~sample per 4~ns),
that samples the waveforms for 10~\micro\second, many times the
1.6~\micro\second\ lifetime of the longest component of scintillation
light.
The second output of the amplifier is passed to a linear
fan-out with one of the two outputs being used for discriminator
triggering and the other for more sophisticated triggers. The
discriminator for each PMT was set at 5~mV, which corresponds to
$\simeq 0.25$ of the mean single photoelectron (SPE) pulse height.  The ``DEAP-1" trigger is
a signal above discriminator threshold within $\pm 20$~ns in both
PMTs, corresponding to approximately 1~\kevee, well below the region
of interest for this study. The remaining output from the linear
fan-out for each PMT is passed to a summing amplifier and then to a
shaping amplifier with an integration time of 6~\micro\second. This
summed signal is passed to a single-channel analyzer (SCA) and used as
a veto for events far larger in energy than the region of
interest. Cutting these high-energy events allows us to increase the
effective data-taking rate by a factor of approximately 3. The DEAP-1
trigger with this veto imposed is called the ``DEAP-1 SCA'' trigger.

\begin{center}
\begin{figure}[htb]
\includegraphics[width=3.0in]{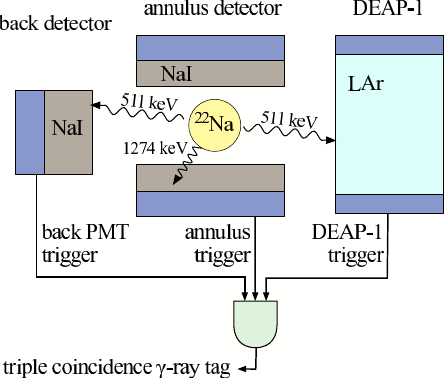}
\caption{Calibration source geometry.  The decay of $^{22}$Na is tagged by $\gamma$'s in the back NaI PMT, the annulus detector, and in DEAP-1, which allows a very low non-$\gamma$ background for calibration of the pulse-shape discrimination.}
\label{fig:annulus}
\end{figure}
\end{center}

The response of DEAP-1 to electromagnetic interactions in the liquid
argon is calibrated by a 10~\micro Ci encapsulated $^{22}\mbox{Na}$
source placed outside of the argon chamber.  The source emits a
positron and a 1274~keV $\gamma$-ray. The positron annihilates in the
source container producing back-to-back 511~keV $\gamma$-rays.  One of
the 511~keV $\gamma$-rays scatters in the argon. 

For the dataset at Queen's University we used two additional detectors to
``tag'' the backwards going 511~keV $\gamma$-ray and the additional
1274~keV $\gamma$-ray.  The source is placed on the axis of an annular
NaI detector with dimensions 3.75''$\times$11.5''$\times$12'' (inner
diameter $\times$ outer diameter $\times$ length) to detect the
1274~keV $\gamma$-ray. The annulus is divided into 4 sections along
its axis with a PMT on each section. The signals from the 4 sections
of the annulus are independently amplified (Phillips 778) and
discriminated (Phillips 705) with the threshold set just below the
1274~keV peak. The outputs of the discriminators are passed to a
logical OR (Phillips 775 with coincidence level 1) to generate the
``annulus trigger". The four annulus sections also have a
higher-energy discriminator which is used as a veto on the annulus
trigger to reduce backgrounds from cosmic rays.

A cylindrical NaI crystal of dimension 3.25" $\times$ 3" (diameter
$\times$ length) coupled to a PMT is placed at one end of the
annulus. This PMT signal is amplified and passed to a single-channel
amplifier (Ortec 420) which is centered on the 511~keV peak to detect
the ``backwards" going gamma ray. The ``back PMT" trigger is thus
generated. The ``global tag" is generated when an annulus trigger and
a back PMT trigger are coincident.  The geometry of this
triple-coincidence tag is shown in Fig.~\ref{fig:annulus}.  The
location of the source and back PMT are optimized to maximize the
coincidence rate with the annulus. The distance between the source and the
center of the DEAP-1 detector was $\simeq$17''. 

For background runs and AmBe neutron calibration runs the DEAP-1
trigger was used. (Some AmBe data was also taken with the DEAP-1 SCA
trigger to ensure that this trigger did not bias the acceptance
of high \fprompt\ events below 300~PE.) For $^{22}$Na data, runs were taken with either the
DEAP-1 trigger in coincidence with the tag, or the
DEAP-1 SCA trigger in coincidence with the tag. Runs using the DEAP-1
trigger with the tag allowed the measurement of the light yield at the
full-energy 511~keV gamma peak whereas runs with the DEAP-1 SCA
trigger took advantage of higher rate acquisition for PSD studies in
the region of interest~\cite{paradorn}. The timing of the triggering is such
that approximately 1~\micro s of the waveform is recorded before the
leading edge of the event. Furthermore, in tagged events, the leading
edge of the waveform measures the relative timing of the tag and the
measurement of light in the detector.  

For the datasets at SNOLAB we used only the cylindrical NaI crystal to
tag the ``backwards'' going gamma ray. Because the mechanical
constraints of the annulus were removed we were able to position the
source $\simeq 13$'' from the center of the argon cylinder.
In the first underground dataset the location of the
tagging PMT was varied from run to run as the tagging was optimized. The
tagging NaI crystal was then kept at approximately 2'' from the source for the
V1720 dataset.

\subsection{Data Flow}

For the dataset at Queen's University and the first data at SNOLAB,
the signal from each PMT is recorded with a Lecroy digitizing
oscilloscope (1 sample per na\-no\-second) at a ``high-gain" setting of
50~mV/division. When higher-energy signals are of interest the signals
are re\-corded at both the high-gain setting and a low-gain setting of
500~mV/division on a second channel.

A Linux data-acquisition computer runs an event builder that reads the
files in real time and converts them to a ROOT-based~\cite{root}
format. The root-based data files are automatically transferred to a
disk farm on the High-Performance Computing Virtual Laboratory
(HPCVL)~\cite{hpcvl} for offline analysis.  For the triple-coincidence
calibration, the rate of scattering events in the argon is
approximately 10~kHz, the rate of back-PMT and annulus coincidences is
approximately 100~Hz, and the triple-coincidence rate is approximately
20~Hz.

For the first dataset at SNOLAB, the oscilloscope was used with 1
sample per 1, 2, or 4 ns depending on the run. The V1720 dataset
at SNOLAB was recorded with a CAEN V1720 waveform digitizer recording
1 sample per 4~ns for 16~\micro\second.

\subsection{Analysis} \label{sec:baseline}

\begin{center}
\begin{figure}[h]
\includegraphics[width=3.5in]{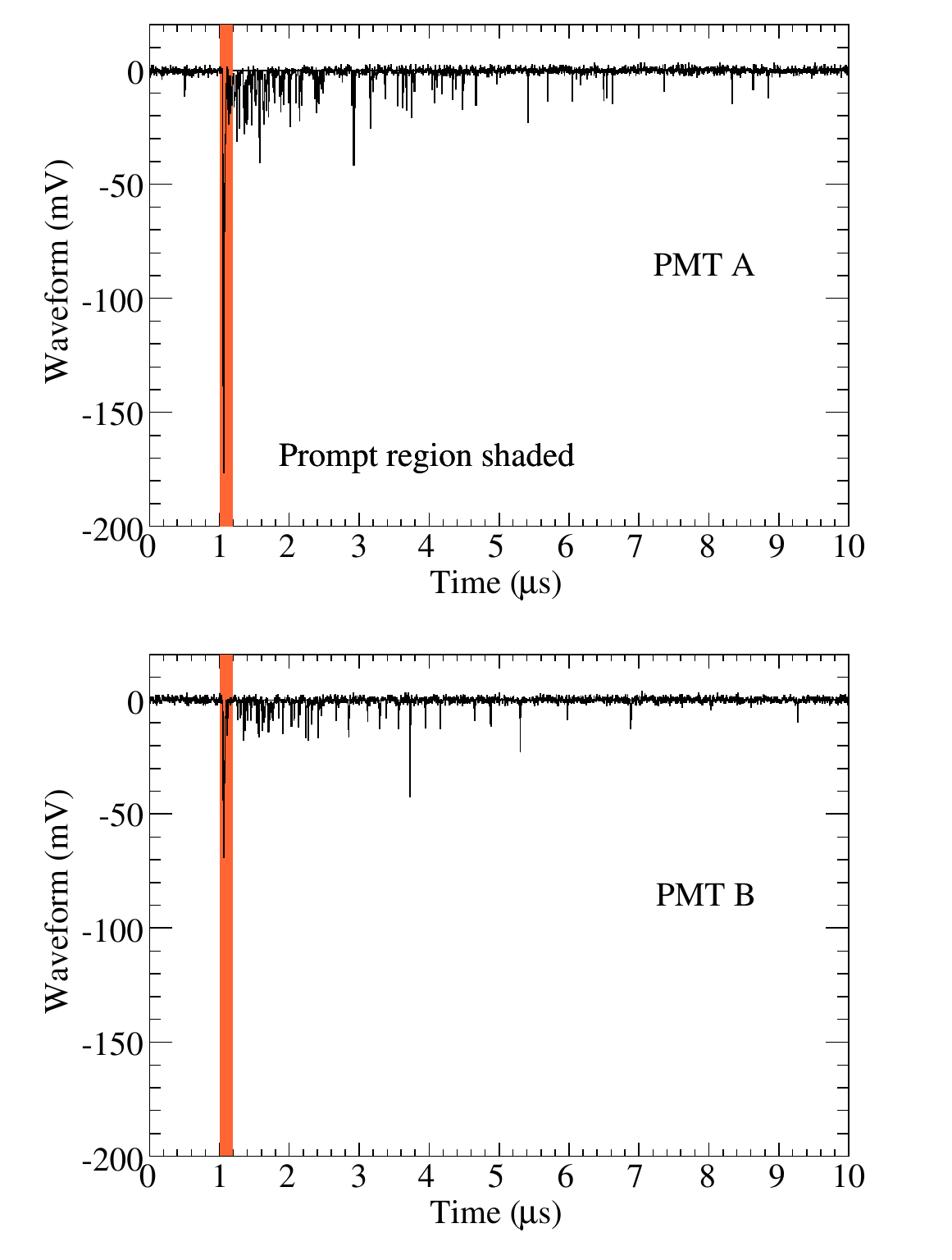}
\caption{A sample event from a $^{22}$Na PSD calibration run. Shown are the recorded traces for each of the two PMTs, labeled A and B. The shaded area shows the prompt region which begins 50 ns before the leading edge of the pulse and ends 150 ns after the leading edge.}
\label{fig:sample_event}
\end{figure}
\end{center}

Figure~\ref{fig:sample_event} shows PMT traces from a sample
$\gamma$-ray event in the region of interest from PSD calibration.  A
linear baseline correction is applied waveform\--by\--waveform: the
baseline is found in the first 0.5~\micro s of the waveform
trace. (This is done with an iterative procedure to remove regions
where random pulses occurred.)  That baseline is used to search for
signals above threshold in the final 4~\micro s of the pulse. Areas
around these signals are removed and a baseline is calculated at the
end of the pulse. A linear function is found from the two baseline
regions and subtracted from the total trace. The mean of the
waveforms in the pre-signal and late-signal region were calculated and
used as cuts in the analysis.

The leading edge for each signal is found using a voltage threshold of 5 mV, which corresponds to approximately one quarter of a photoelectron.  The prompt region of the pulse is defined from 50 ns before the edge to 150 ns after the edge. The late region is defined from 150 ns to the end of the waveform, $\simeq9\,\micro\second$ after the leading edge. The total charge for each PMT for both prompt and late regions is found and converted to units of photoelectrons using the single photoelectron charge calibration described below. Defining $\mbox{Q}^{\mbox{\tiny A,B}}_{\mbox{\tiny prompt,late}}$ as the number of photoelectrons (PE) in PMT A or B that are prompt or late, and $\mbox{TotalPE}$ as the total number of photoelectrons in the entire waveform for both PMTs, we write the fraction of prompt light as
\begin{equation}\label{eq:fpdef}
\mbox{F}_{\mbox{\tiny prompt}} \equiv \frac{  \mbox{Q}^{\mbox {\tiny A}}_{\mbox{\tiny prompt}} +  \mbox{Q}^{\mbox{\tiny B}}_{\mbox{\tiny prompt}} }{ \mbox{TotalPE}  }.
\end{equation}
The relative signals in the two PMTs are used to reconstruct the position of the event, \zfit, along the cylindrical axis of the detector,
\begin{equation}
\mbox{Z}_{\mbox{\tiny fit}} \equiv 35.2\,\mbox{cm} \times \frac{\mbox{Q}^{\mbox {\tiny A}} -\mbox{Q}^{\mbox{\tiny B}} }{ \mbox{TotalPE}},
\end{equation}
where $\mbox{TotalPE} \equiv \mbox{Q}^{\mbox {\tiny A,B}}_{\mbox{\tiny prompt}} + \mbox{Q}^{\mbox {\tiny A,B}}_{\mbox{\tiny late}}$ and 35.2 cm is the distance from the center of the cylinder to the front face of the PMT.
\begin{center}
\begin{figure}[h]
\includegraphics[width=3.2in]{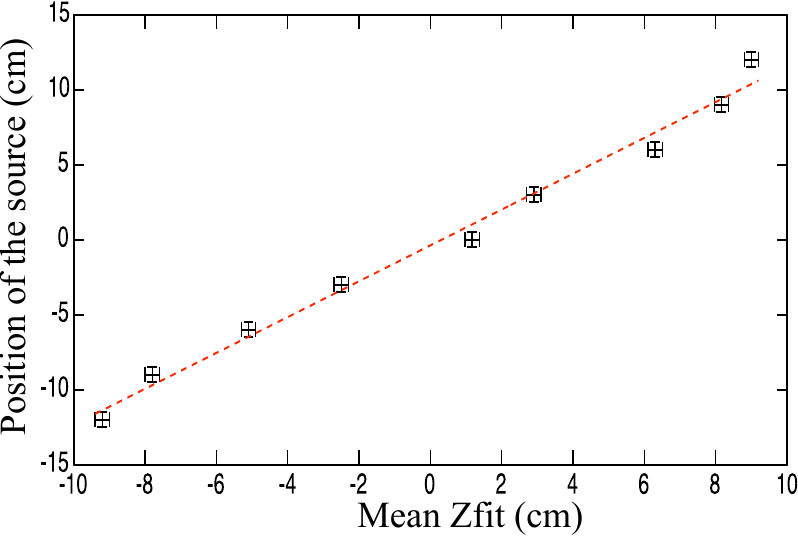}
\caption{Position of the 511~keV $\gamma$ source along the detector versus mean Zfit. 
The error bar on the source position corresponds to the physical size of the source. Horizontal error bars on mean Zfit 
come from the Gaussian fit to the data. Linear regression is shown in red.}
\label{fig:xfitcal}
\end{figure}
\end{center}
\begin{center}
\begin{figure}[h]
\includegraphics[width=3.5in]{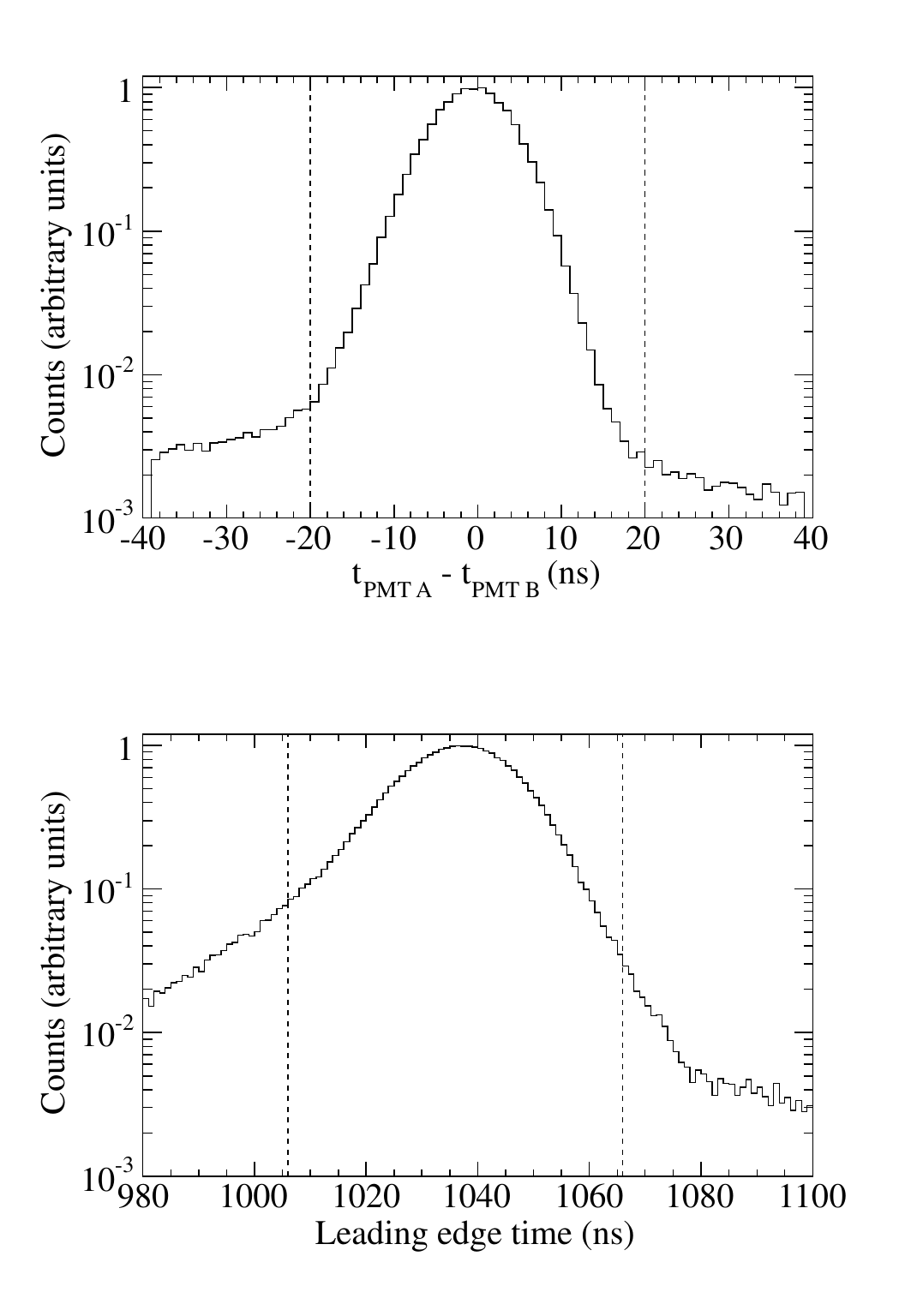}
\caption{The timing distributions for a typical surface dataset run for events with $120<\mbox{TotalPE} < 240$ and \zfit\ in the central 20~cm of the detector. The vertical lines indicate the cut values. For the top plot the cuts are always imposed at $\pm~20~\mbox{ns}$. For the bottom plot the mean value is determined by a Gaussian fit and the cuts are imposed at the mean $\pm~30~\mbox{ns}$.}
\label{fig:timing}
\end{figure}
\end{center}

The \zfit\ variable was calibrated using only
the back PMT and the DEAP-1 detector with the back PMT pulled farther
back to precisely define the direction of the forward gamma ray. The
$^{22}$Na source and back PMT were scanned along the axis of the
detector, with a typical calibration result shown in Fig.~\ref{fig:xfitcal}.
The spatial resolution depends on energy, and ranges from
$\sigma\approx1~\centi\metre$ for high-energy alphas, through $5~\centi\metre$
at 511~keV peak (approx. 1400~PE), to broader resolution of up to $10~\centi\metre$ in 120--240~PE window.

Because background rates were highest near the ends of the
detector, a cut was performed in this analysis to look only at events
in the middle of the detector, thus reducing the probability of an
accidental coincidence between a tag and a high-\fprompt\ event.
Example \zfit\ distributions and efficiency of the \zfit-based cut will be discussed
later.

The time between leading edges $\delta t$ for the two PMTs is
found. All data are
passed through a cut requiring $|\delta t|<20\,\mbox{ns}$. This cut
eliminated a small number of events in which the hardware trigger was
satisfied but the waveform was not characteristic of a single
scintillation event. This distribution is centered at a small negative
value that may be due to differing transit times in the two PMTs or
cable length.

The average leading edge time of the two PMTs, $t_{\mathrm{edge}}$,
is used in $^{22}$Na data to define a time window in software for accepting
events in DEAP-1 with respect to the time of the back-PMT signal. For
PSD analysis, a cut is imposed that accepts only events whose leading edge
time is within 30~ns of the mean time as found by a Gaussian fit to
the distribution. This cut was defined to include almost all
legitimate coincidences while reducing the time window for random
pileup of tags with high-\fprompt\ backgrounds, described later.
Figure~\ref{fig:timing} shows the timing distributions for a typical
PSD run. 

\begin{center}
\begin{figure}[h]
\includegraphics[width=3.5in]{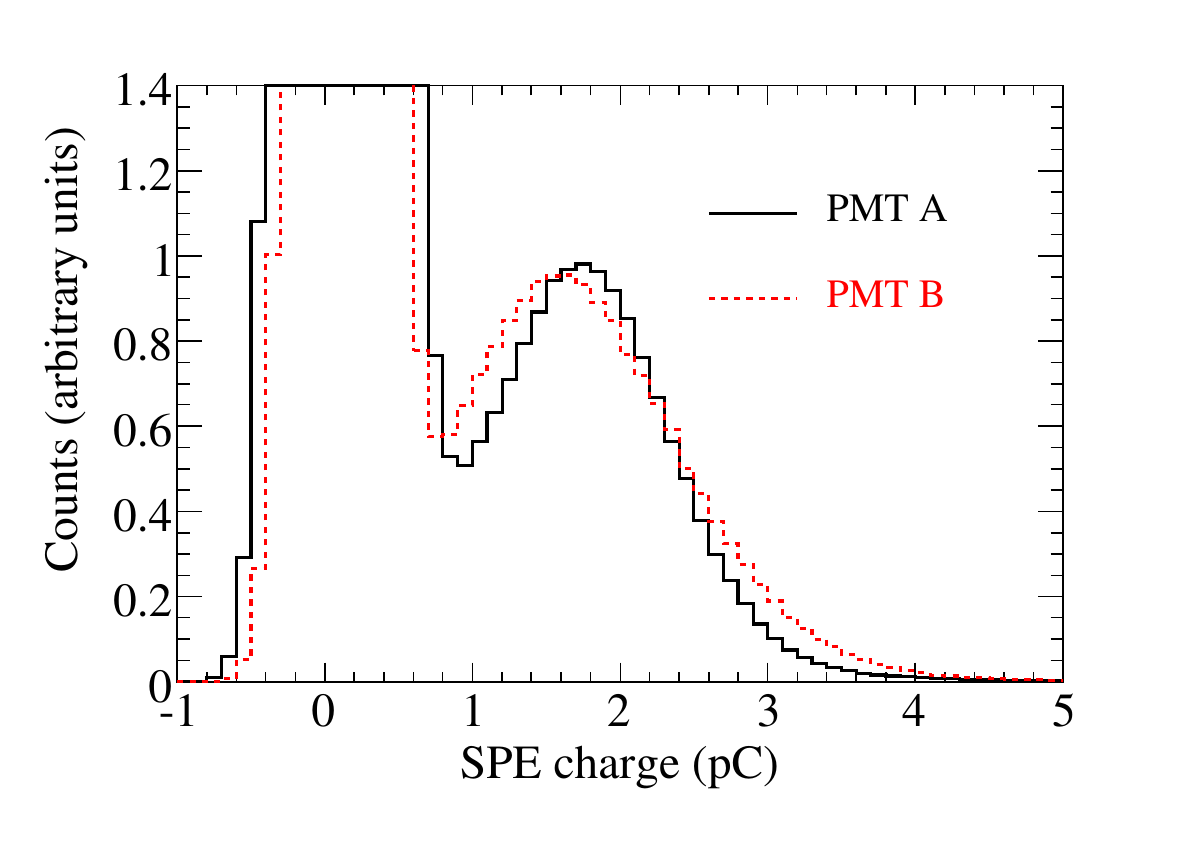}
\caption{Single photoelectron spectra of the DEAP-1 PMTs, used to calibrate the absolute light yield and number of detected photons.}
\label{fig:spe_hists}
\end{figure}
\end{center}

The mean charge from single photoelectrons (SPE) is derived from these data. Using only pulses with a low number of photoelectrons (below 200), we scan the final $4\,\micro\second$ of the pulses (high-gain channel only) looking for signals crossing a threshold of 5~mV, compared to a typical 20~mV height of an SPE pulse. These signals are integrated from 5~ns before the threshold crossing to 20~ns after the threshold crossing (a total of 26 samples). On a run-by-run basis these baseline corrected SPE charges are collected in a histogram. The mean charge is determined by the iterative procedure outlined in Refs~\cite{ford,sno_spe} that finds the mean between 0.4 and 2.5 times the previous value until convergence is achieved.
Figure~\ref{fig:spe_hists} shows histograms for a typical run. The mean charges for the two PMTs are 1.73 and 1.82~pC, with a variation throughout the running period of less than 4\%.

The main advantage of the above method is speed and robustness in tracking run-to-run variations even on low statistics or noisy data,
without any assumptions about the shape of the distribution.  For more accurate absolute SPE charge calibration, additional information
about the SPE charge distribution is needed, in particular about the under-amplified component of the SPE spectrum, induced by photons
hitting directly the first dynode, which manifests as a low-charge shoulder on the SPE peak, and generally overlaps with the pedestal.
As a cross-check, we exercised fitting the charge spectrum with an SPE model function chosen as a double peak gamma distribution or a Gaussian function\footnote{In later generations of DEAP-1~\cite{Pollmann:2012ad}, supplied with larger 8~in. Hamamatsu R5912-HQE PMTs, an LED pulser was installed to address the systematic uncertainties in the SPE charge calibration.}.
Ultimately, we assign 10\% systematic uncertainty to the absolute value of the mean SPE charge, which is motivated by comparison with
results returned by other SPE models.

\begin{center}
\begin{figure}[h]
\includegraphics[width=3.5in]{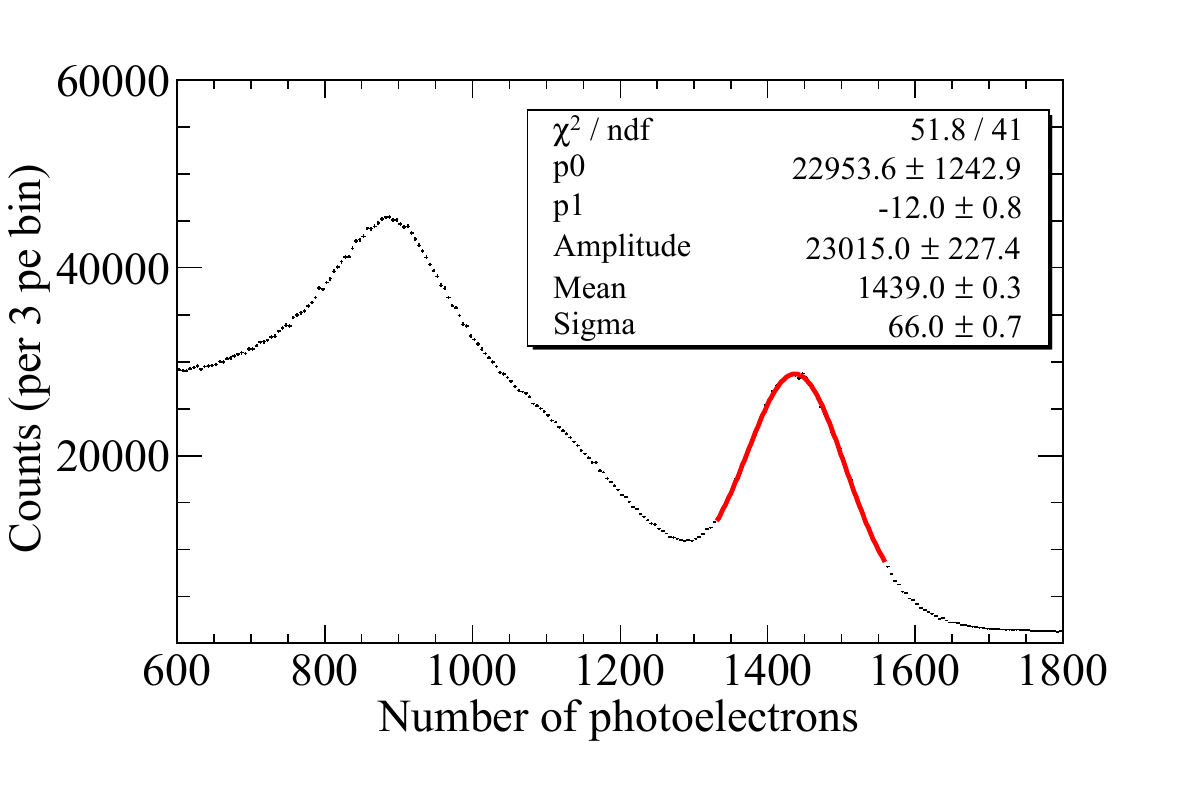}
\caption{The spectrum from the $^{22}$Na source used for energy calibration, zoomed in to show both the 511~keV annihilation peak and its Compton edge. A model function consisting of the first order polynomial
($p0+p1*x$) and a Gaussian function is used to fit the 511~keV peak and thus obtain an energy calibration.}
\label{fig:ecal}
\end{figure}
\end{center}

The energy response is calibrated throughout the data taking mainly with the 511~keV peak from $^{22}\mbox{Na}$. The position and 
the width of the 511~keV annihilation peak is determined from a fit to a superposition of a linear and a Gaussian function, as shown in Fig.~\ref{fig:ecal}.
The average light yield is approximately $2.8 \pm 0.1$ and $2.7 \pm 0.1$ photoelectrons per \kevee\ for the surface and 
underground datasets, respectively, where \kevee\ is the electron-equivalent energy. These numbers and uncertainties were estimated from weighted average of the $^{22}$Na peak positions over the entire dataset (for further details on the energy response stability, see Sec.~\ref{sec:stability}). The light yield has an additional
10\% systematic uncertainty from the SPE charge calibration, as discussed earlier, and a downward systematic uncertainty of $-0.2$~PE/\kevee\ at 59.5~keV (decreasing toward higher energies) due to a likely systematic over-counting of the number of photoelectrons (see Sect.~\ref{sect:meanfprompt}).

Light yield from $^{22}$Na calibrations is consistent with results of routinely performed calibrations with the 59.5~keV gamma from an Americium-Beryllium (Am-Be) source, with the position of the Compton edges associated with 1461 and 2615~keV gammas from natural 
radioactivity in the detector surroundings~\cite{Eoin}, and with 81~keV and 
356~keV $^{133}$Ba lines~\cite{Lidgard}.

\subsection{Effect of Data-Quality Cuts}\label{sec:cuts}
Data-quality cuts for the dataset taken underground at SNOLAB with the V1720 digitizers were studied in detail to ensure there was no bias
in the \fprompt\ distributions. 

The cuts were optimized to decrease the relative rate of accidental coincidences between
tags and random backgrounds based on detailed study of the backgrounds
in the detector. The cut conditions are given below (some cuts used in the surface dataset differ from the ones used in the V1720 underground dataset):
\begin{enumerate}
\item $120<\mbox{TotalPE}<240$
\item $-12~\mbox{cm}<\mbox{Z}_{\mathrm{fit}}<10~\mbox{cm}$ in the V1720 underground dataset or $\mathrm{abs}(\mbox{Z}_{\mathrm{fit}}+1.8)<10~\mbox{cm}$ in the surface dataset.
\item  $|\delta t|<20~\mbox{ns}$ (after time-zero calibration).
\item Leading edge time within $\pm$20~ns of mean in the V1720 underground dataset or within $\pm$30~ns of mean in the surface dataset.
\item Prompt and late charges for each PMT correspond to a positive number of photoelectrons.
\item The mean value of the leading edge baseline is
within 1~mV of the mean for the run (eliminates events with signal in
the leading baseline from pileup). The mean value of the
trailing-edge baseline is within 6~mV of the mean for the run.
\end{enumerate}

\begin{center}
\begin{figure}[t]
\includegraphics[trim=0 0 0 40, clip=true, width=0.97\columnwidth]{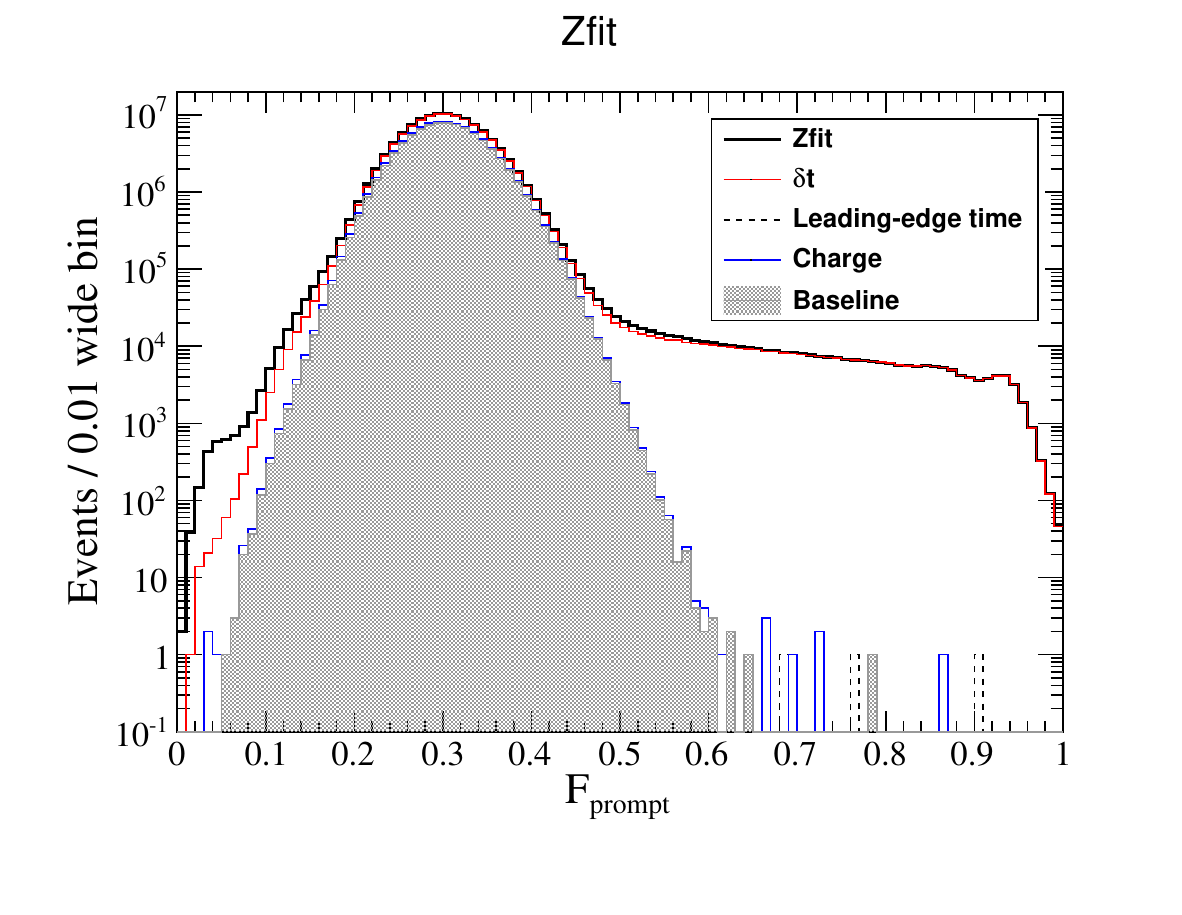}
\includegraphics[trim=0 0 0 40, clip=true, width=0.97\columnwidth]{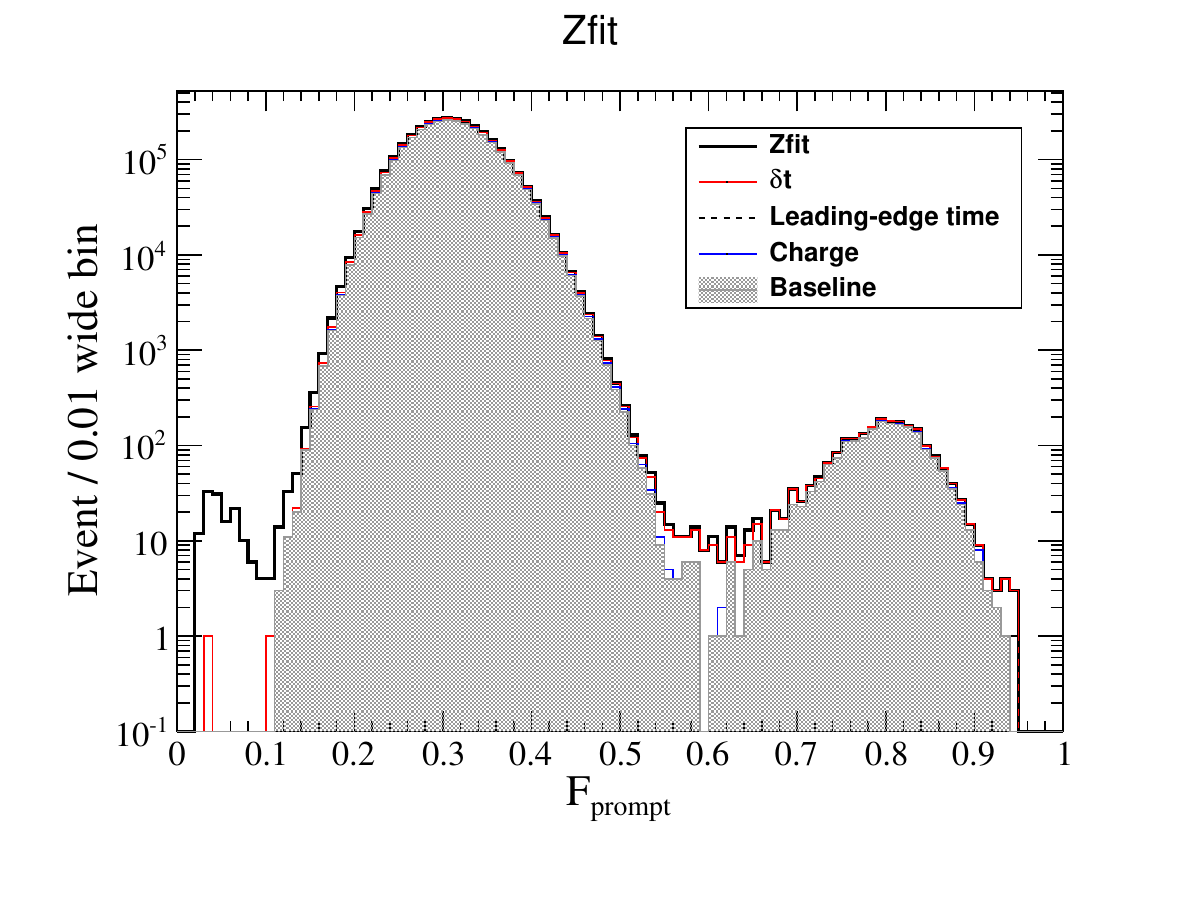} 
\caption{ The \fprompt\ distributions for all SNOLAB $^{22}$Na data (top)
  taken with the V1720 and for all Am-Be runs (bottom) taken with the
  V1720 shown after the incremental imposition of the cuts (enumerated in the
  text). For PSD running most high \fprompt\ events are eliminated by
  ensuring that the timing between the leading edge of the waveforms
  and the tag was as expected. The dashed entries were cut by ensuring
  all charges had physical values. The remaining unfilled entries were
  removed with cuts on the baseline values. One event, consistent with
  accidental coincidence, remains above 0.7. For neutron data the cut
  efficiencies are used in determining the neutron acceptance.} 
\label{fig:cuts_ambe}
\end{figure}
\end{center}
\begin{center}
\begin{figure}[h]
\includegraphics[width=3.5in]{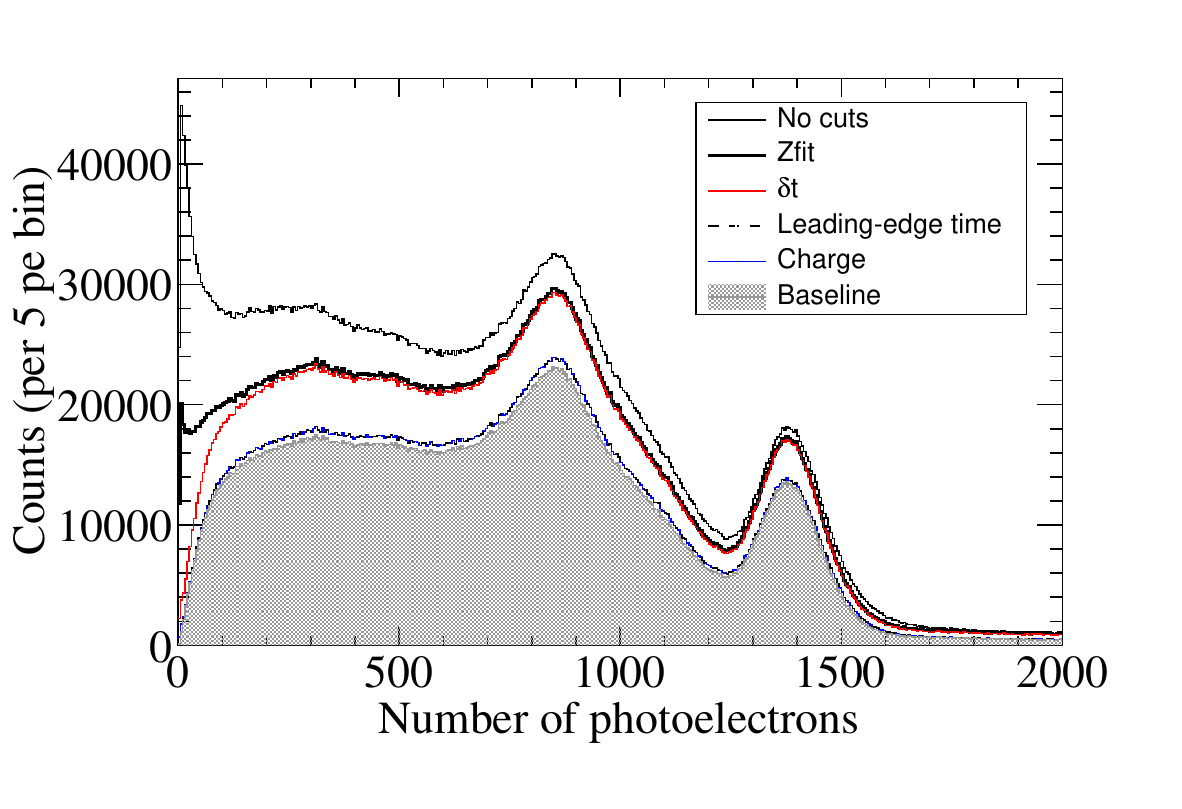}
\caption{ The TotalPE distributions for a full-spectrum $^{22}$Na run taken at SNOLAB
   with the V1720 electronics. Cuts are applied incrementally, as in Fig.~\ref{fig:cuts_ambe}.} 
\label{fig:cuts_pe}
\end{figure}
\end{center}
The effects of cuts are shown in Figure~\ref{fig:cuts_ambe} for $^{22}$Na and Am-Be neutron events, and in Figure~\ref{fig:cuts_pe} for one
full-spectrum $^{22}$Na run, both recorded with the V1720 digitizers.  The numerical values of events after all cuts for the Am-Be runs are shown in Tab.~\ref{tab:cuts_ambe}.
\begin{table}[h]
\begin{center}
\begin{tabular}{l|c|c}
 & \fprompt$<0.7$ & \fprompt$>0.7$ \\ \hline
TotalPE & $4.20\times 10^{6}$ & $2.54\times 10^{3}$ \\
\zfit & $3.23\times 10^{6}$ & $1.99\times 10^3$ \\
$\delta t$ & $3.15\times 10^{6}$ & $1.99\times 10^3$ \\
Leading-edge time & $3.01\times 10^{6}$ & $1.86\times 10^3$ \\
Charge  & $3.01\times 10^{6}$ & $1.86\times 10^3$ \\
Baseline  & $2.99\times 10^{6}$ & $1.84\times 10^3$ \\
\end{tabular}
\caption{The number of events surviving each cut for Am-Be calibration events is tabulated.}
\label{tab:cuts_ambe}
\end{center}
\end{table}

\section{Results}

\subsection{Surface run}
Data were collected with the DEAP-1 detector between August 20th and October 16th, 2007.  A total of 63,072,900 triple-coincidence $^{22}$Na calibration events were recorded for the PSD analysis.

\begin{center}
\begin{figure}[htb]
\includegraphics[width=3.5in]{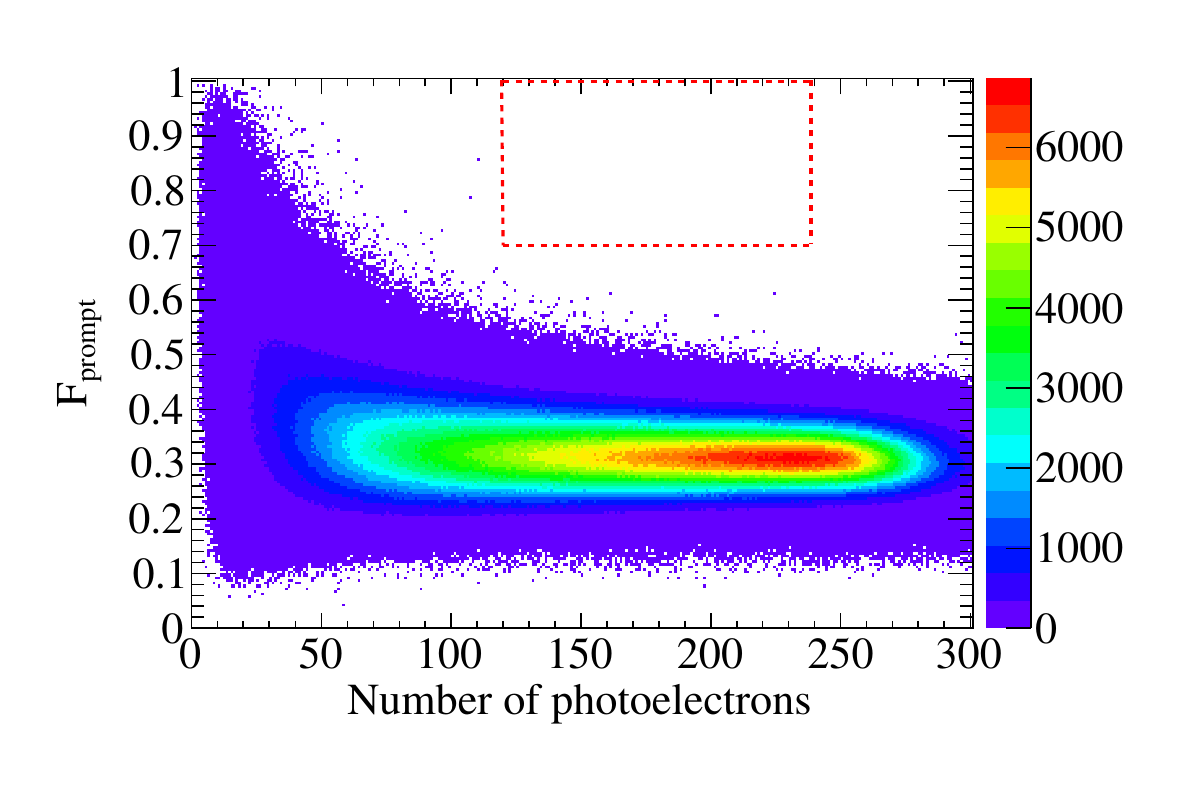}
\caption{\fprompt~ versus energy distribution for the triple-coincidence $\gamma$-ray events. The linear color scale (color online) reflects the number of counts per bin of 0.005 by 1~PE size.  The region between 120--240 photoelectrons for \fprompt~$>~0.7$ contains no events. The threshold was set based on the projected discrimination of $^{39}$Ar in the large detector (not based on this data set).}
\label{fig:data_fprompt}
\end{figure}
\end{center}

Figure~\ref{fig:data_fprompt} shows for these events the distribution of \fprompt, defined as in Eq.~(\ref{eq:fpdef}), versus number of photoelectrons. In this data set a hardware threshold imposed with the SCA restricts the high-photoelectron events to approximately 300 photoelectrons. Analysis of neutron calibration data shows that this does not remove high-\fprompt\ (0.7 $<$ \fprompt $<$ 1) nuclear recoil events.

\begin{center}
\begin{figure}[htb]
\includegraphics[width=3.5in]{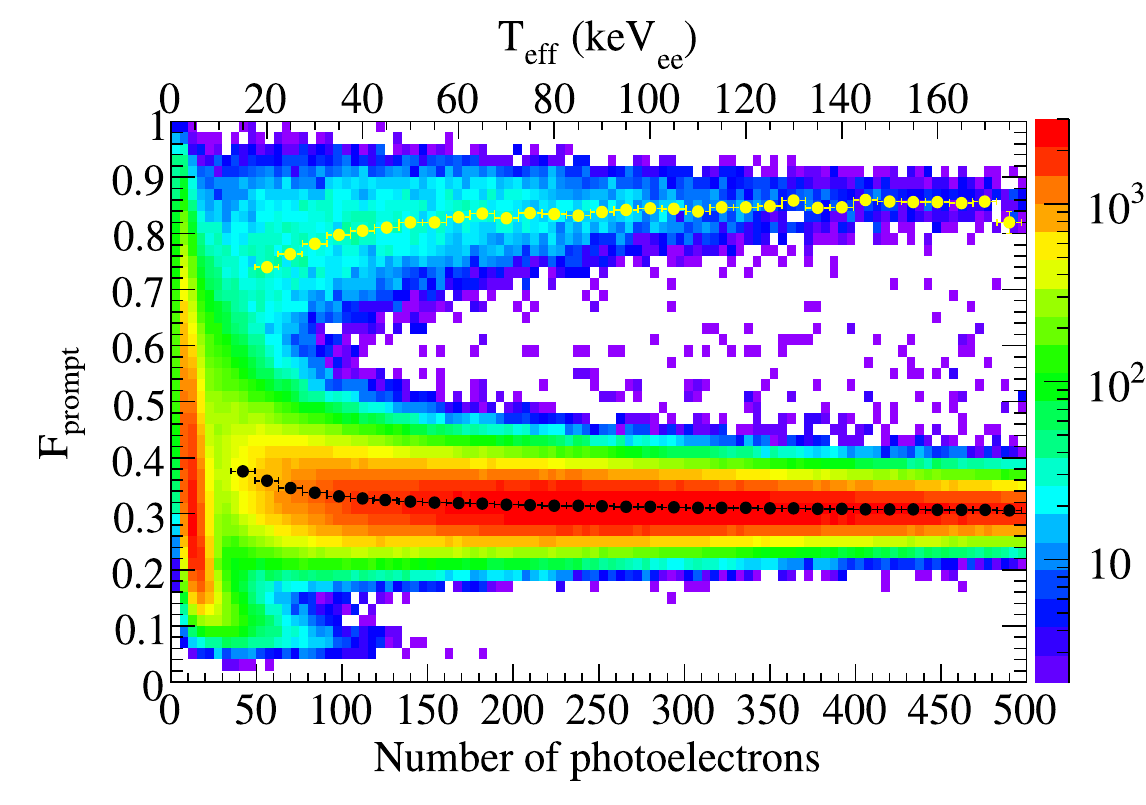}
\caption{\fprompt~ versus energy distribution for neutrons and $\gamma$ rays from an Am-Be calibration source. The logarithmic color scale (color online) reflects the number of counts per bin of 0.02 by 5~pe size.  The upper band is from neutron-induced nuclear recoils in argon. The lower band is from background $\gamma$-ray interactions. Gaussian fits to the \fprompt\ distribution  are performed separately for both bands, in each of 5~\kevee\ slices of the spectrum from 15--200~\kevee\ range for $\gamma$-ray and 20--200~\kevee\ range for neutron-induced events.  Fits are constrained to 0.2--0.5 and 0.65--1.0 \fprompt\ ranges for electronic and nuclear recoils, respectively, to avoid introducing significant biases due to the bands overlap. Resulting most likely \fprompt\ values are shown as black and yellow points for the lower- and upper-band, respectively. The change in the mean \fprompt\ values when going to lower energies likely dominated by a systematic effect (see Sec.~\ref{sect:meanfprompt}).
}\label{fig:neutron_fprompt}
\end{figure}
\end{center}

Figure~\ref{fig:neutron_fprompt} shows the \fprompt\ versus photoelectron distribution for neutrons and $\gamma$-rays from an Am-Be calibration source. We used this calibration dataset to determine the region of interest in \fprompt\ for nuclear recoil events. The position of the mean from the Gaussian fit (with \fprompt$>$0.65 restriction) in each bin is taken as the 50\%~recoil acceptance threshold, while higher acceptance thresholds are derived from fitted mean and standard deviation using Gaussian quantiles. This approach was chosen, as opposed to calculating the nuclear recoil band quantiles directly from the data, in order to avoid biases due to overlap with the electronic recoil band in the lowest energy bins. Using the fitted mean and sigma results in \fprompt\ corresponding to a given acceptance that is: (1) consistent to 0.01 with quantiles from the data for the entire 120--240~PE range and (2) conservative, i.e. always lower than from the direct method.

\begin{center}
\begin{figure}[htb]
\includegraphics[width=3.5in]{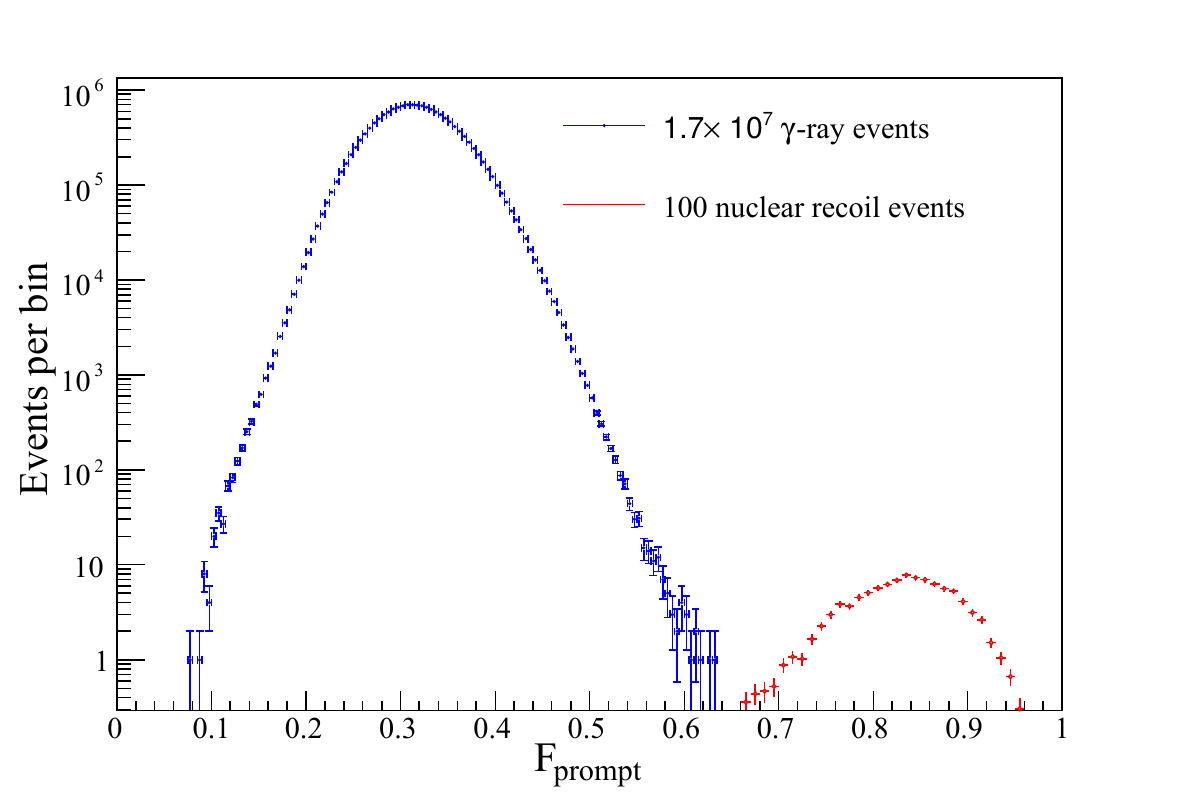}
\caption{\fprompt~ distribution for 16.7 million tagged $\gamma$-ray events from the $^{22}$Na calibration between 120 and 240 photoelectrons, i.e. approximately 43--86~\kevee\ (blue squares).  Also shown is the distribution of nuclear recoil events from the Am-Be calibration ($\gamma$-ray events from this source are not shown), normalized to the total number of 100 events (red crosses). Bin widths for $\gamma$-ray and nuclear recoil event distributions are 0.005~pe and 0.01~pe, respectively. No $\gamma$-ray events are seen in the nuclear recoil region.}
\label{fig:fprompt_gamma_n}
\end{figure}
\end{center}

As shown in Section~\ref{section:analytic}, we expect to use pulse-shape discrimination for signals of approximately 120~photoelectrons and above.  We therefore pre-determined the region of interest for demonstration of PSD to be 120--240 photoelectrons and 0.7$<$\fprompt$<$1.0, corresponding in DEAP-1 to approximately 43--86~\kevee\ and a nuclear recoil acceptance of not less than 90\% (uncertainties on the nuclear recoil band position and width are taken into account conservatively).  Figure~\ref{fig:fprompt_gamma_n}, which shows the \fprompt\ distribution for $\gamma$-rays and  nuclear recoils for events between 120 and 240 photoelectrons, was used to evaluate the PSD performance.  We find that none of the 16.7 million $\gamma$-ray events leak into the nuclear recoil region, and infer from these data that the PSD in liquid argon\footnote{Despite the fact that the $^{22}$Na deposited energy spectrum is softer than $^{39}$Ar $\beta$ spectrum, PSD for both event sources in DEAP-1 is nearly the same (see Sec.~\ref{sec:syst}).} at 90\%\ nuclear recoil acceptance is less than $6 \times 10^{-8}$.

We have also repeated this analysis with a lower photoelectron threshold and higher \fprompt\ region.  For a neutron detection efficiency of 50\%, there is one contamination event observed between
70--240 photoelectrons (approximately 25--86~\kevee), for an observed pulse-shape discrimination of $4.7 \times 10^{-8}$ in that range.

\subsection{Detector stability}\label{sec:stability}
Several tests of detector stability were performed.  

\begin{center}
\begin{figure}[htb]
\includegraphics[width=3.5in]{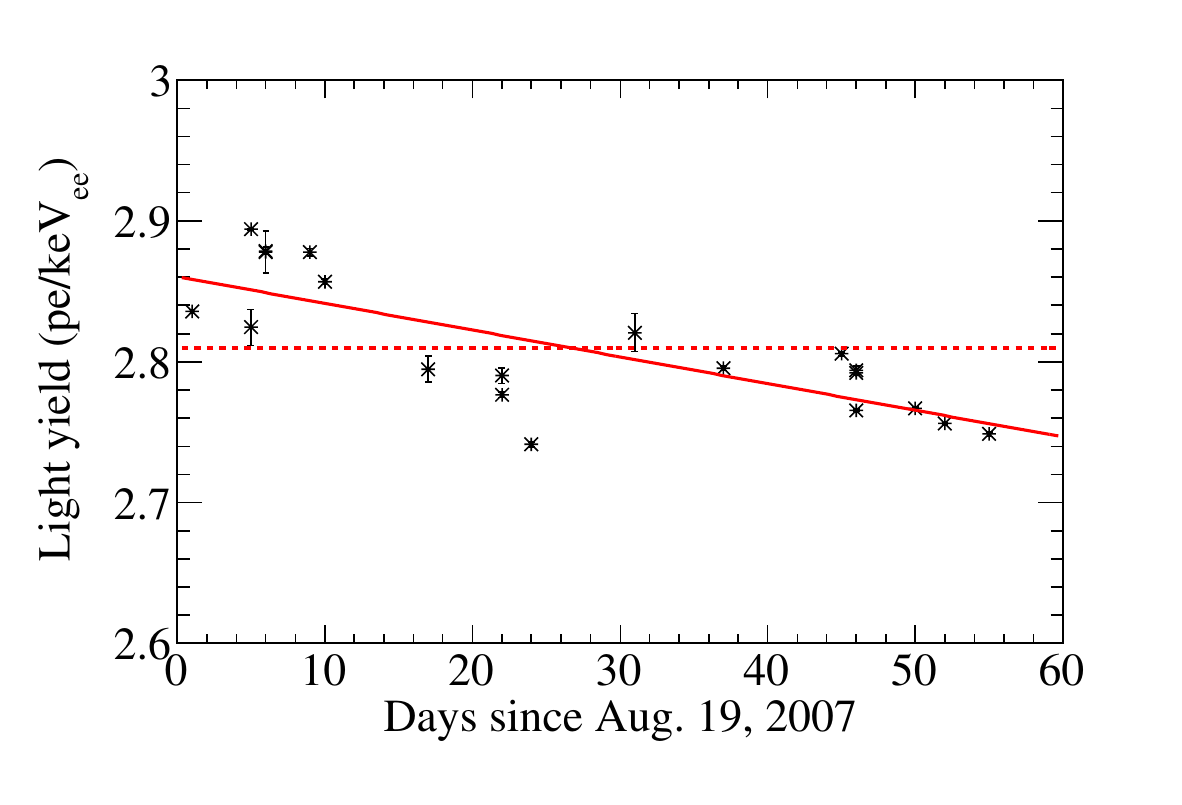}
\caption{Light yield stability during the surface run.  The average light yield is $2.8\pm0.1$~photoelectrons/\kevee, indicated with the dashed red line.  For each point shown are only the statistical uncertainties coming from the fit to the $^{22}$Na peak and used to calculate the weighted average. The scatter of the data points is within the systematic uncertainty of the single photoelectron calibration.  An indication of a degrading trend equivalent to light yield loss of (1.9$\pm$0.4)$\times 10^{-3}$ photoelectrons/\kevee/day is seen (red solid line).}
\label{fig:lightyieldtime}
\end{figure}
\end{center}

\begin{center}
\begin{figure}[htb]
\includegraphics[width=3.5in]{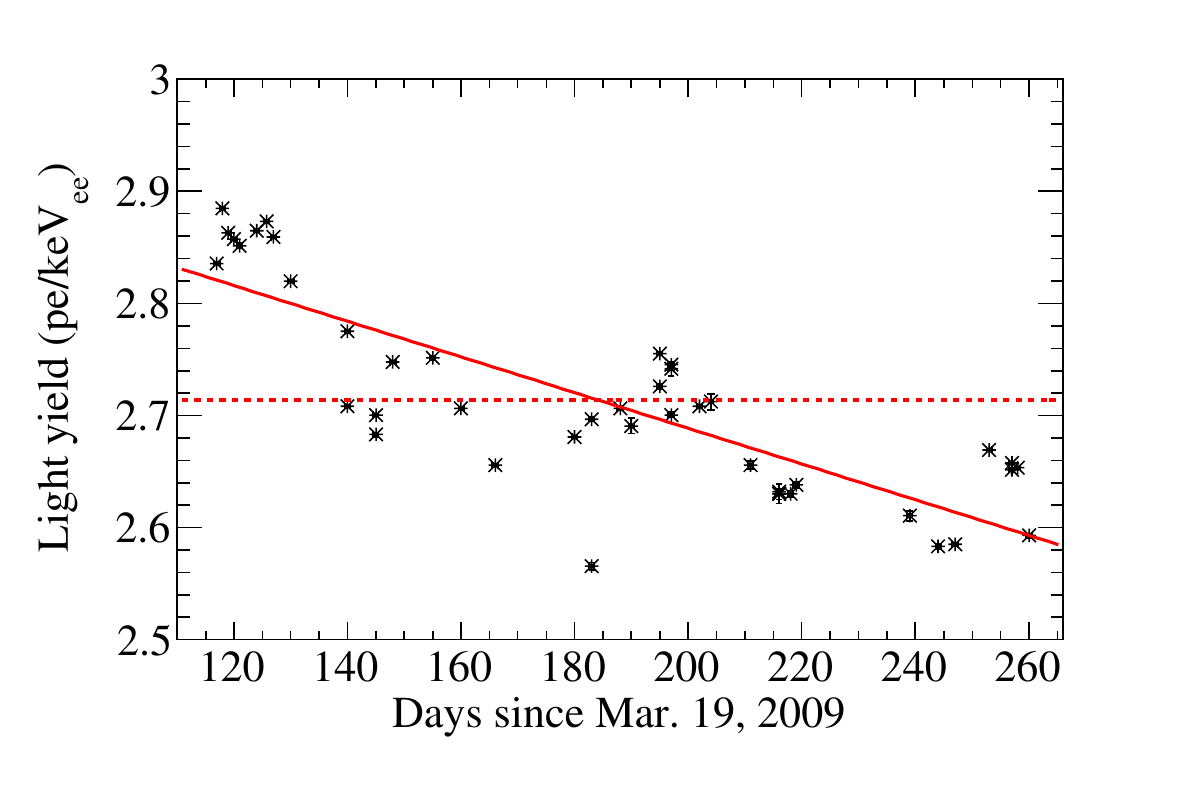}
\caption{Light yield stability during the run at SNOLAB. The average light yield is $2.7\pm0.1$~photoelectrons/\kevee, indicated with the dashed red line.  For each point shown are only the statistical uncertainties coming from the fit to the $^{22}$Na peak and used to calculate the weighted average.  The scatter of the data points is within the systematic uncertainty of the single photoelectron calibration.  An indication of a degrading trend equivalent to light yield loss of (1.5$\pm$0.2)$\times 10^{-3}$ photoelectrons/\kevee/day is seen (red solid line).}
\label{fig:lightyieldtimev2}
\end{figure}
\end{center}

The light yield is approximately $2.8 \pm 0.1$ and $2.7 \pm 0.1$ photoelectrons per \kevee\ for the surface and underground run, respectively (see Figures~\ref{fig:lightyieldtime} and \ref{fig:lightyieldtimev2}). It was stable within a few per cent throughout both running periods, although a slight degrading systematic trend can be seen, which is factored into the uncertainty through the weighted average.

Impurities in the argon can decrease the light yield by absorbing UV photons or by quenching argon dimers before they decay, thereby also decreasing the observed lifetime of the triplet state in liquid argon~\cite{Acciarri:2008kv, Acciarri:2008kx}. Therefore, we measured the triplet lifetime in DEAP-1 over the course of the run to check that impurities did not build up in the detector over time and to look for possible correlations with the light yield.

We use $^{22}$Na calibration data to measure the triplet lifetime. For each calibration run, we find all events that pass the data cleaning cuts and contain over 200 photoelectrons. The raw traces for these events are aligned according to the measured trigger positions and summed. We then fit the following model to the average trace between 500 and 3000~ns from the trigger:
\begin{equation}
f(t) = A \exp(-t/\tau_3) + B,
\end{equation}
where $A$ is a normalization factor, $\tau_3$ is the triplet lifetime and $B$ is a constant baseline term.  There are systematic effects associated with the linear baseline correction discussed in Section~\ref{sec:baseline}, which motivated including only the 500--3000~ns region in the fit, as in the later part of the average trace the exponential decay trend becomes skewed.

As a consistency check, we measured $\tau_3$ for photoelectron bins of size 200 between 200 and 1600 photoelectrons and did not observe any systematic effect from the signal size.  We estimated the size of the error associated with both the baseline correction and the fit window to be 40~ns by changing the start and end times of the fit by 500~ns. We performed the fit for both corrected and uncorrected traces and estimated the size of the error associated with the baseline to be 50~ns. We added the two estimated systematic errors to determine a combined systematic error of 60~ns.

\begin{center}
\begin{figure}[htb]
\includegraphics[width=3.5in]{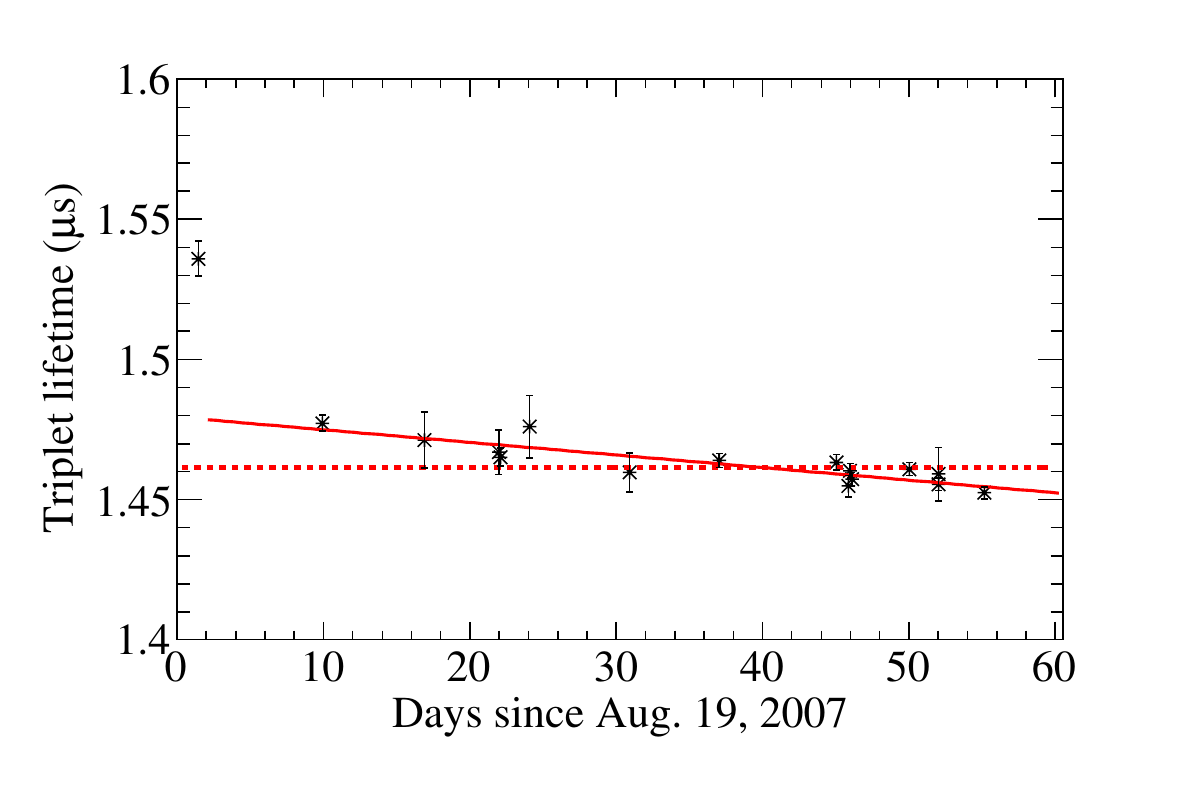}
\caption{Measured lifetime of the triplet component in liquid argon during the run on surface. Uncertainties shown are statistical only.  The averaged long time constant is 1.46$\pm$0.06~\micro\second~(dashed line).  The first data point lies significantly above the average value, which can be attributed to initial instability during the detector start up period.  The scatter in the data points is consistent with the systematic uncertainty of the measurement. A slight decreasing trend is seen, corresponding to the time constant decrease of 0.45$\pm$0.07~ns per day, if excluding the first data point (solid line).}
\label{fig:TripletLifetime}
\end{figure}
\end{center}

\begin{center}
\begin{figure}[htb]
\includegraphics[width=3.5in]{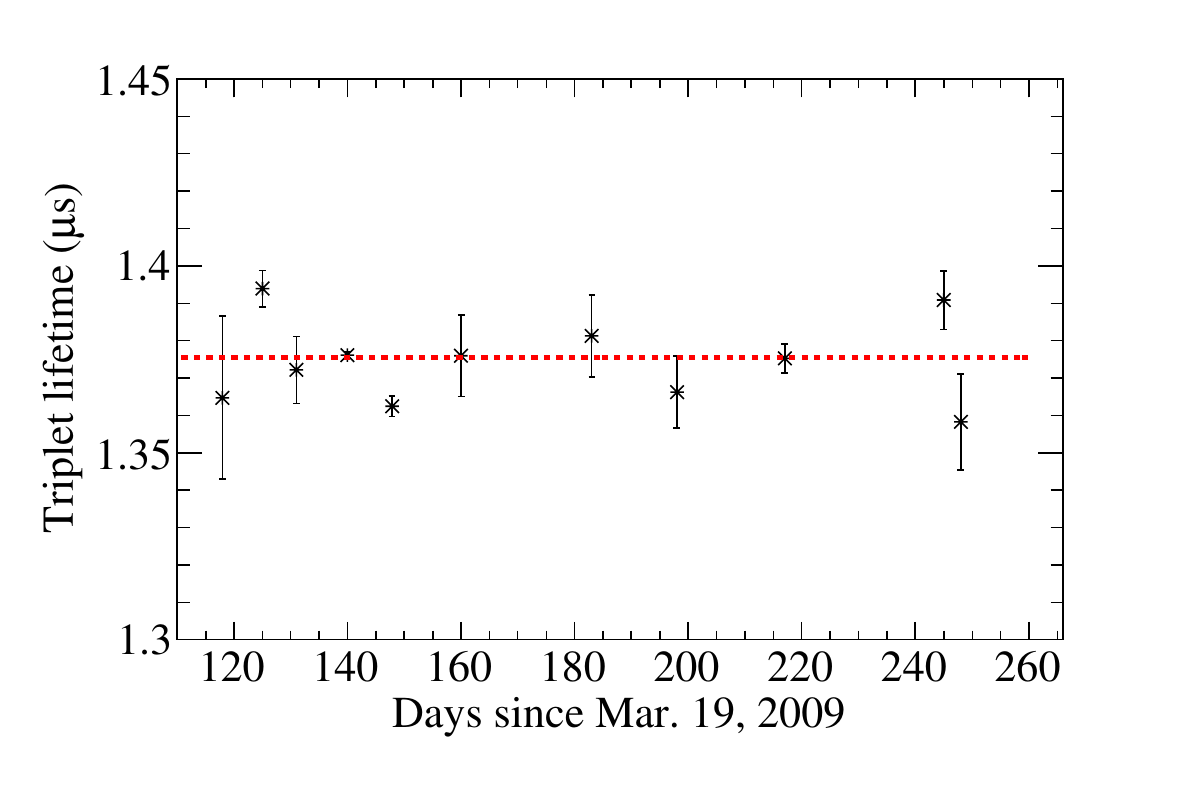}
\caption{Measured lifetime of the triplet component in liquid argon during the run at SNOLAB.  Uncertainties shown are statistical only.  The averaged long time constant is $1.38\pm0.06$\micro\second~(dashed line). }
\label{fig:TripletLifetime_v2}
\end{figure}
\end{center}

The measured lifetimes over the course of the run for traces without the baseline correction are shown in Figures~\ref{fig:TripletLifetime} and ~\ref{fig:TripletLifetime_v2}, in which the error bars shown are statistical only.  We measure the average long time constant of $1.46\pm0.06~({\mathrm{syst.}})~\micro\second$ and $1.38\pm0.06~({\mathrm{syst.}})~\micro\second$ in the surface and the V1720 underground datasets, respectively, consistent with other reported values~\cite{hitachi, Acciarri:2008kv, Lippincott:2008ad}. 

The overall increase of the impurity level throughout the run would manifest as the triplet lifetime decrease.
We do see a slight trend in the surface run, which is below the absolute systematic accuracy of the measurement, therefore does not affect the uncertainty on the global run average quoted above. It is consistent with the time constant decrease of 0.45$\pm$0.07~ns per day. Based on a simple relation
\begin{equation}\label{eq:ex_ratio}
\fpromptmath = \frac{I_1+\alpha' I_3}{I_1+I_3},
\end{equation}
where $\alpha'=0.09$ is the fraction of the triplet light measured in the prompt region of the pulse and $I_1$ and $I_3$ are singlet and triplet intensities, respectively, the expected mean \fprompt\ shift due to this effect is approximately 2\% over 60 days of running and has no consequences for further analysis, as (a) it would not change the number of leakage events in the PSD region of interest  (b) in the analytical model a dispersion parameter, $b$, accounts for this instrumental effect, and (c) we verified that the mean \fprompt\ for electronic and nuclear recoils within each dataset remains stable within $\pm$1\% with respect to the global run average and that the PSD cut (0.7$<$\fprompt$<$1.0) corresponds consistently to the nuclear recoil acceptance of not less than 90\%.

Light yield reduction caused by the time constant shortening observed in the surface run is not expected to exceed 2\%, and can only partially explain the reduction in light yield for that run.  The above, together with good triplet lifetime stability in the underground run, suggests additional sources of the light yield reduction, such as the deterioration of detector optics or drifts in the PMT gains, imperfectly corrected for with the single photoelectron charge calibration.

The main purpose of the energy calibration with the $^{22}$Na source is determination of the absolute energy window in \kevee, where
the limit on pulse-shape discrimination (always derived for the nominal 120--240 photoelectrons region) is valid. The appropriate average 
light yield value is used for each dataset and, in addition to this, in order to account for the hints of light yield drift, we assign
a systematic uncertainty to the absolute energy scale, which reflects the difference in light yield between the beginning and the end
of the data taking period.

Thus, for the surface run we establish the position of the absolute energy window as 43--86~\kevee\ with
4\% systematic uncertainty and for the underground run as 44--89~\kevee\ with 6\% systematic uncertainty. 
When reporting the PSD limit from the combined dataset we will use the energy window position relevant for the underground run, as it
conservatively covers both cases.

\subsection{High-$F_{\mathrm{\it prompt}}$ backgrounds in the surface run and at SNOLAB}\label{sec:hfprates} 
There is a a non-zero probability of a random coincidence between genuine nuclear recoils and double- or triple-coincidence tags. A second source of high-fprompt backgrounds which are linked to the calibration source will be discussed in Sec.~\ref{sect:pileuprandom}. The nuclear-recoil 
backgrounds are dominantly coming from alpha decays occurring on the inner detector surfaces (with the nuclear recoils depositing energy in 
liquid argon and alphas entering the chamber wall) and from alphas scintillating in liquid argon in areas of the detector where the light
collection is poor (gaps between acrylic sleeve and windows, neck of the detector). Significant efforts were made to investigate and reduce both 
types of high-\fprompt\ events, which is discussed in detail in Ref.~\cite{Pollmann:2012ad}.

\begin{center}
\begin{figure}[htb]
\includegraphics[width=3.5in]{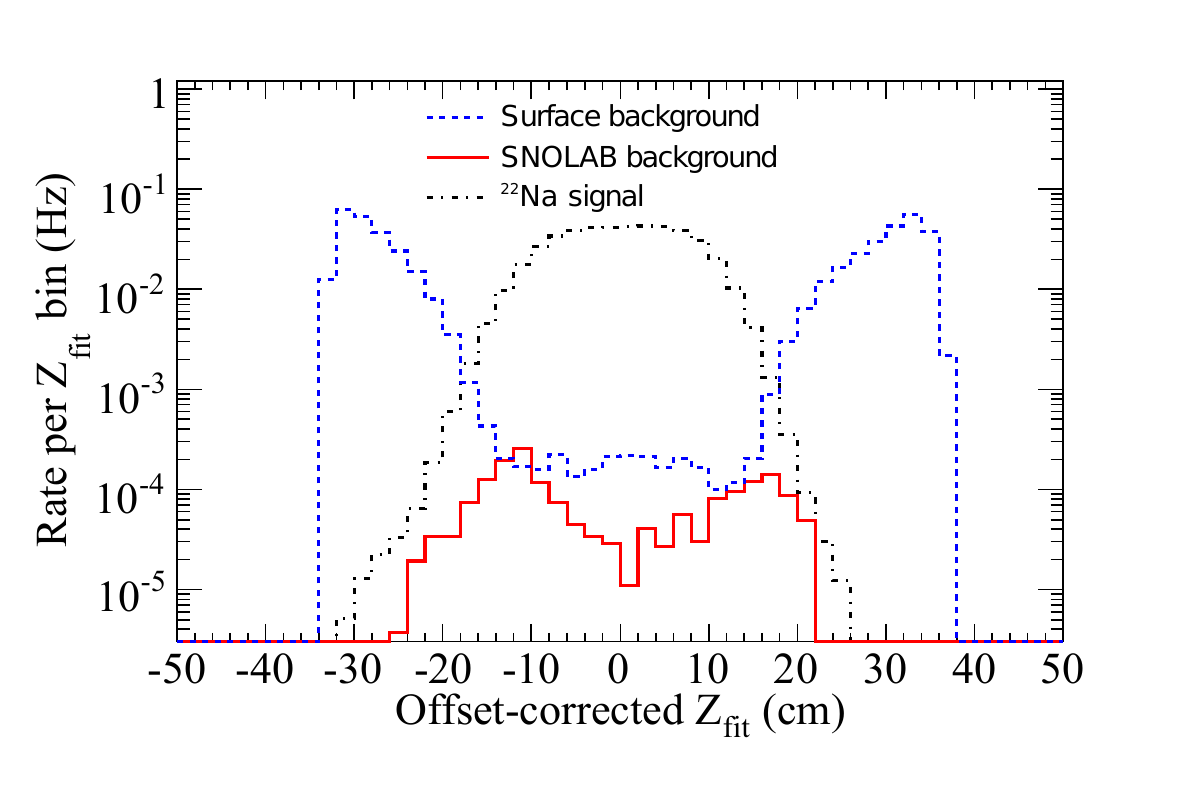}
\caption{Comparison of \zfit\ distribution for $\gamma$-rays from the $^{22}$Na data, and for high-\fprompt\ backgrounds during the surface run (labeled ``Surface background''), where the offset between \zfit\ value and the position in the detector has been subtracted. Also shown, for reference, is the distribution of high-\fprompt\ background events with the detector operating underground at SNOLAB.}
\label{fig:zfigbgcom}
\end{figure}
\end{center}

\begin{center}
\begin{figure}[htb]
\includegraphics[width=3.5in]{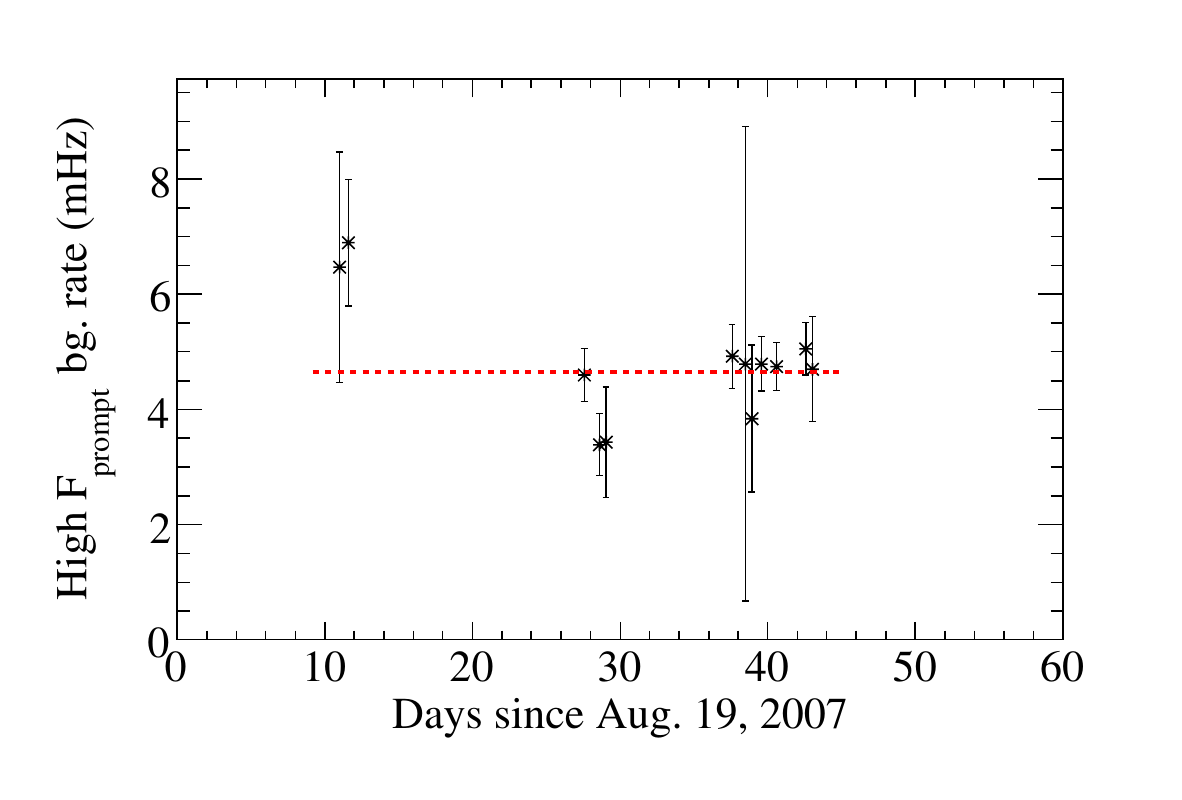}
\caption{High-\fprompt\ background event rate versus time. The average background rate is $4.6 \pm 0.2$ mHz.}
\label{fig:bgdate}
\end{figure}
\end{center}

High-\fprompt\ backgrounds are evaluated using the DEAP-1 trigger
without any calibration sources present.  Figure~\ref{fig:zfigbgcom}
shows the \zfit\ distribution of high-\fprompt\ background events,
compared to that of $^{22}$Na calibration $\gamma$-ray events.  Also
shown for reference is the \zfit\ distribution of
high-\fprompt\ background events with the DEAP-1 detector operating
underground at SNOLAB.  The reduced backgrounds underground allowed a
more sensitive measurement of PSD in argon, described in
Section~\ref{sec:snolab}.  The average background rate in the region
of interest (120--240 photoelectrons) measured for the surface data is
$4.6 \pm 0.2$~mHz, constant throughout the run as shown in
Fig.~\ref{fig:bgdate}. The rate was $1.01\pm0.1$~mHz for the first underground dataset and $0.29\pm0.04$~mHz for the V1720 dataset.


With the triple-coincidence tag, the measurement on surface is limited by random coincidences between the global tag (back-PMT and annulus) and high-\fprompt\ background events in DEAP-1.  The rate of high-\fprompt\ pileup during the $\gamma$-ray calibration run is
\begin{eqnarray}
\label{eq:pileup}
R_{\mathrm{bkg}}=R_{\mathrm{tag}} R_n \Delta t,
\end{eqnarray}
where $R_{\mathrm{tag}}$ is the global tag rate, $R_n$ is the rate of high-\fprompt\ background events measured in DEAP-1 and $\Delta t$ is the width of the time window from the software data-quality cut.

The achievable PSD can be found by dividing $R_{\mathrm{bkg}}$ by the rate of data acquisition in the energy region of interest (ROI), $R_{\mathrm{ROI}}$:
\begin{eqnarray}
  \label{eq:psd}
  D_{\mathrm{bkg}} = \frac{ R_{\mathrm{bkg}}}{R_{\mathrm{ROI}}}=\frac{R_{\mathrm{tag}} R_n \Delta t}{R_{\mathrm{ROI}}}.
\end{eqnarray}
$D_{\mathrm{bkg}}$ is thus the discrimination level where we would expect to find one background event.
Table~\ref{table:psd} shows the relevant parameters and the result from calculating $D_{\mathrm{bkg}}$, and it is below the discrimination level demonstrated in this work.

\begin{table}[htbp]
\centering
\caption{Background from random coincidence between the
  global tag and high-\fprompt\ background events in the
  argon. $R_{\mathrm{tag}}$ is the coincidence rate between the
  annulus and back-PMT with the $^{22}$Na source in place (back-PMT
  rate alone for the underground data), $\Delta t$ is the coincidence
  time window imposed by the analysis, $R_{n}$ is the rate of
  high-\fprompt\ background events in the liquid argon,
  $R_{\mathrm{ROI}}$ is the rate of triple-coincidence events
  (double-coincidence for the underground data) and $D_{\mathrm{bkg}}$
  is the PSD level where one radon-coincidence background event is
  expected. The numbers for the first underground dataset (labeled UG scope) are approximate in this table as discussed in the text. }
\begin{tabular}{l | c | c | c}
\hline
Variable & Surface & UG scope & UG V1720\\
\hline
$R_{\mathrm{tag}}$ (Hz) & 1000 & 7200 & 6000 \\
$\Delta t$ (ns) & 60 & 40 & 40 \\
$R_{n}$ (mHz) & 4.6 & 1.01 & 0.29 \\
$R_{\mathrm{ROI}}$ (Hz) & 18 & 74 & 40 \\
$D_{\mathrm{bkg}}$ & 1.53$\times 10^{-8}$ & $\simeq$4$\times 10^{-9}$ & 1.74$\times 10^{-9}$ \\
\hline
\end{tabular}
\label{table:psd}
\end{table}

\subsection{Underground run at SNOLAB} \label{sec:snolab}
After completion of the data runs at surface the detector was
reassembled underground at SNOLAB. The data presented here was from
two datasets: a smaller set collected in October 2008 and a larger set
collected between February and December 2009. The larger set
corresponds to the generation 1 detector described in
\cite{Pollmann:2012ad}.

The high-\fprompt~background rate in the larger dataset was reduced by
a factor of approximately 4.5~in the first underground dataset and a
factor of 15 in the second,  relative to surface
values. This allowed for a more sensitive measurement of PSD in argon.
The detector response including light yield for the surface, short
underground and long underground datasets were similar.

In the first underground dataset the setup was similar to surface with
three exceptions. First, the tag for the $^{22}$Na source was only
the back PMT and the annulus was not used. Second the $^{22}$Na data-taking
runs were performed with the PMT waveform sampling frequency
varied between 1000, 500 and  250~MHz  throughout the dataset. Third, the
distance between the tagging PMT and the source was varied within the
dataset to optimize signal to noise.

In the large underground dataset the waveform digitization was
performed with CAEN V1720 250~MHz waveform digitizers, which allowed
to reduce dead time and significantly increase the data taking
rate. The V1720 dataset underground was collected with a new inner
acrylic detector chamber and new TPB coatings which reduced background
rate.

\begin{center}
\begin{figure}[htb]
\includegraphics[width=3.5in]{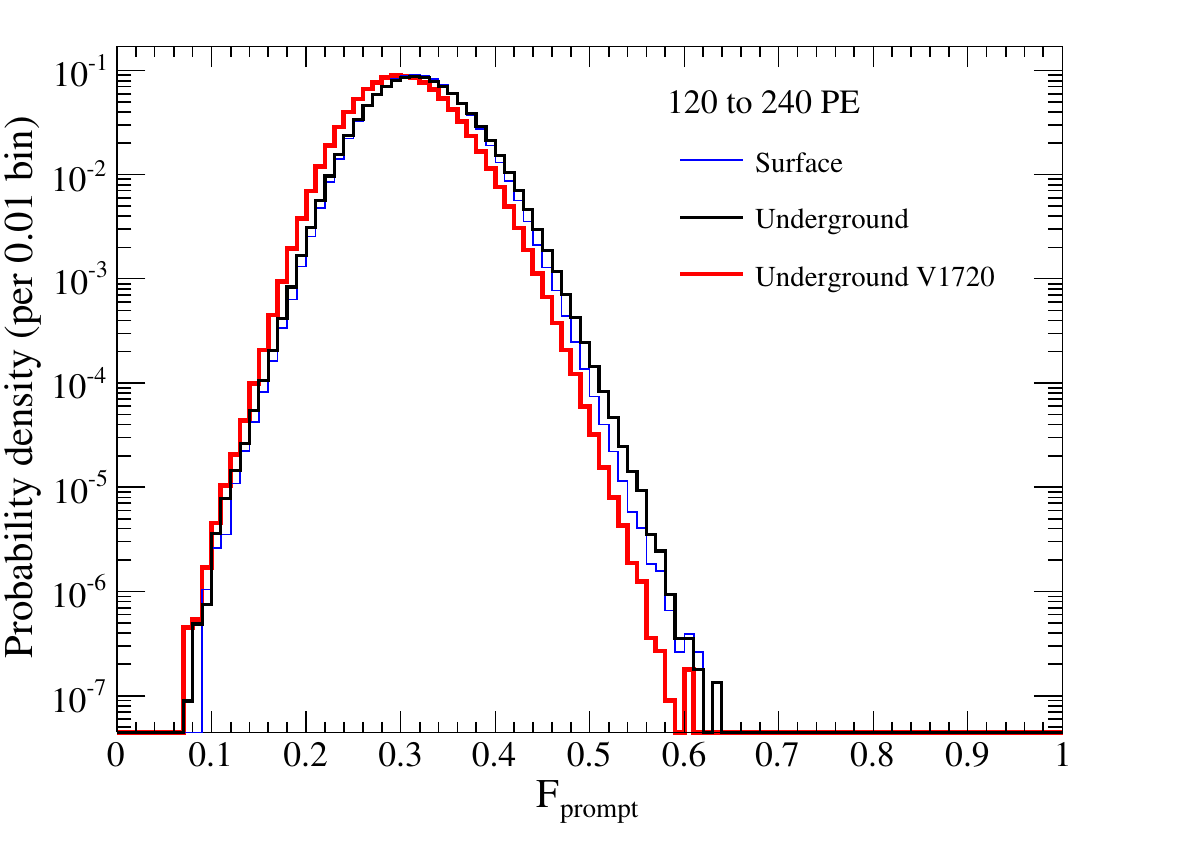}
\caption{\fprompt\ distribution, normalized to unity, for representative runs on surface (blue line), underground using the
original DAQ (black line) and underground using the V1720 (red line).
\label{fig:ug1}}
\end{figure}
\end{center}

Because the light yields are similar, the expected PSD response should
be the same from 120 to 240~pe.  Figure~\ref{fig:ug1} shows the
\fprompt\ distribution for representative runs on surface, underground
using the original DAQ and underground using the V1720.  The
distributions look similar, however the V1720 data is shifted by a few
per cent towards lower \fprompt\ values.  This shift is attributed to
different systematic late and prompt pe counting offsets, which are
discussed in Sec.~\ref{sect:meanfprompt}.  Additionally, about 1\%
shift can be attributed to improved energy resolution of the detector
used for the V1720 run derived from an improved baseline algorithm,
which effectively reduces the number of low energy events measured in
the region of interest.

The short underground dataset was taken with some variation in data
acquisition. It is included in the histogram of the combined
dataset but it is not used for detailed analysis of the shape of the
\fprompt\ distribution in subsequent sections.

The PSD cut (0.7$<$\fprompt$<$1.0) for all three runs corresponds
consistently to the nuclear recoil acceptance of not less than 90\%,
which allows us to combine them as below in order to extract the PSD
limit. In the next section they will be considered separately for the
sake of comparison with the analytic model.

Background runs using the DEAP-1 trigger only were used to measure
rates of high-\fprompt\ events between 120 and 240~pe.
Based on Eq.~(\ref{eq:pileup}), from the background and tagging rates 
one can determine the number of accidental coincidences of
high-\fprompt\ events in the data set (see Table~\ref{table:psd}).

\begin{center}
\begin{figure}[htb]
\includegraphics[width=3.5in]{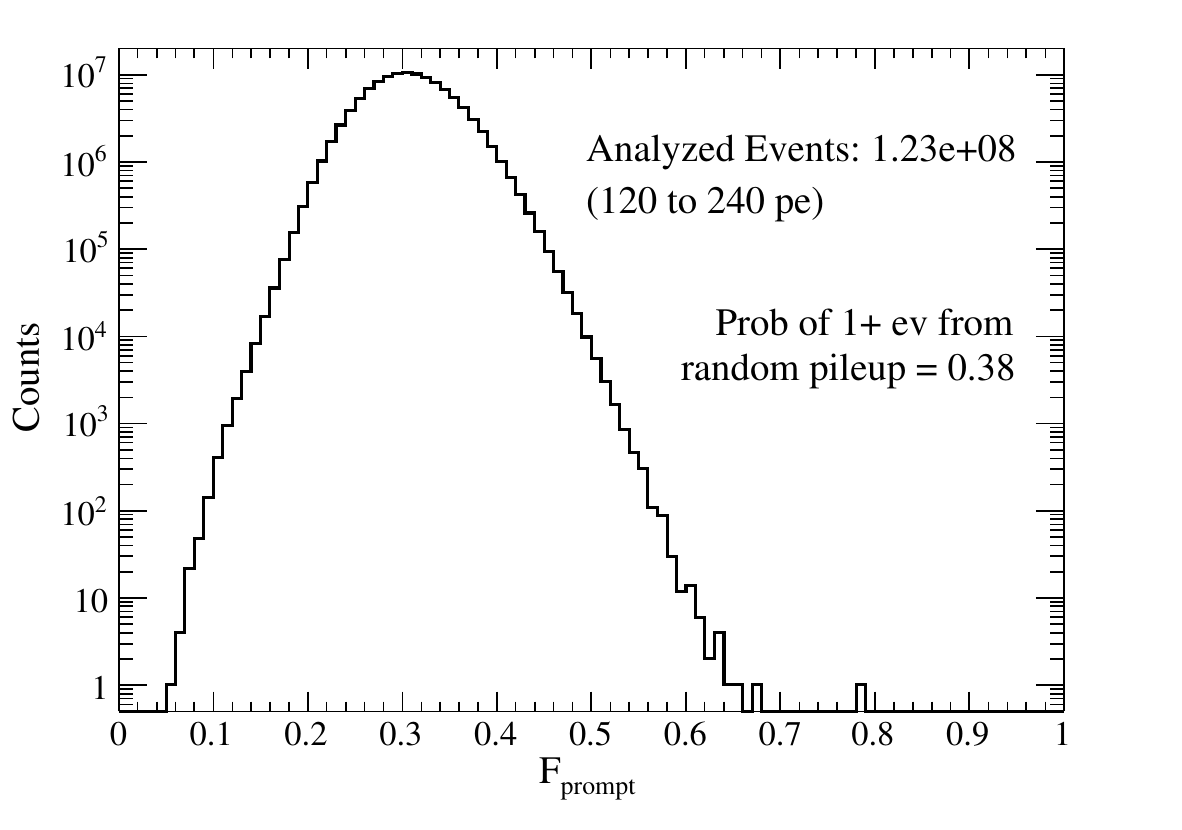}
\caption{The combined PSD data set from surface and underground runs at $\sim$2.7~pe/keV.
\label{fig:ug2}}
\end{figure}
\end{center}

\begin{center}
\begin{figure}[htb]
\includegraphics[width=3.5in]{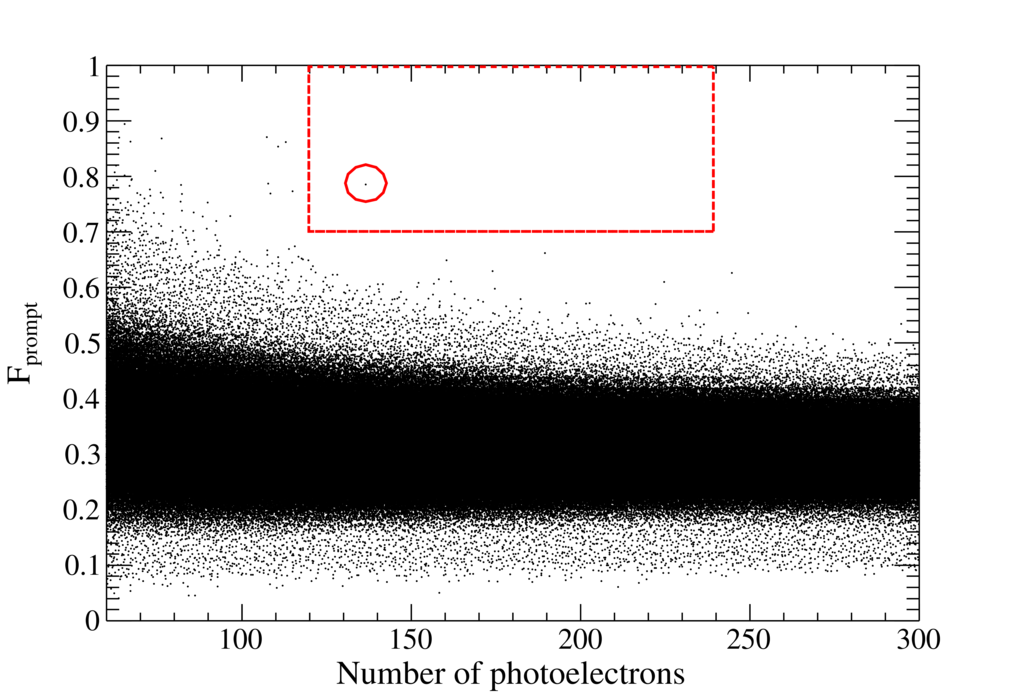}
\caption{\fprompt~versus energy distribution for the combined PSD data set containing triple- and double- coincidence $\gamma$-ray events, in surface and underground runs, respectively, at $\sim$2.7~pe/keV. The region between 120--240 photoelectrons for \fprompt$>0.7$ contains one contamination event, marked with a red circle. 
\label{fig:ug3}}
\end{figure}
\end{center}

The surface data set comprised $1.7 \times 10^7$ events between 120
and 240 pe with an expected pile-up background of 0.26. The first
underground data set comprised $2.2 \times 10^7$ events with an
expected background of 0.07 events. The second underground run with
the V1720 digitizers comprised $8.5\times 10^7$ events with an
expected background of 0.15 events. These numbers are slightly
different than in Table~\ref{table:psd} because they account for run-to-run variation
in tagging and trigger rates within the datasets. Thus the total expected pile-up
background is 0.48 events and the probability of obtaining one or more
pile-up events is 38\%. The entire PSD data set at $\sim$2.7~pe/keV
(surface and underground) is shown in Fig.~\ref{fig:ug2} and
Fig.~\ref{fig:ug3}. There is one event at high \fprompt, which is
consistent with random pile-up with a background event (such as a
nuclear recoil coming from a decay on the detector surface, or a
degraded alpha).

The above values are used together with the background expectations to
evaluate 90\%~C.L. upper limits assuming the number of measured
leakage events is Poisson distributed (using a standard method
described in Ref.~\cite{Cowan}). The results relevant for
90\%\ nuclear recoil acceptance (0.7$<$\fprompt$<$1.0) are:
$<$1.4$\times$10$^{-7}$ and $<$3.3$\times$10$^{-8}$ for surface and
underground runs, respectively, or $<$2.7$\times$10$^{-8}$ for both
runs taken together. While this result is based on $^{22}$Na calibration,
no significant difference in PSD power is expected for $^{39}$Ar $\beta$ events (see Sec.~\ref{sec:syst}).

\section{Analytic model for discrimination power} \label{section:analytic}

We developed a model that describes the probability  $R(f;N_{pe})$ to find \fprompt\ value $f$ at TotalPE $N_{pe}$, outlined in \ref{sect:fpromptgeneral} through \ref{sect:energymiscalibration}.

The model takes as inputs the probability distributions for the number of prompt and the number of late PE detected in response to an interaction with a particle of a given energy. For our dataset, we use beta-binomial distributions as per Eq.~(\ref{eq:betabin}), convolved with a Gaussian.

The beta-binomial distribution is a binomial distribution where the binomial probability $p$ is itself a random variable distributed as the beta distribution. Another way to interpret this is that the mean of the binomial distribution governing how many prompt or late PE are recorded for an event is not fixed, but picked randomly from a beta distribution for each event. The binomial probability in this case is the probability to turn a photon emitted in the interaction into a photoelectron, or in other words to detect it. The expectation value for $p$ must be $\bar{p} \simeq 2.7/40$ - the detector light yield divided by the photon yield in argon. By randomly varying $p$ as a beta distribution we account for changes in the light detection probability with event position and with time. Since in DEAP-1 the detection probability is so small, this step dominates the distribution over the original binomial distribution of the number of prompt and late photons emitted. We refer the reader to \ref{sect:parentdistr} for further discussion of this distribution.

The Gaussian distribution convolved into the beta-binomial models the noise affecting the PE-counts in the prompt and late windows due to the read-out electronics and uncertainty in the SPE charge calibration, with the combined standard deviations of $\sigma_{es,p}$ or $\sigma_{es,l}$. The final form of the uncorrelated prompt-PE and late-PE distributions is then
\begin{multline}
P_E(n_p)=\sum_{n'=0}^\infty \text{BetaBin}(n';\mu_p,\bar{p},b) \\
                  \times\frac{1}{\sqrt{2\pi}\sigma_{es,p}} \, \mathrm{e}^{-\frac{1}{2}\left(\frac{n'-n_p}{\sigma_{es,p}}\right)^2} \label{eq:binomqe}
\end{multline}
\begin{multline}
L_E(n_l)=\sum_{n'=0}^\infty \text{BetaBin}(n';\mu_l,\bar{p},b) \\
                  \times\frac{1}{\sqrt{2\pi}\sigma_{es,l}} \, \mathrm{e}^{-\frac{1}{2}\left(\frac{n'-n_l}{\sigma_{es,l}}\right)^2} \label{eq:binomle}
\end{multline}
where $\mu_p$ and $\mu_l$ are the mean values of the uncorrelated distributions and $b$ is a shape paramter (a measure of the dispersion beyond the pure binomial case).

The total variances of these distributions are, using Eq.~(\ref{eq:sigmabetabin}):
\begin{align}
\sigma_p^2 &= \sigma_{es,p}^2 + (1+1/b)\mu_p \label{eq:sigp} \\
\sigma_l^2 &= \sigma_{es,l}^2 + (1+1/b)\mu_l \label{eq:sigl}
\end{align}

From these uncorrelated distributions, we build the conditional distribution to have $n_p$ prompt PE given the event has $N_{pe}$ TotalPE, for events of energy $E$:

\begin{align}
P'_E(n_p;N_{pe}) &= P_E(n_p) \cdot L_E(N_{pe} - n_p) \label{eq:qprimemain} 
\end{align}

Since our calibration source is not mono-energetic, we build the sum over all energies, Eq.~(\ref{eq:qpe}), where we approximate the probability to detect an event of energy $E$ at $N_{pe}$ TotalPE, $T_E(N_{pe})$, as a Gaussian with width 
\begin{equation}
\sigma_t^2 = \sigma_p^2 + \sigma_l^2 \label{eq:eres} 
\end{equation}
and mean
\begin{equation}
\mu_t = \mu_p + \mu_l = E \cdot Y
\end{equation}
where $Y$ the light yield.

Window noise of width $\sigma_w$ is folded into the distribution as explained in \ref{sect:windownoise}.

The conditional distribution for $n_p$ prompt PE is then
\begin{multline}
P'(n_p;N_{pe}) = \sum_{n_p'=0}^\infty \sum_{\mu_t=0}^\infty  P_E'(n_p';N_{pe})\cdot N(\mu_t) \\
		\times \frac{1}{\sqrt{2\pi}\sigma_t} \, \mathrm{e}^{-\frac{1}{2}\left(\frac{N_{pe} - \mu_t}{\sigma_t}\right)^2} 
		\frac{1}{\sqrt{2\pi}\sigma_w} \, \mathrm{e}^{-\frac{1}{2}\left(\frac{n_p' - n_p}{\sigma_w}\right)^2} \label{eq:lastconvolution} 
\end{multline}

Eq.~(\ref{eq:lastconvolution}) is turned into a distribution for the \fprompt\ variable by variable transformation $n_p = f\cdot N_{pe}$:
\begin{eqnarray}               
R(f;N_{pe}) &=& N_{pe} \cdot P'\left(n_p=f \cdot N_{pe};N_{pe}\right) \label{eq:fpromptmodel}
\end{eqnarray}

It is a function of the means $\mu_{p},\mu_{l}$ and widths $\sigma_{p},\sigma_{l}$ of the uncorrelated prompt-PE and late-PE distributions, modified by the energy spectrum of the source and by window noise.

\subsection{Comparison with data} \label{sect:datacomparison}
The model from Sec.~\ref{section:analytic} was fit to the \fprompt\ vs TotalPE histogram (Fig.~\ref{fig:data_fprompt}) for the underground and surface data sets. Due to the large computational resources required to perform the 2D fit over all TotalPE bins, 12 1-PE wide slices every 20~PE, from 60 to 280 PE, were selected from the 2D distribution and the 2D fit included only those slices. In addition to the 8 overall fit parameters, each TotalPE slice has an individual normalization parameter in the fit.  To describe the \fprompt\ distribution for a range of TotalPE bins, the distributions of the individual TotalPE bins in the range were added together, where the normalization parameter for each bin was taken from the bin in the range that was included in the fit.

The start values and parameter ranges of the fit were determined using further instrumental inputs and constrains described in this section.

\subsubsection{Noise terms and energy resolution}
The uncertainty in the counting window of approximately 30~ns leads to an uncertainty in the late photoelectron count of up to $\epsilon_{\textrm{win}} = 2.5$\%, hence in Eq.~(\ref{eq:qprimemain}) ${\sigma_{\textrm{win}} \leqslant 0.025 \cdot \mu_t (1 - \bar{F_{p}})}$.

The detector hardware's non-uniformity contribution, $b$ from Eq.~(\ref{eq:binomqe}), is unconstrained.

The SPE noise contribution
for 1~PE is estimated to be $\epsilon_\text{spe} = 0.34$ from the SPE histogram shown in Fig.~\ref{fig:spe_hists} by fitting a Gaussian to the SPE spectrum and taking $\epsilon_\text{spe} = \sigma^P_{\textrm{spe}}/\mu^P_{\textrm{spe}}$. It goes with the measured number of photoelectrons such that $\sigma^2_{\textrm{spe}} = \epsilon^2_\text{spe} \cdot N_{pe}$.
\begin{center}
\begin{figure}[h]
\includegraphics[width=3.25in]{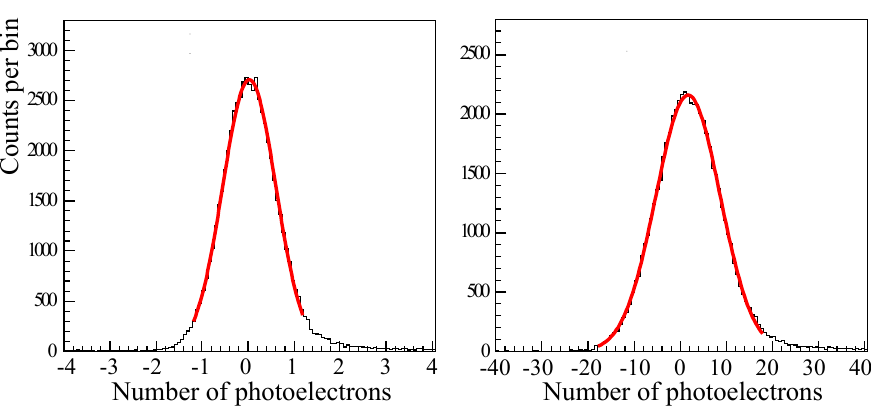}
\includegraphics[width=3.25in]{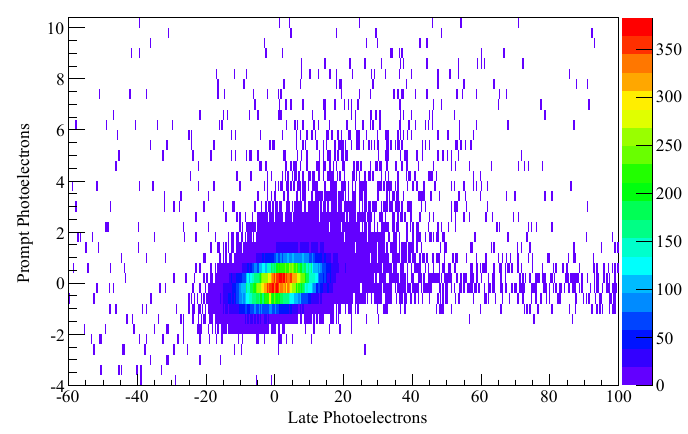}
\caption{Prompt (top left panel) and late (top right panel) electronic noise components from a run triggered with a random pulser, each fit with a Gaussian function (red line), with 0.07~PE and 0.7~PE wide bins, respectively. These contributions are representative of the noise generated by the hardware electronics and analysis algorithms. The late light, integrated over a 10~\micro s window is more susceptible to electronic noise than the short prompt window. Prompt and late electronic noise components are correlated (bottom panel).}
\label{fig:pulser}
\end{figure}
\end{center}

Electronic noise of $\sigma_{\textrm{elec,p}}$=0.58~photoelectrons in the prompt window and $\sigma_{\textrm{elec,l}}$=7.0~photoelectrons in the late window are measured by triggering the detector with an external pulse generator and analyzing the PMT traces, as shown in Fig.~\ref{fig:pulser}. Both noise contributions are significantly correlated (with correlation factor of $\rho^{LP}_{\textrm{elec}}\approx$0.5). The correlation in the noise introduces a small correlation between $P_E$ and $L_E$, which is disregarded. Since this correlation is positive and generally moves events between TotalPE bins, it is not of the same quality as the window noise and cannot be treated in the same framework.

In the V1720 underground dataset, an additional zero-sup\-pres\-sion step was used in order to remove periodic low-amplitude noise picked up by the electronics and not
seen in the previous dataset. The zero-sup\-pres\-sion algorithm also significantly suppressed electronic noise in general, however, a discontinuous profile of the residual
noise does not translate into a meaningful measurement of $\sigma_{\textrm elec}$ to use as the analytic model input and benchmark against $^{22}$Na data.

The electronic and SPE noise add up to the noise in Eqs.~(\ref{eq:binomqe}) and (\ref{eq:binomle})
\begin{equation}
\sigma_{es}^2 = \sigma_{\textrm{elec}}^2 + \sigma^2_{\textrm{spe}}
\end{equation}
and the total variance of these distributions is then
\begin{equation}
\sigma^2 = \sigma_{\text{BetaBin}}^2 + \sigma^2_{\textrm{spe}} + \sigma^2_{\textrm{elec}}
\end{equation}
or written out
\begin{align}
\sigma_p^2 &= (1+1/b+\epsilon^2_\text{spe}) \bar{F_{p}} \mu_t + \sigma_{\textrm{elec,p}}^2 \label{eq:sigmap}  \\
\sigma_l^2 &= (1+1/b+\epsilon^2_\text{spe}) (1 - \bar{F_{p}}) \mu_t+ \sigma_{\textrm{elec,l}}^2 \label{eq:sigmal} 
\end{align}
where we made use of Eq.~(\ref{eq:sigmabetabin}).

The energy resolution function employed in the model is composed from the prompt and late noise as shown in Eq.~(\ref{eq:eres}).
\begin{center}
\begin{figure}[h]
\includegraphics[width=3.5in]{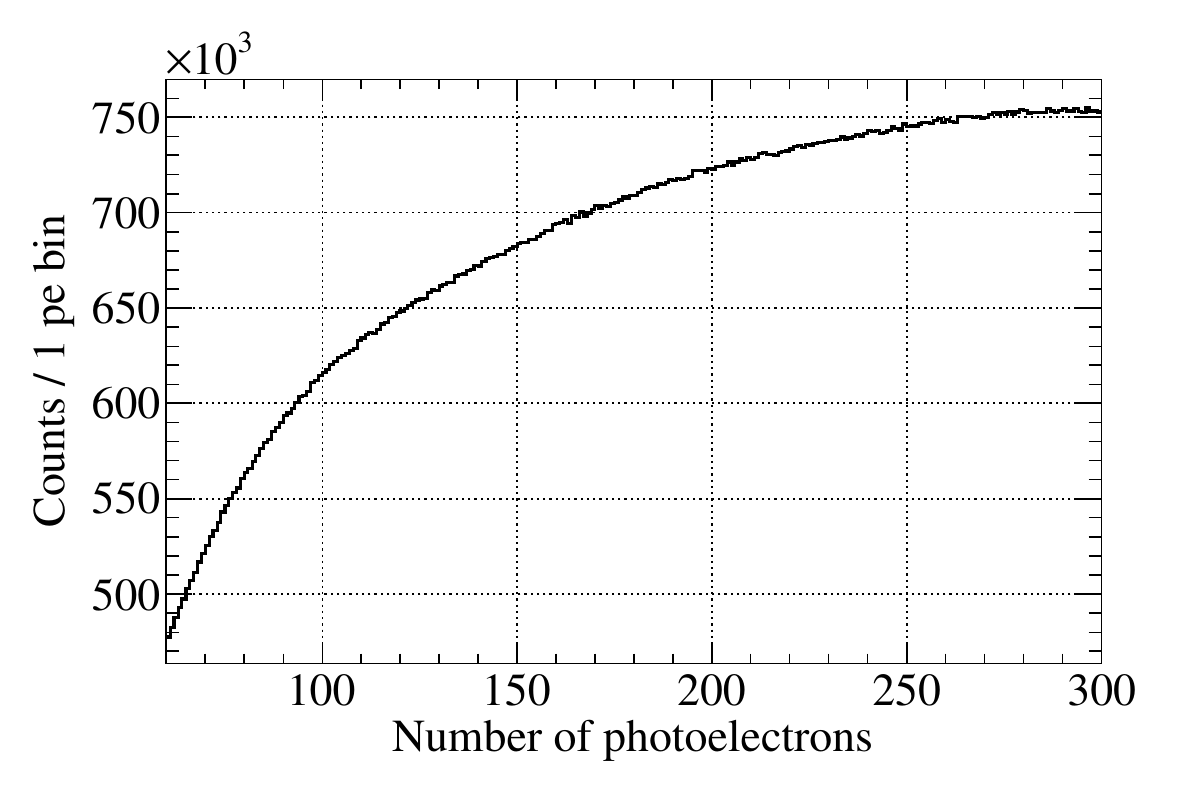}
\caption{Energy spectrum from the $^{22}$Na dataset taken at SNOLAB with V1720.}
\label{fig:pespectrum}
\end{figure}
\end{center}

\subsubsection{Energy spectrum}
The measured spectrum over the energy scales of interest is shown in Fig.~\ref{fig:pespectrum}. The simulated spectrum of raw energy deposit 
in the detector from $^{22}$Na calibration source gammas is to a good
approximation flat over the 120--240 photoelectron region. The measured spectrum varies by only about 13\% for events of 120--240 
photoelectrons, though is heavily dependent at lower energies. This is partially due to a large fraction of low energy events not passing 
the data cleaning cuts~\cite{Lidgard}. A flat spectrum was used in the model Eq.~(\ref{eq:lastconvolution}). The systematic difference
introduced in this step was determined by comparing the analytic model results when using the empirical and when using the flat spectrum for $N(E)$ and found to be negligible, see Sec.~\ref{sec:syst}.

\subsubsection{Systematic PE counting error} \label{sect:meanfprompt}

We observe an upturn of the mean of the \fprompt\ distribution for electron-recoil events toward lower TotalPE, as seen in Fig.~\ref{fig:neutron_fprompt}. This upturn can be explained by a number of effects including: argon scintillation physics, a Prompt~PE-dependent trigger efficiency, or a systematic error in PE counting. We built models for each of these three effects and found that the model where all of the observed upturn is accounted for by a systematic mistake in counting the number of prompt and the number of late PE describes our data best. We however include the model where the upturn is due to argon scintillation physics for comparison in Fig.~\ref{fig:syst}.

\begin{figure}[htbp] 
	\includegraphics[width=3.0in]{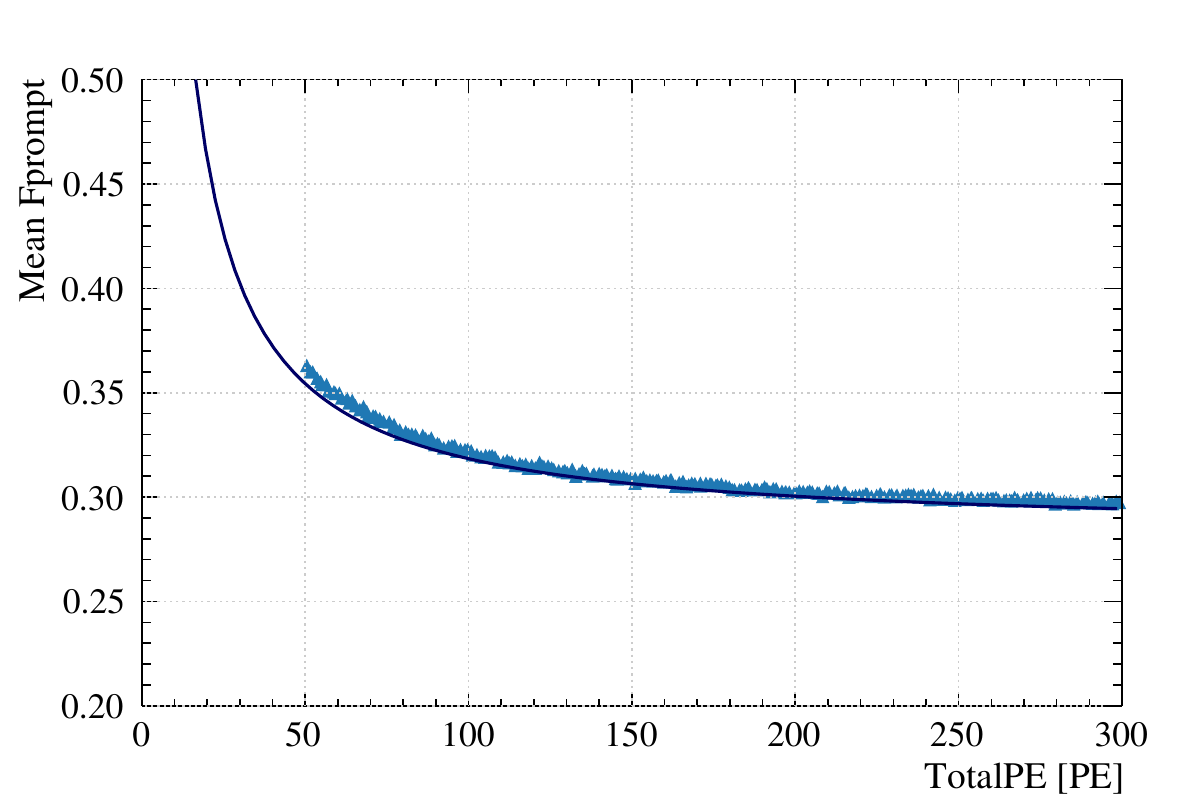}
     \includegraphics[width=3.0in]{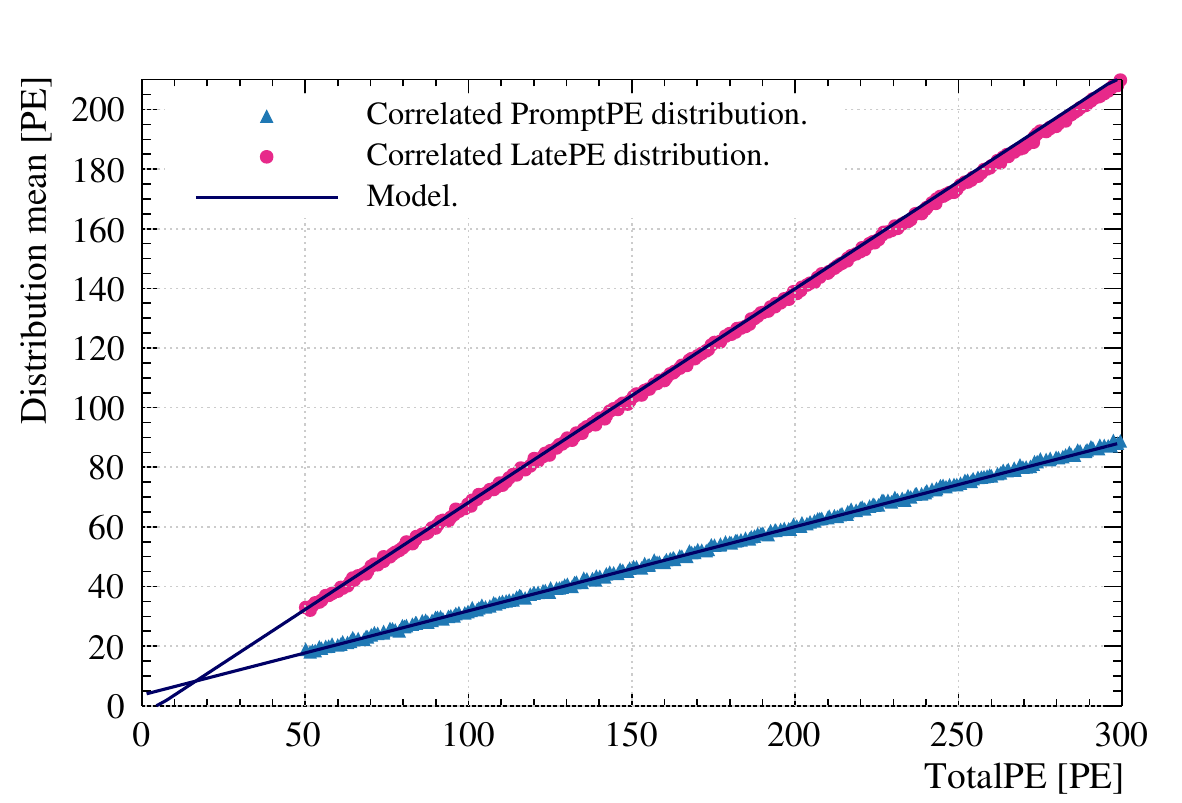}
     \caption{The mean of the \fprompt\ distribution (top) and means of the prompt-PE and late-PE distributions (bottom) in each bin of TotalPE (V1720 data). The expectation based on model parameters used to describe the \fprompt\ distribution, using Gaussian approximations and without convolution over energy, is drawn as a solid line.}
     \label{fig:MeasuredNoiseFp}
\end{figure}

The systematic counting error hypothesis is motivated by the behaviour of the means, $\mu'_p$ and $\mu'_l$, of the conditional prompt-PE and late-PE distributions\footnote{The correlated distributions, Eq.~\ref{eq:lastconvolution} for prompt and equivalent equation for late-PE, are what is directly measured in the experiment, and thus their means can be directly calculated from the data. For Gaussian free parent distributions, the relation between the means of the correlated and free distributions is calculated in \ref{sect:gausparents}.}, which as shown in Fig.~\ref{fig:MeasuredNoiseFp} is linear between 50 and 300~PE with a non-zero x-axis intercept. This is expected from a model, detailed in \ref{sect:energymiscalibration}, in which the observed number of prompt and late PE are equal to a true number plus an offset\footnote{We assume the SPE charge calibration is correct and include the possibility that it is slightly wrong as a systematic uncertainty in the model, so it is not considered here.}, while the true underlying mean of the \fprompt\ distribution is constant with energy.
\begin{figure}[htbp] 
	\includegraphics[width=3.0in]{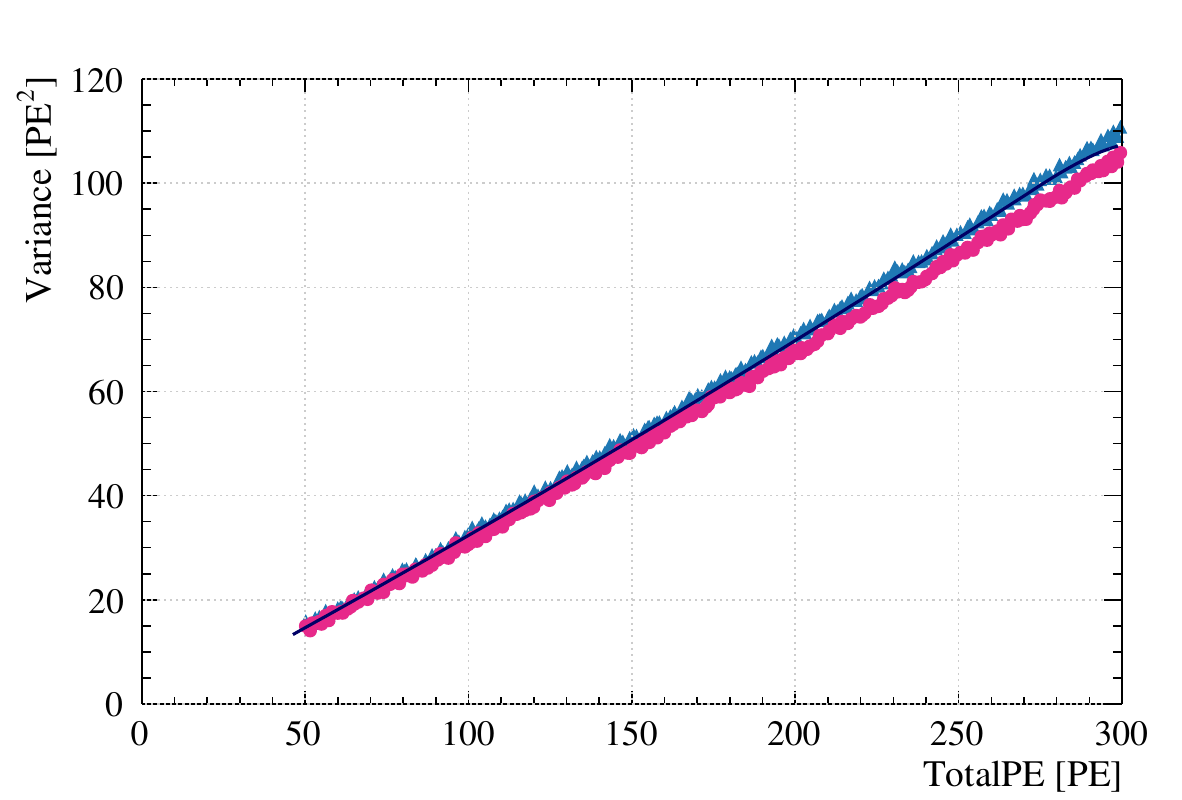}
     \includegraphics[width=3.0in]{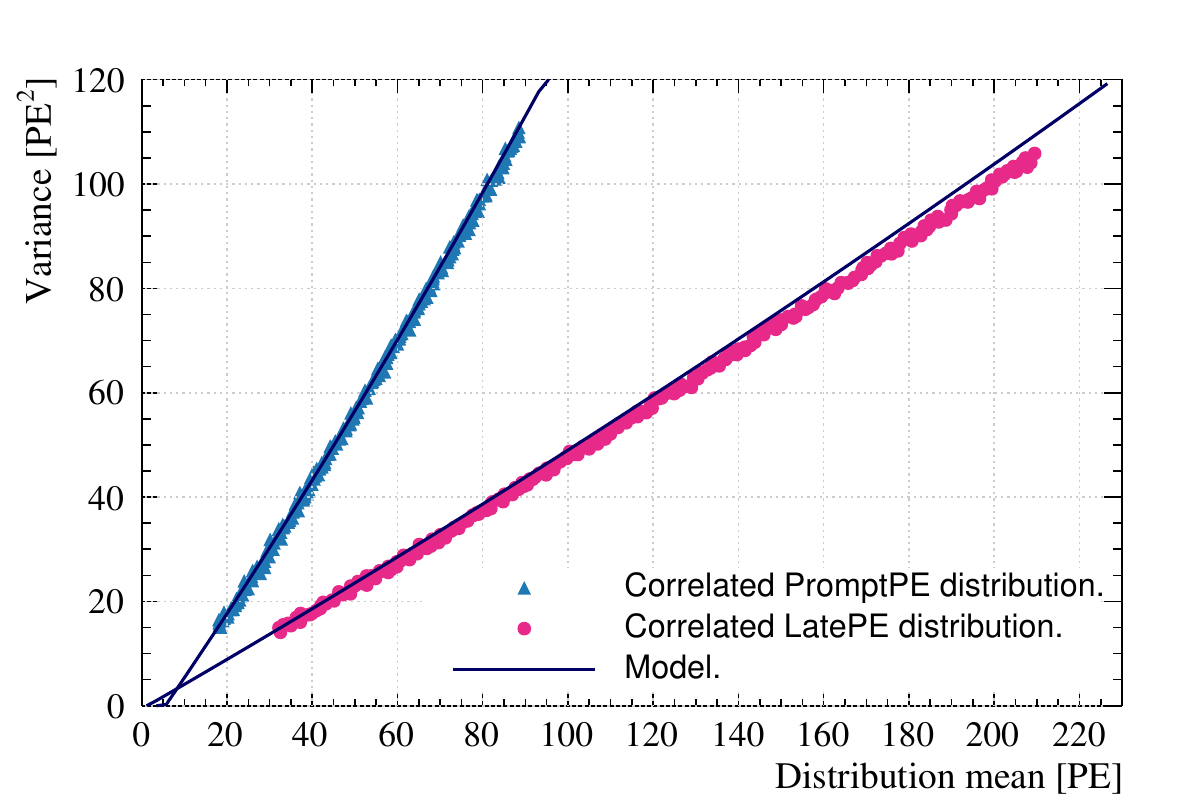}
     \caption{The variances of the prompt-PE and late-PE distributions in each bin of TotalPE (top) and versus the mean of the distribution (bottom). The expectation based on model parameters used to describe the \fprompt\ distribution, using Gaussian approximations, is drawn as a solid line. The top plot includes the effect of a convolution over the energy spectrum.}
     \label{fig:MeasuredNoisePrompt}
\end{figure}

The systematic over-counting hypothesis is further cross-checked by studying how the variances of the correlated distributions behave with the distribution mean and with TotalPE, shown in Fig.~\ref{fig:MeasuredNoisePrompt} overlaid with the model prediction from Eq.~(\ref{eq:latevar}). 
A Gaussian approximation to the uncorrelated
distributions is used and no energy convolution was applied except in the top panel of Fig.~\ref{fig:MeasuredNoisePrompt}. This explains some 
of the observed discrepancy with the model.
We also tried to describe the TotalPE-dependence of the distributions' variances using a general polynomial noise function, while assuming the measured values of prompt and late PE, and thus TotalPE, are correct. This function required an unphysical negative noise value for at least one of the polynomial terms to adequately fit the data, indicating that the data is significantly influenced by an instrumental effect. The over-counting model fits the observed variances without any unphysical noise terms.

The model lines are drawn with the same parameters used to describe the \fprompt\ distributions and summarized in Tab.~\ref{tab:psdfit}, and are consistent with our data. We henceforth assume that in the energy range of interest here, the true mean \fprompt\ is a constant, while the observed mean \fprompt\ has a TotalPE-dependent bias due to a systematic counting error in the number of prompt and late PE.

\subsubsection{Pileup and $^{22}$Na induced backgrounds} \label{sect:pileuprandom}
In Sections~\ref{sec:hfprates} and \ref{sec:snolab} the probabilities of random coincidences between the global tag and high-\fprompt\ ROI events in argon were discussed for both runs. Also present in the data are prompt events with energies lower than $\sim$60~PE, which originate mainly from \v{C}erenkov emission in PMT glass and lightguide acrylic, dark noise and, potentially, from discharges in the PMTs. 

Random coincidences of genuinely tagged $^{22}$Na events with such low energy events add extra photoelectrons to the signal in one or both PMTs, can pass the basic cuts, and can measurably affect the low-\fprompt\ tail of the \fprompt\ distribution. This effect is not included in the analytic model and can lead to discrepancies with the data at lower \fprompt\ values. It is negligible in the high-\fprompt\ tail.

The 1.27~MeV $\gamma$ from the $^{22}$Na source itself can Compton scatter in acrylic or glass creating prompt \v{C}erenkov light simultaneous with genuine $\gamma$-induced scintillation event in liquid argon, effectively shifting the event towards higher \fprompt\ values. This has negligible effect on the low-\fprompt\ tail, but can influence the leakage probability at higher \fprompt\ values.

A two-stage simulation was used to compute the effect on the \fprompt\ distribution due to the radioactive source-related coincidences. The first stage was a simulation of $^{22}$Na decays in a realistic Geant4 model of DEAP-1, which included the NaI crystal used for tagging, shielding configuration, as well as all details of the detector itself, in particular the wavelength-dependent optical attenuation of UVA acrylic, measured ex-situ\footnote{This effect was first recognized in a later DEAP-1 detector generation which was built using UV-transmitting acrylic so that the \v{C}erenkov light yield was increased~\cite{tcald_thesis,tcald_paper} and the effect therefore more apparent.}. From this simulation, the spectrum of those \v{C}erenkov events was produced which occur in coincidence with tagged events of 120--240~PE that still pass data selection cuts. Deposition of approximately 511~keV energy in NaI was set as the tagging criterion, consistent with discriminator settings in the experiment.

The second stage was a fast toy Monte Carlo with logic equivalent to that of the analytic model, i.e. using the model spectrum and beta binomial probability distributions to draw numbers of detected prompt and late~PE\footnote{A more elaborate scintillation model taking into account partition of deposited energy between excitation and ionization quanta and recombination was also tried, following the approach used in NEST~\cite{NEST}. It agreed with the simpler model inside of the parameter space explored in this work.} and then smearing them with respective Gaussian noise terms, parameterized as in Tab.~\ref{tab:psdfit}. PE counts were then randomly generated with the \v{C}erenkov spectrum from the first step and added to the number of detected prompt~PE.

Simulation results are compared with the data in the next section.

\subsection{Results}
The model parameters that best describe both the behaviour of the prompt-PE and late-PE distribution means and variances as well as the \fprompt\ distribution versus TotalPE, are summarized in Tab.~\ref{tab:psdfit}, together with their expected ranges from independent measurements. The $\sigma_\text{elec,p}$ was kept fixed at $0.57$~PE (surface) or $0.5$~PE (underground) since a small change here has little effect and numerical stability of the model was better that way. The SPE calibration was also fixed in the fit. 
For the V1720 dataset the systematic effect of varying the SPE calibration in the fit was studied, and the maximum resulting change to the fit parameters is quoted in brackets.
\begin{table}[htbp]
   \centering
   \small
   \begin{tabular}{@{} l|ll|ll @{}} 
      \toprule
                						& \multicolumn{2}{c|}{Surface} & \multicolumn{2}{c}{Underground} \\
                						& {\footnotesize Fit} & {\footnotesize Expected} & {\footnotesize Fit} & {\footnotesize Expected} \\
      \midrule
      $\bar{F_{p}}$ 					& 0.300 & -- & 0.282 (+0.001) & -- \\
      $\delta_p$ [PE] & 6.1  & --  &  6.57 (+0.2) & 6.0--7.0\\
      $\delta_l$ [PE]   &  5.3 & -- & 3.96 ($^{+1.3}_{-0.2})$ & 2.0-3.8 \\
      $\epsilon_\text{spe}$  			& 0.26 & 0.34  & 0.25 (+0.14) & 0.34 \\
      $\epsilon_{\textrm{win}}$ 		& 0.016 & $<$0.025 &  0.016 (-0.008)& $<$0.025 \\
      $\sigma_\text{elec,p}$ [PE] 			& 0.57 & 0.57 & 0.5 & $<$0.57 \\
      $\sigma_\text{elec,l}$ [PE] 				& 6.6 & $<$7.0 & 2.26(-0.7) & $<$7.0 \\
      $b$  								& 1.86 & -- & 1.65 ($^{+0.25}_{-0.35}$)  & -- \\
      \bottomrule
   \end{tabular}
   \caption{Fit results and independent measurement expectation for the surface and underground \fprompt\ versus TotalPE data. Numbers in brackets are the maximum amount of systematic shift that comes with changing the SPE calibration by $\pm$10\%.}
   \label{tab:psdfit}
\end{table}

\begin{center}
\begin{figure}[htb]
\includegraphics[width=3.5in]{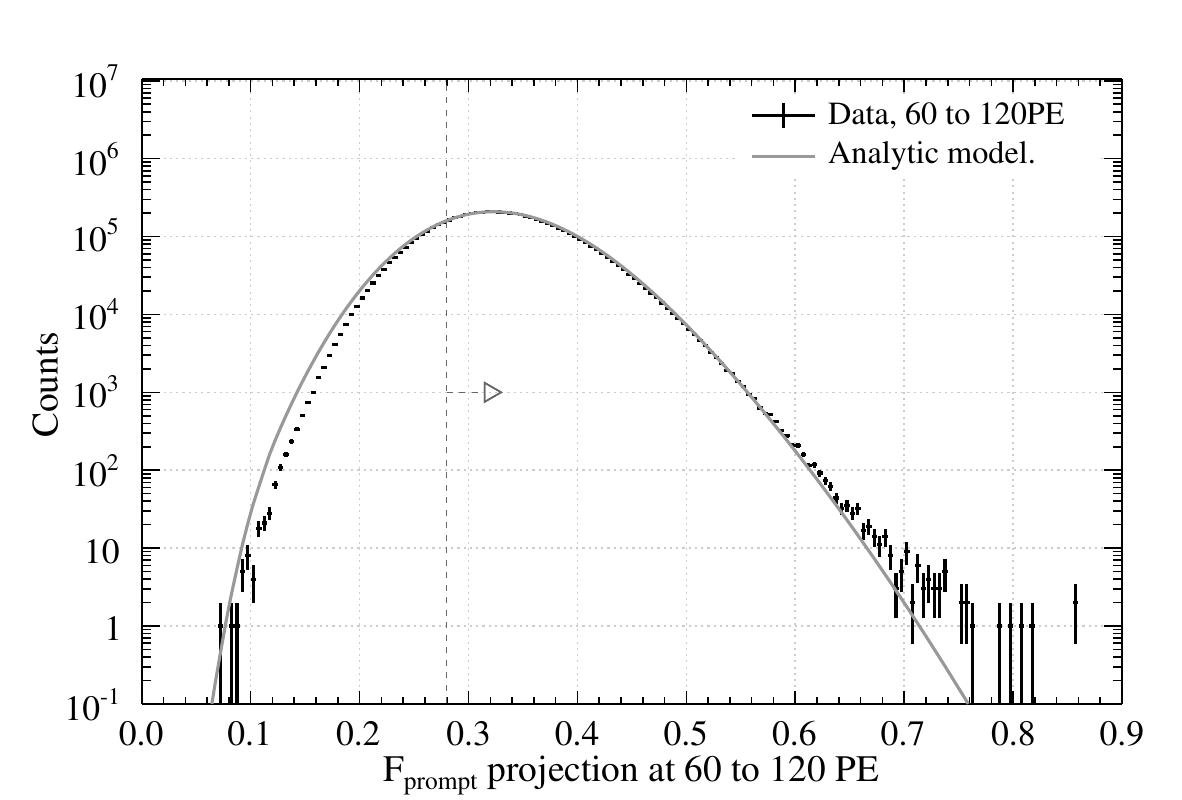}
\caption{Comparison of the surface run data and the analytic model in the region 60--120 photoelectrons (approximately 21--43~\kevee). Dashed vertical line with an arrow indicates the fit range.}
\label{fig:analyticmodel60}
\end{figure}
\end{center}

\begin{center}
\begin{figure}[htb]
\includegraphics[width=3.5in]{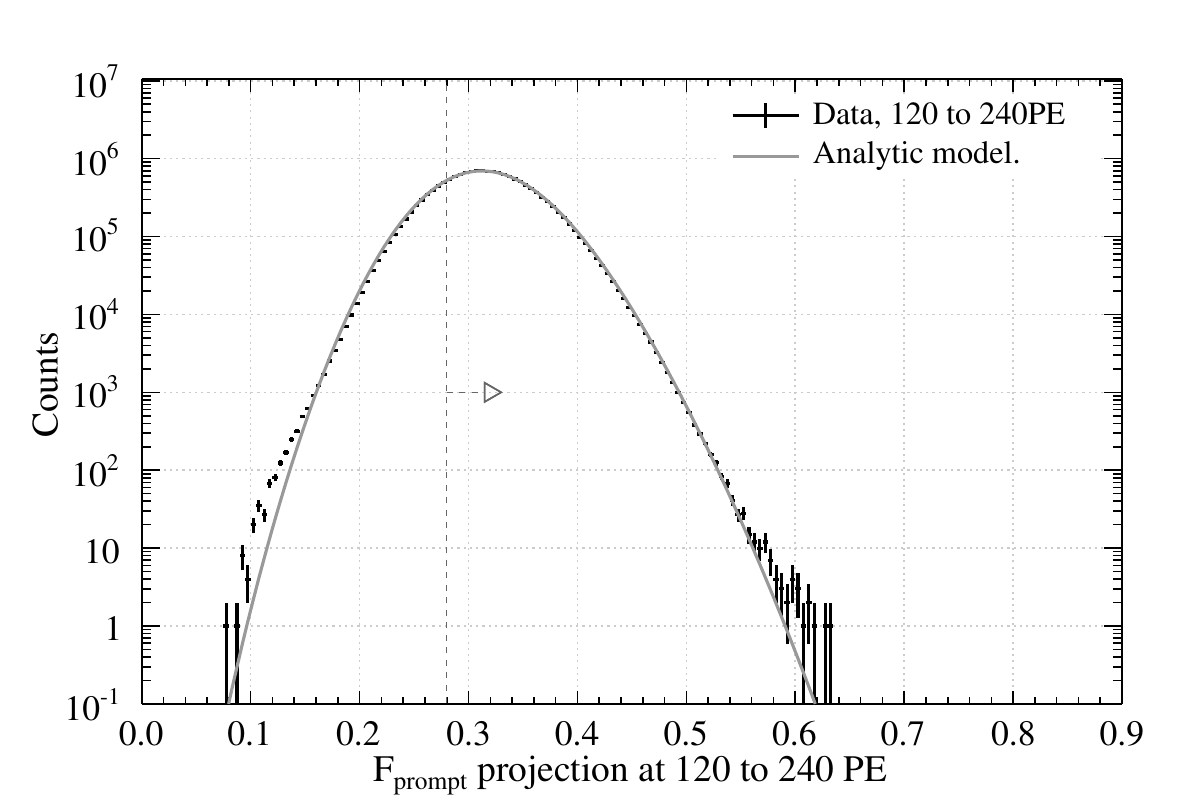}
\caption{Comparison of the surface run data and the analytic model in the region 120--240 photoelectrons (approximately 43--86~\kevee). Dashed vertical line with an arrow indicates the fit range.}
\label{fig:analyticmodel120}
\end{figure}
\end{center}

\begin{center}
\begin{figure}[htb]
\includegraphics[width=3.5in]{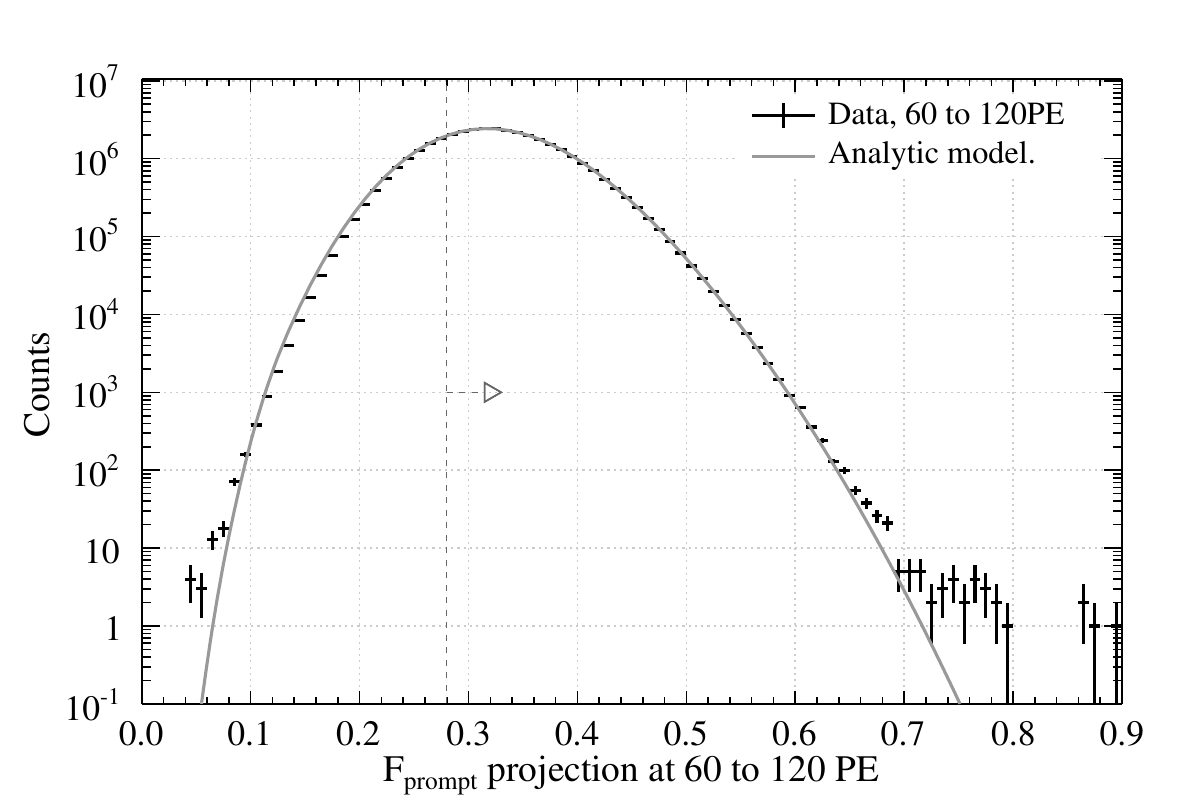}
\caption{Comparison of the SNOLAB V1720 data and the analytic model in the region 60--120 photoelectrons (approximately 22--44~\kevee). Dashed vertical line with an arrow indicates the fit range.}
\label{fig:analyticmodel60ug}
\end{figure}
\end{center}

\begin{center}
\begin{figure}[htb]
\includegraphics[width=3.5in]{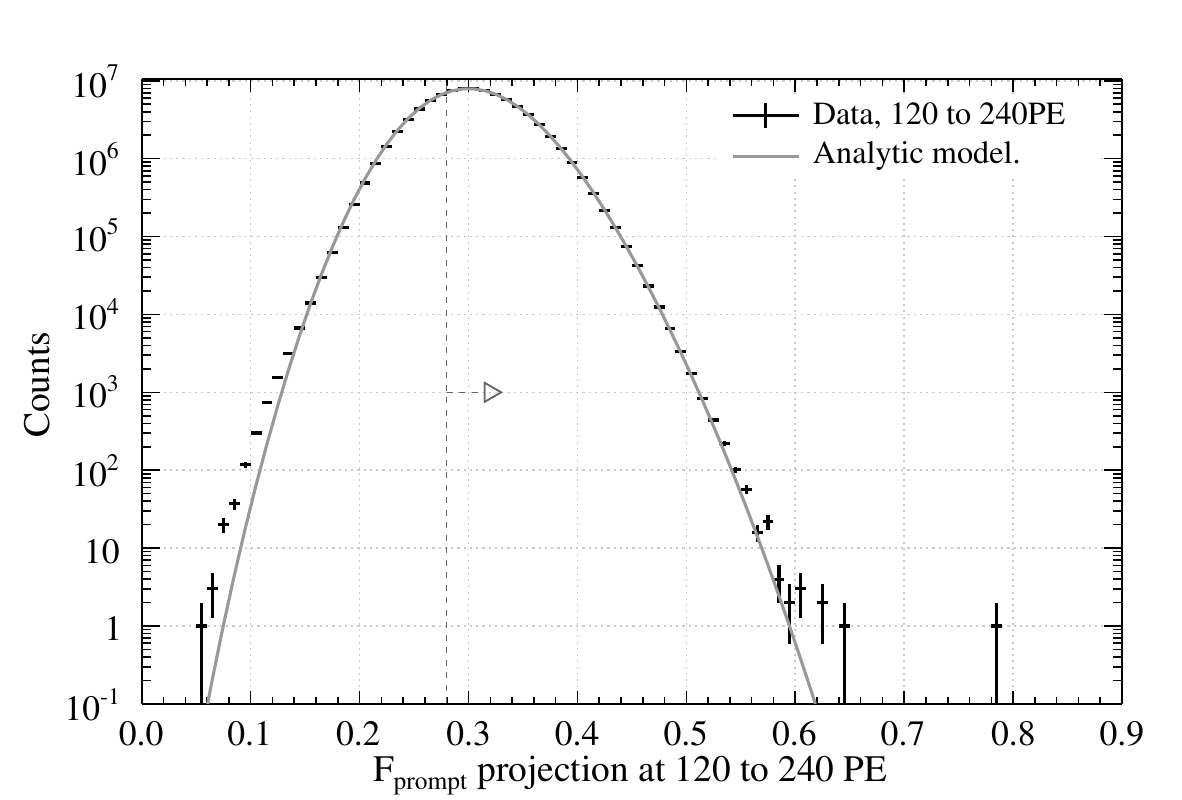}
\caption{Comparison of the SNOLAB V1720 data and the analytic model in the region 120--240 photoelectrons (approximately 44--89~\kevee). Dashed vertical line with an arrow indicates the fit range.}
\label{fig:analyticmodel120ug}
\end{figure}
\end{center}

\begin{center}
\begin{figure}[htb]
\begin{overpic}[trim=10 52 9 20, clip=true, width=0.48\columnwidth]{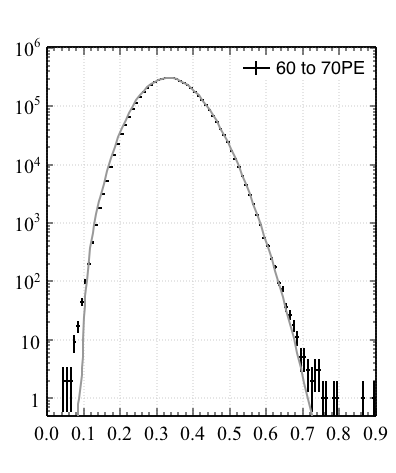}\linethickness{0.5pt}\multiput(34.66,2.5)(0, 2){46}{\color{light-gray}\line(0,-1){1}}\end{overpic}
\begin{overpic}[trim=10 52 9 20, clip=true, width=0.48\columnwidth]{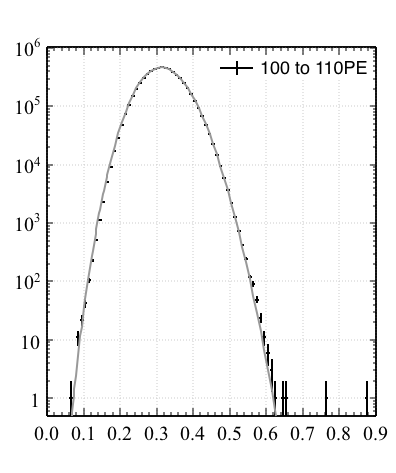}\linethickness{0.5pt}\multiput(34.66,2.5)(0, 2){46}{\color{light-gray}\line(0,-1){1}}\end{overpic}
\begin{overpic}[trim=10 52 9 20, clip=true, width=0.48\columnwidth]{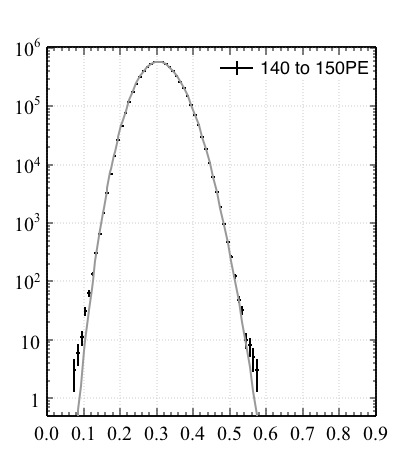}\linethickness{0.5pt}\multiput(34.66,2.5)(0, 2){46}{\color{light-gray}\line(0,-1){1}}\end{overpic}
\begin{overpic}[trim=10 52 9 20, clip=true, width=0.48\columnwidth]{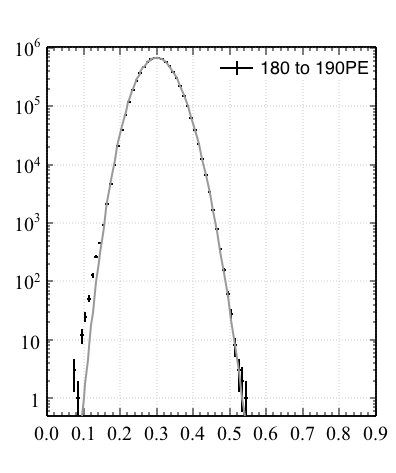}\linethickness{0.5pt}\multiput(34.66,2.5)(0, 2){46}{\color{light-gray}\line(0,-1){1}}\end{overpic}
\begin{overpic}[trim=10 30 9 20, clip=true, width=0.48\columnwidth]{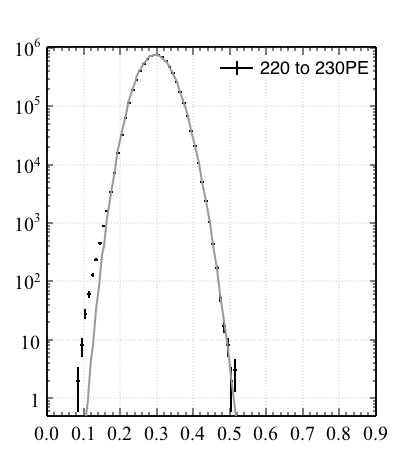}\linethickness{0.4pt}\multiput(32.85,7.5)(0, 2){44}{\color{light-gray}\line(0,-1){1}}\end{overpic}
\begin{overpic}[trim=10 30 9 20, clip=true, width=0.48\columnwidth]{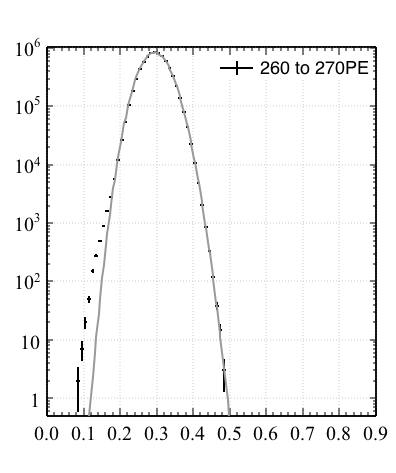}\linethickness{0.4pt}\multiput(32.85,7.5)(0, 2){44}{\color{light-gray}\line(0,-1){1}}\end{overpic}
\includegraphics[trim=0.75cm 0cm 0cm 13.21cm, clip=true, width=\columnwidth]{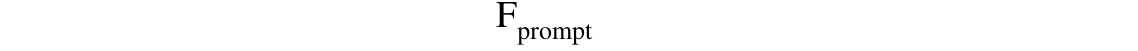}
\caption{The \fprompt\ distribution from the underground run data shown separately for each of six energy bins, as indicated on the plots. Dashed vertical lines indicate the lower limit of the fit range.}
\label{fig:4fpug}
\end{figure}
\end{center}

The model overlaid on the data is shown in Figs.~\ref{fig:analyticmodel60} and \ref{fig:analyticmodel60ug} for the TotalPE region 60--120 photoelectrons, and in Figs.~\ref{fig:analyticmodel120} and \ref{fig:analyticmodel120ug} for the TotalPE region 120--240 photoelectrons, for the surface and V1720 underground datasets respectively. Fig.~\ref{fig:4fpug} shows the model and data in 10~PE bins between 60 and 270~PE for the underground dataset. 

The analytic model describes the measured \fprompt\ versus TotalPE distribution from 60 to 280~photoelectrons and 0.28 to 1.0~\fprompt\ with reduced $\chi^2$ of 3.1 (surface) and 2.0 (underground) for a fit performed as explained in Sect.~\ref{sect:datacomparison}.


The region below \fprompt\ of 0.28 was excluded from the fit. The distribution in this region is affected by cut and trigger efficiencies, as well as pileup. 
At lower energies, as in Figures~\ref{fig:analyticmodel60} and \ref{fig:analyticmodel60ug}, this may remove events with comparatively low number of photoelectrons in the prompt window or add additional pileup events, with non-trivial dependence on run conditions (such as background rate and spectrum) as well as on cuts. As this was not studied in detail, it is not clear which of the two effects should dominate.
At higher energies, the effect of pileup with low energy backgrounds adds additional events below \fprompt\ of 0.28. It is noticeable around 0.1~\fprompt\ above 180~PE in Fig.~\ref{fig:4fpug}, where the PSD distribution is narrow enough that the additional events separate out clearly.
This low-\fprompt\ region is not further studied as the position of the distribution peak and the high-\fprompt\ tail are the relevant measures for determining the leakage into the nuclear-recoil region.

The fit between model and data is not equally good in all energy bins. The individual p-values above 0.28 \fprompt\ for a number of 1~PE wide TotalPE bins are shown in Tab.~\ref{tab:fitresult} for the V1720 underground dataset.
The model does not account for coincidences from events caused by the $^{22}$Na source, so it is not expected to match the data perfectly. The model adequately describes the data in the region of interest 120--240~PE, and diverges toward lower and higher TotalPE. The region below about 100~PE is very sensitive to effects from photoelectron counting errors and to the event energy spectrum, and the region above 200~PE is very sensitive to the absolute value of the noise terms, which are strongly correlated with the photoelectron counting error at lower PE. The noise contributions as a function of TotalPE are parametrized in the noise model which may not be adequate anymore at these PE ranges.
\begin{table}[htbp]
\caption{Model match above 0.28 \fprompt\ in individual 1~PE wide bins for the V1720 underground dataset (see text).}
\begin{center}
\begin{tabular}{|c|c|}
TotalPE bin & p-value \\
\hline 
60   & $2\times10^{-18}$ \\ 
80   & $4\times10^{-4}$ \\
100  & $2\times10^{-5}$ \\
120  & $6\times10^{-4}$ \\
140  & 0.02 \\
160  & 0.23 \\
180  & 0.02 \\
200  & 0.16 \\
220  & 0.02 \\
240  & $5\times10^{-4}$ \\
260  & $1\times10^{-4}$ \\
280  & $4\times10^{-8}$ \\
\end{tabular}
\end{center}
\label{tab:fitresult}
\end{table}%

Despite fair p-values for individual TotalPE bins inside of the 120--240~PE region, when all these bins are stacked together there is a clear excess of events above \fprompt=0.58, which is visible in Fig.~\ref{fig:analyticmodel120ug}. In Sec.~\ref{sect:pileuprandom} a \v{C}erenkov type background was discussed, which, as shown below, included in a simulation reproduces this excess. Because of computational complexity the \v{C}erenkov contribution was not included in the analytic model and could not be used for fitting.

\begin{center}
\begin{figure}[htbp]
\includegraphics[width=3.5in]{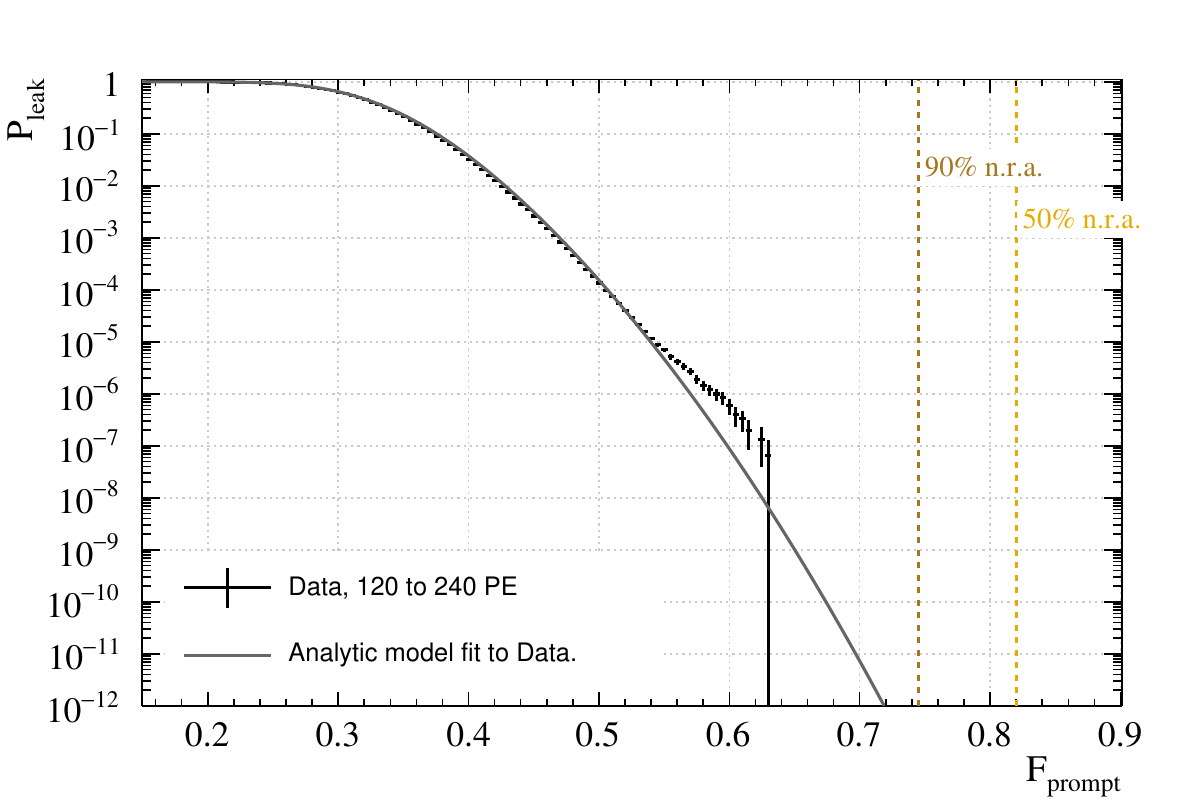}
\caption{\pleak\ distributions from $^{22}$Na calibration data from DEAP-1 on surface with analytic model fit. Also indicated are \fprompt\ values for 50\% and 90\% nuclear recoil acceptances (brown and orange dashed lines). Nuclear recoil acceptances (n.r.a.) are determined from neutron calibration data.}
\label{fig:pleakna22}
\end{figure}
\end{center}
\begin{center}
\begin{figure}[htbp]
\includegraphics[width=3.5in]{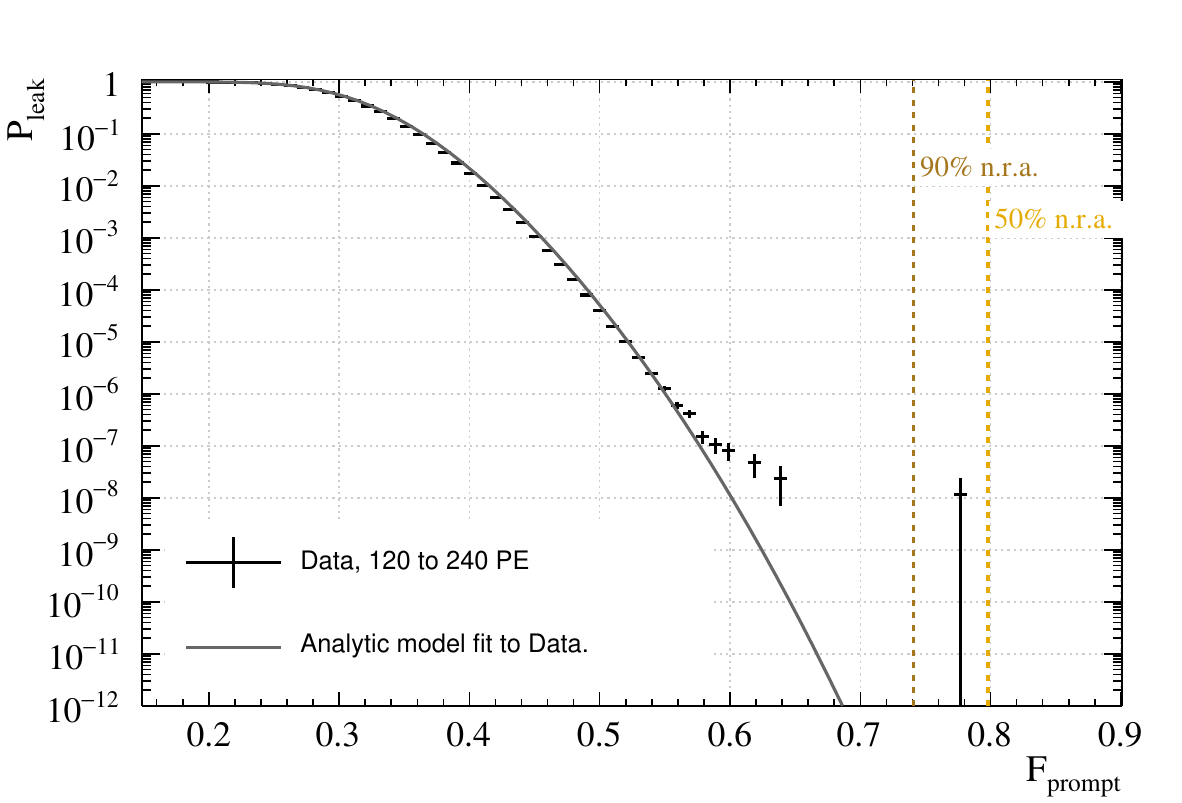}
\caption{\label{fig:pleakna22ug}\pleak\ distributions from $^{22}$Na calibration data from DEAP-1 underground at SNOLAB (V1720 data) with analytic model fit.
Also indicated are \fprompt\ values for 50\% and 90\% nuclear recoil acceptances (brown and orange dashed lines).  Nuclear recoil acceptances are determined from neutron calibration data.}
\end{figure}
\end{center}
To represent the discrimination power graphically, we define the variable \pleak\ as
\begin{eqnarray}
\pleakmath(f') & = & \frac{\int_{f'}^{1}\fpromptmath(f)df}{\int_{0}^{1}\fpromptmath(f) df} \label{eqn:pleak}
\end{eqnarray}
\noindent for a given \fprompt\ distribution.  The value of \pleak\ for a given \fprompt\ value thus represents the pulse-shape discrimination.  Figures~\ref{fig:pleakna22} and \ref{fig:pleakna22ug} show the \pleak\ distributions derived from the $^{22}$Na calibration data (the same data shown in Figures~\ref{fig:analyticmodel120} and \ref{fig:analyticmodel120ug}), along with the predictions from the analytic PSD model. Also shown are reference acceptances for nuclear recoil events estimated directly from neutron calibration data as shown in Fig.~\ref{fig:neutron_fprompt} for the surface dataset.

\begin{figure}[htbp] 
     \includegraphics[width=3.5in]{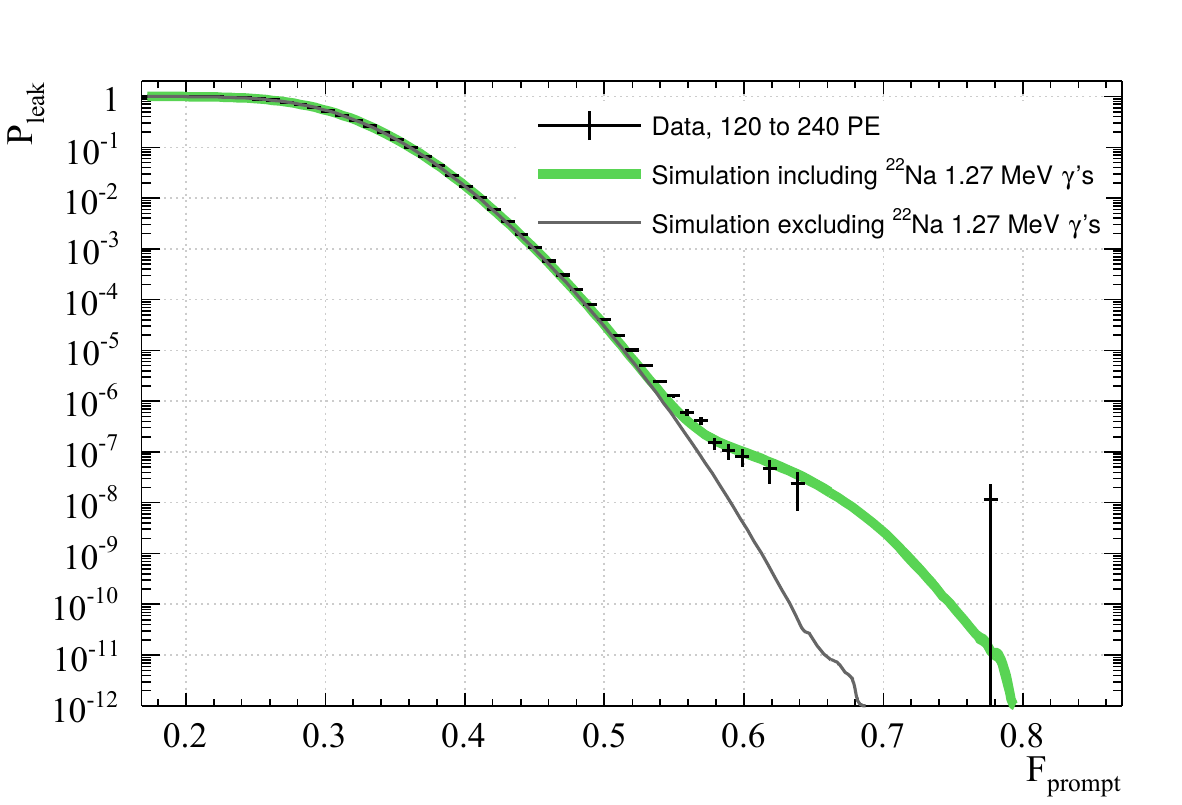}
     \caption{Leakage probability for simulated electromagnetic interactions. Prompt and late~PE are drawn from distributions mimicking the analytic model discussed, including all sources of noise, with parameter values from the underground dataset (grey). In the second simulation (green), PMT counts are additionally drawn from a spectrum of high \fprompt\ \v{C}erenkov events, which was generated with a full Geant4 simulation of the detector. This results in an increased leakage probability above a certain \fprompt, as observed in the data. Simulated curves have the statistics of $\sim$1 at 10$^{-12}$. Based on the simulated spectrum we expect 1.9 events above Fprompt=0.65, while no events are observed (excluding the outlier at \fprompt$\simeq$0.78, which is most likely unrelated to 1.27~MeV $\gamma$'s, as explained in Sec.~\ref{sec:snolab}). The probability of such an occurence is 15\%.}
     \label{fig:explaintail}
\end{figure}

The result of the Monte Carlo simulation including \v{C}erenkov emission induced by $^{22}$Na gammas (see Sec.~\ref{sect:pileuprandom}) is shown in Fig.~\ref{fig:explaintail} for parameters chosen for the underground dataset, overlaid with the data. The simulation reproduces the measured distribution including the excess in the high-\fprompt\ tail.

\subsection{Discussion}\label{sec:lepileup}

Considering the systematic uncertainties in the photoelectron counting, both due to the offsets and possible scaling of up to 10\%, and considering that effects such as trigger efficiency and pileup are not included in the model, the model describes the data reasonably well, especially when considering the region above around 0.28~\fprompt\, which is no longer effected by trigger efficiency and pileup with low-energy events, and when including the effect from pileup with source induced \v{C}erenkov events through Monte Carlo simulation.

Fig.~\ref{fig:explaintail} shows that after accounting for \v{C}erenkov light induced by gammas from the $^{22}$Na source, the deviation of the leakage probability from the analytic PSD model above 0.55~\fprompt\ is in agreement with the measurement from Fig.~\ref{fig:pleakna22ug}.

The noise terms from the analytic model must add up to the total energy resolution. The total energy resolution was measured near the 59.5~keV peak from the Am-Be source to be $\sigma_t=11.1$~keV (surface) and $\sigma_t=8.8$~keV (underground).
As per Eqs.~(\ref{eq:eres}), (\ref{eq:sigmap}), and (\ref{eq:sigmal}), the model resolution is about $\sigma_t = 6.7$~keV (surface) and $\sigma_t = 6.5$~keV (underground). From a full Monte Carlo simulation of the detector response in the underground configuration, we expect a $\sim$25\% discrepancy between the measured resolution and the model resolution because the calibration source full energy peak in the liquid argon has a small low-energy tail component from gammas which reach the target with energies already degraded via Compton scattering in the detector materials.
Smearing the raw energy deposit spectrum obtained from a Geant4 simulation of DEAP-1 Am-Be and $^{22}$Na calibration runs with the energy resolution from the analytic model gives the overall full energy peak widths consistent with the data for the underground dataset. Since the parameters for this dataset will be used for extrapolations, we did not in detail study the additional discrepancy for the surface dataset.
 
In our model, we describe the upturn of the mean \fprompt\ value at lower energies, in the energy range under consideration here, by an instrumental effect, namely a systematic over-counting of the number of PE in the prompt and late windows. This upturn has been seen in many experiments~\cite{darkside,Lippincott:2008ad,Regenfus} and the cause may not be the same each time. We for example note that the measurement in \cite{Regenfus} shows an essentially flat mean \fprompt\ in the energy range considered in this work, as assumed in our model. It is likely that the  mean \fprompt\ for electromagnetic events truly increases toward lower energies due to the underlying scintillation physics, at energies below what is considered in this work. The discrepancy between the model line and the data in Fig.~\ref{fig:MeasuredNoiseFp} (top) hints at such a true change in the mean \fprompt\ value below about 80~PE. 

The reason for the systematic over-counting of prompt and late PE is currently not well understood. A possible mechanism is trigger-induced cross-talk on the electronics, which would effectively shift the baseline for a certain amount of time and due to our simple baseline algorithm and charge-based PE counting, this adds a fixed amount of charge to both the prompt and the late window, with more charge added in the prompt window as it is closer to the trigger time. Dark noise and afterpulsing also add extra counts, though preferentially in the late window. 

The relatively big change in the \fprompt\ distribution mean from comparatively small errors in the PE count highlights the need to carefully understand the behaviour of the electronics, including the baseline and the trigger efficiency, and to count PEs accurately in a future experiment, especially toward lower energies. We also note that a prompt~PE-dependent trigger efficiency can introduce a significant energy-dependent bias in the mean and width of the measured \fprompt\ distributions, especially at lower energies. This effect when modelled did not describe our data as well as the PE-shift, but may also add to the discrepancy between the data and the model curve in Fig.~\ref{fig:MeasuredNoiseFp} (top).

\subsection{Systematic uncertainties}\label{sec:syst}
We estimate how systematic uncertainties on input parameters of the analytic model affect the predicted discrimination power by calculating the \pleak\ distributions
\begin{itemize}
\item With flat and with measured energy spectrum.
\item With the mean SPE charge increased or decreased by 10\%.
\item Using a running mean \fprompt\ and assuming no counting error in the number of prompt and late PE.
\end{itemize}

\begin{center}
\begin{figure}[htbp]
\includegraphics[width=3.5in]{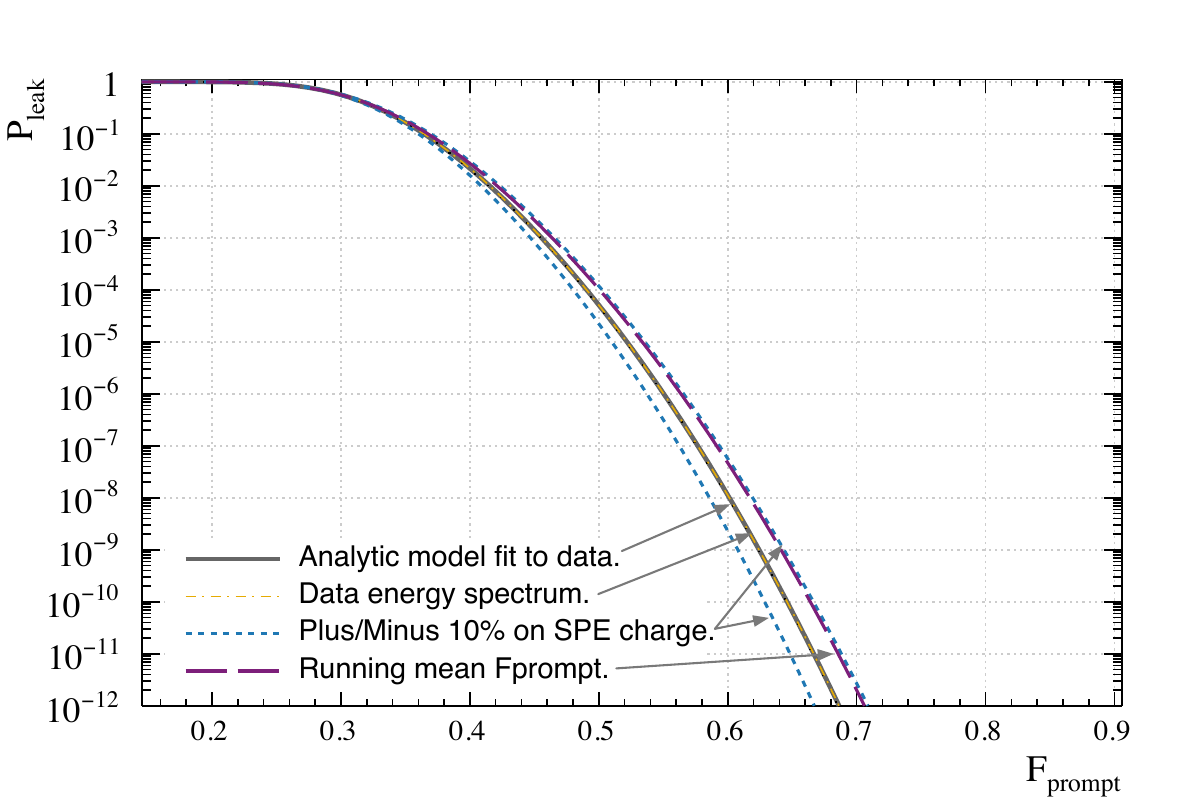}
\caption{Systematic uncertainties on the model \pleak\ distribution from Fig.~\ref{fig:pleakna22ug}, for V1720 data taken at SNOLAB (gray bold line). Drawn in orange dash-dot and indistinguishable from the former is the model using the measured energy spectrum in the convolution. The blue dotted lines show uncertainty due to 10\% shift in the spe charge calibration when keeping all other model parameters the same. The purple dash-dot line is drawn using the same parameter values as the nominal model (grey line) but no prompt or late offset, and a running mean \fprompt\ value, behaving as shown Fig.~\ref{fig:MeasuredNoiseFp} top, instead. 
\label{fig:syst}}
\end{figure}
\end{center}

Fig.~\ref{fig:syst} shows the curves for model parameters from Fig.~\ref{fig:pleakna22ug} and with the above three parameters changed. The dominant systematic uncertainty is due to the uncertainty in the SPE charge calibration. The uncertainty due to the over-counting of charge slightly smaller than that from the SPE charge calibration, and it is over-estimated since no further fits were performed to try and adjust the noise parameters for a better fit.

At 0.65 \fprompt\, these systematic effects change the leakage probability by about an order of magnitude in either direction.

\section{Dark Matter Sensitivity of a Tonne-Scale Detector}
In the previous section we demonstrated that the analytic model based on Eq.~(\ref{eq:lastconvolution}) adequately describes the data from the DEAP-1 detector in the region of \fprompt\ relevant for leakage of electromagnetic events into the nuclear recoil region, after adding source-induced pileup effects by Monte Carlo simulation.

Using the model but not the source-induced pileup contribution, we next estimate the expected discrimination power in liquid argon versus energy threshold assuming a large target mass detector. We have designed DEAP-3600, a large spherical detector consisting of 255 PMTs surrounding a spherical target with a mass of 3600~kg of liquid argon~\cite{BoulayTAUP}.  Geant4 simulations benchmarked against the light yield in the DEAP-1 detector predict that in this geometry a light yield of approximately 8~PE/\kevee\ could be achieved.  Assuming the detector is constructed of clean materials and appropriately shielded so that genuine nuclear recoil backgrounds have been mitigated, the dominant detector background in argon will be from $\beta$ decays of $^{39}$Ar. Argon that is condensed from the atmosphere is known to contain cosmogenically-produced $^{39}$Ar, with a rate of decays of approximately 1~Bq per kg~\cite{Looslipaper}.
\begin{figure}[htbp] 
     \includegraphics[width=3.5in]{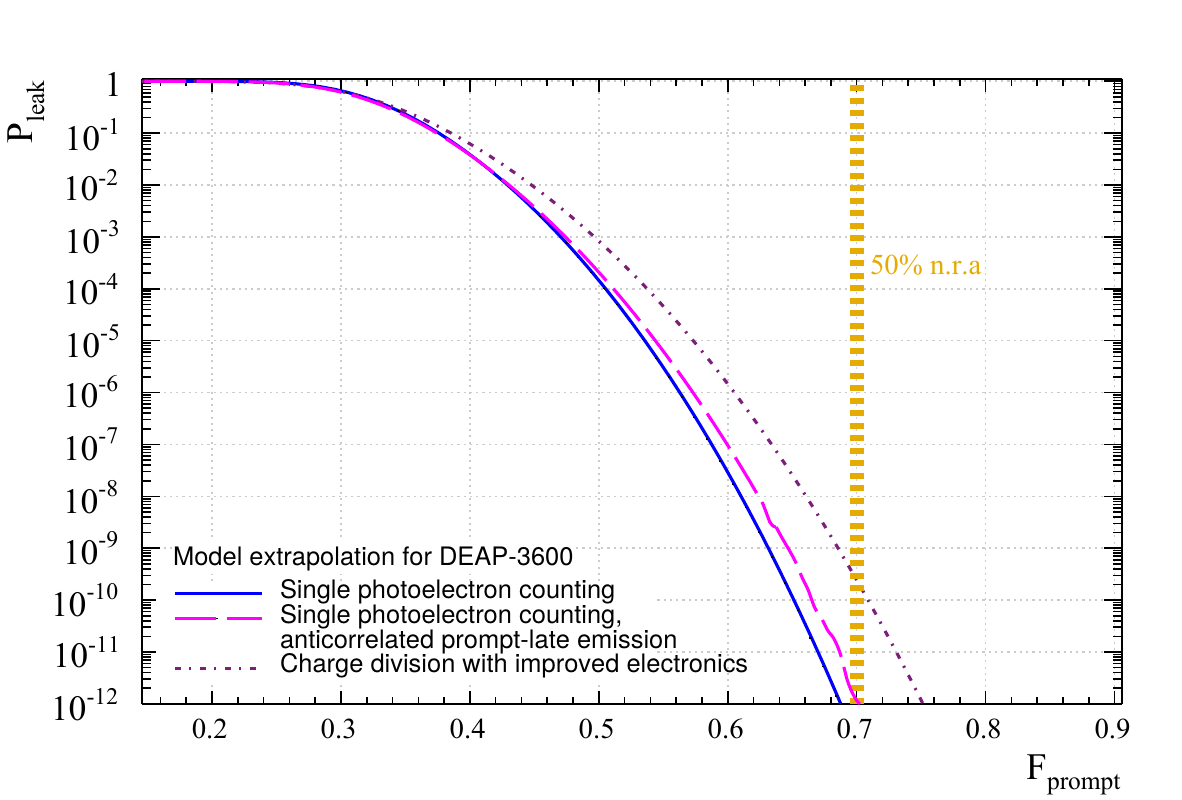}
     \caption{Leakage estimation at 120 to 240~PE for a detector with 255 PMTs and 8~PE/\kevee\ light yield (15--30~\kevee\ energy window). The underlying energy spectrum is that of the $^{39}$Ar beta decay, PE counting is assumed to be accurate, and the energy dependence of the mean of the \fprompt\ distribution is taken from~\cite{Regenfus}. The solid line shows the analytic model without SPE or electronics noise, and the dotted line has the same SPE noise as the DEAP-1 measurement, but electronics noise per PMT is reduced to account for better read-out electronics and then scaled up to 255 PMTs. The dashed line was generated with a toy simulation following the logic of the analytic model and also assuming anticorrelation from binomial partition between prompt and late scintillation photons at a fixed energy. The window noise is taken from the DEAP-1 measurement. Nuclear recoil acceptance median is taken from the SCENE measurement~\cite{Cao2014} and corrected for differences in integration window definitions.}
     \label{fig:d3pleak}
\end{figure}

Since the model developed does not describe the data perfectly, and several of the input parameters will be different in the large detector, we calculate model predictions for the large detector using conservative estimates.
The \pleak\ distribution given by Eq.~(\ref{eqn:pleak}) is shown in Fig.~\ref{fig:d3pleak} with the binomial probability centred at $\bar{p} = 8/40$, corresponding to 8~PE/\kevee\ light yield, for the energy region of 15--30~\kevee. We use the $^{39}$Ar energy spectrum in the energy convolution, and assume correct PE counting, i.e. the prompt and late-PE offsets are zero. To conservatively include a possible upturn in the true mean of the \fprompt\ distribution at lower energies, we implemented the energy dependence of mean \fprompt\ as observed in \cite{Regenfus}. The leakage is calculated for the following cases: (a) assuming that SPE identification and counting analysis will make the electronics and SPE noise negligible, (b) using a simple DEAP-1 style charge division method for SPE calibration, i.e. the same SPE noise, with the electronics noise values from the DEAP-1 V1720 underground dataset scaled up to 255 PMTs (summed in quadrature), and late noise reduced by a factor of 6 to account for low-noise electronics and, additionally, (c) as an extreme case meaningful for a high light yield detector, a toy simulation discussed earlier in Sec.~\ref{sect:pileuprandom} was extended to include Fano fluctuations in the total number of scintillation photons as well as a binomial fluctuation between the initial number of prompt and late scintillation photons, effectively adding maximal anticorrelation between both populations at a fixed energy (see \ref{sect:parentdistr} for details). 
The window noise $\sigma_{w}$ was in each case taken from the DEAP-1 model.

In order to suppress the $^{39}$Ar background in the large detector to the required level of less than 0.2 events in a three-year run, PSD at the level of $10^{-10}$ is required. The model estimation in Fig.~\ref{fig:d3pleak}, using conservative model parameters, indicates that PSD at this level is possible in the energy region of 15--30~\kevee\ at 50\% nuclear recoil acceptance if the electronic and SPE noise can be kept sufficiently low. The uncertainty on this extrapolation is necessarily large, and is again dominated by the accuracy with which photoelectrons are counted. 

In the large detector, each of the 255 PMTs will detect on average less than a single PE per region-of-interest event, so that PMT hits are going to be well separated. Single PE identification and counting will then bring substantial improvement over the simple charge integration and division method used in DEAP-1 analysis, so that the electronic and SPE noise parameters will likely be negligible compared to the statistical and detector noise in Eqs.~(\ref{eq:sigmap}) and (\ref{eq:sigmal}).

The leakage in DEAP-1 diverges from the model curve due to pileup events. These are much easier to identify in the large detector, again because each PMT here will have on average less than 1~PE per region-of-interest event, so a sensitive statistical likelihood test can be constructed by considering the number of PE per PMT and using the expected time structure of LAr scintillation.

Recent work has shown that there is a possibility of obtaining argon that has been sequestered underground and is depleted in $^{39}$Ar by more than a factor of 1000~\cite{GalbiatiPaper,UAr}.
\begin{center}
\begin{figure}[htbp]
\includegraphics[width=3.5in]{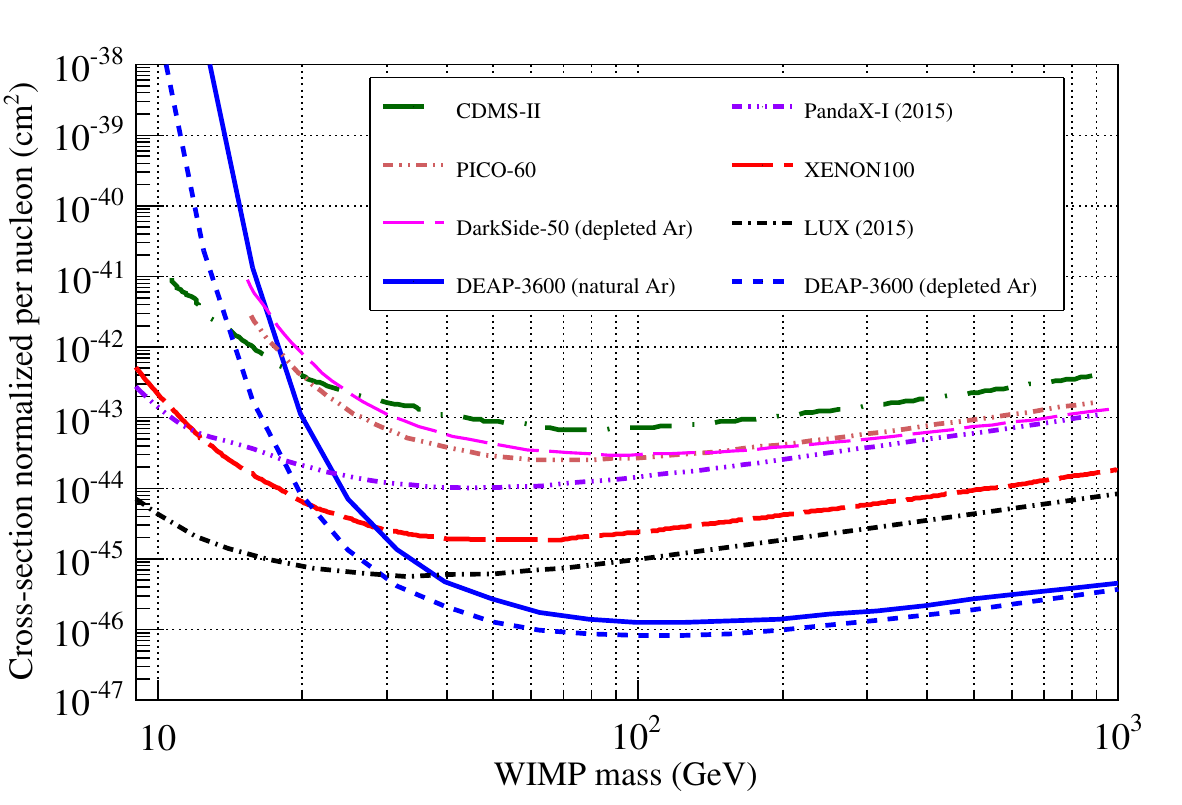}
\caption{Dark matter sensitivity of liquid argon.  Shown are the current experimental limits from the CDMS-II, DarkSide-50, PICO-60, PandaX-I, XENON100 and LUX collaborations, and the expected sensitivity for 3 tonne-years of liquid argon with a 15~\kevee\ threshold, and with a 12~\kevee\ threshold for argon that has been depleted in $^{39}$Ar by a factor of 100. The DEAP-3600 curves do not take into account the sensitivity reduction due to the nuclear recoil acceptance less than 100\%.}
\label{fig:wimplimits}
\end{figure}
\end{center}

Figure~\ref{fig:wimplimits} shows the 90\%~C.L. sensitivity versus WIMP mass for both natural and depleted argon cases, compared to the current experimental limits of the CDMS-II~\cite{cdms}, DarkSide-50~\cite{ds50}, PICO-60~\cite{pico60}, PandaX-I~\cite{pandax}, XENON100~\cite{xenon100} and LUX~\cite{lux} experiments.  We used standard assumptions for the galactic halo outlined in reference~\cite{Lewin:1995rx} and assumed a nuclear recoil scintillation efficiency of 25\% compared to electrons~\cite{gastleretal}.

\section{Summary}
We have shown that pulse-shape discrimination in liquid argon can be used to separate electron and nuclear recoil events, respectively induced by $\gamma$'s and by neutrons, in the energy region 44--89~\kevee\ known with 6\% systematic uncertainty, with a leakage of less than $2.7 \times 10^{-8}$ (90\% C.L.) for 90\% nuclear recoil detection efficiency.  

This is the highest discrimination factor reported for liquid argon and
improves by an order of magnitude the result reported by microCLEAN~\cite{Lippincott:2008ad} (for 60--128~keV of nuclear recoil energy and for a nuclear recoil acceptance of 50\%) and by a factor of $\sim$5 the result reported by DarkSide-50~\cite{darkside} ($1.5 \times 10^7$ events collected in 8.6--65.6~\kevee\ energy range with no leakage for nuclear recoil acceptance of 90\%).

An analytic model, which assumes beta-binomial distributions for the singlet and triplet components of the scintillation signal and accounts for statistical and systematic sources of noise in these distributions, describes the main features of the observed \fprompt\ distribution above 21~\kevee\ and above 0.28~\fprompt .  This model, using conservative estimates for parameters not well known at lower energies, projects a discrimination power in argon of approximately $10^{-10}$ at 15~\kevee\ and 50\% nuclear recoil acceptance, for a detector with 8~photoelectrons per \kevee\ light yield, which allows for sufficient background rejection of $^{39}$Ar in a 1000-kg liquid argon dark matter search experiment.

\section*{Acknowledgements}
We thank the members of the MiniCLEAN and DEAP-3600 collaborations for fruitful discussions and comments to the draft of this paper. We thank David Bearse for invaluable technical support. We are grateful to SNOLAB and its staff for on-site support and help during \mbox{DEAP-1} installation and operations.  We also thank Compute Canada and the Center for Advanced Computing, formerly the High Performance Computing Virtual Laboratory (HPCVL), Queen's University site, for the computational support and data storage.

This work is supported by the National Science and Engineering Research Council of Canada (NSERC), by the Canada Foundation for Innovation (CFI), by the Ontario Ministry of Research and Innovation (MRI), and by the David and Lucille Packard Foundation.

The work of our co-op and summer students is gratefully acknowledged.

\appendix 
\section{Deriving the \fprompt\ distribution - general framework} \label{sect:fpromptgeneral}

The time arrival profile of photo-electrons (PE) is detected for each event. 
The numbers of PE arriving in a given prompt time window, $n_p$, and those arriving in a given late time window, $n_l$, are counted. 
The total number of PE in the event, TotalPE, is $n_p + n_l = n_t$. 
Over the course of an experiment a wide range of events of varying TotalPE are recorded, with associated distributions of prompt and late~PE. 
We denote these as $T(n_t)$, $P(n_p)$ and $L(n_l)$ and label them as ``free parent distributions" for reasons that will become clear in following sections.
The functional form of the free parent distributions are the empirical result of a combination of the microscopic physics and the particularities of the experimental detection, such as detector geometry and electronics.  
Their means are labeled $\mu_t$, $\mu_p$ and $\mu_l$.

Pulse shape discrimination between electron-recoil and nuclear-recoil events is achieved using the \fprompt\ parameter $f \equiv n_p/(n_p+n_l)$ and we are interested in the functional form, 
$R(f,N_{pe})$, that describes the distribution of $f$ at any particular fixed number TotalPE = $N_{pe}$. 
Of particular interest are the ``tails" of the $R(f,N_{pe})$ distribution, that is to say the behaviour of the function far away from its peak, since it is the tails which determine the experiment's 
ability to discriminate between an enormous quantity of electron-recoil events and the rare nuclear-recoil events. 

The functional form of the \fprompt\ distribution is first derived for events from interactions with fixed energy that are detected with a specific number of TotalPE.  
This basic functional form is then extended to distributions for the case where interactions with many energies contribute, or where events with a range of TotalPE are included, obtained through summation. 
\newline
\ \\  
Some general properties and assumptions are: \newline  
Ionizing particles of energy $E$ will create events with a TotalPE distribution whose mean is related to the interaction energy through the detector light yield, $\mu_t = E \cdot Y$, 
and the following derivations assume that the light yield is approximately constant in the energy range of interest. 
The mean of the TotalPE distribution from many events, $\mu_t$, is representative of the energy of the particles measured in PE, while $n_t$ is the detected number of PE of an individual event, and the $n_t$ are distributed about the mean $\mu_t$ due to 
particularities of the apparatus and detection medium that result in a finite energy resolution.

Additionally, the means of the Prompt and Late distributions are always related through the mean of the \fprompt\ distribution at a given energy, $\bar{f_p}(E)$, 
\begin{align}
\mu_p(E) &= \bar{f_p}(E) \mu_t(E) \\
\mu_l(E) &= \left(1-\bar{f_p}(E)\right) \mu_t(E)
\end{align}
Furthermore, at a given energy the probability distribution for TotalPE can be written as the convolution between the Prompt and Late distributions:
\begin{eqnarray}
T_E(n_t) &=& \left(P_E(n_p) \ast L_E(n_l)\right)(n_t) \nonumber \\
         &=& \sum_{n_p=0}^\infty P_E(n_p) \cdot L_E(n_t - n_p)  \label{eq:convtotale}
  \end{eqnarray}
where the subscript $E$ makes explicit that interactions at a particular energy are considered.

\subsection{Mono-energetic events at fixed TotalPE}
Consider a radioactive calibration source that can produce events of a known fixed energy E. 

At a fixed TotalPE, $n_t = N_{pe}$ where $N_{pe}$ is a constant. The probability for a mono-energetic event to be observed with $n_p$ prompt~PE is denoted $P_E'(n_p;N_{pe})$, 
and likewise for $n_l$, $L_E'(n_l;N_{pe})$. 
$P_E'(n_p;N_{pe})$ can be computed by taking the original probability to observe any event with $n_p$ prompt~PE and multiplying this by the probability to find the matching number of $n_l = N_{pe} - n_p$ late~PE.

\begin{eqnarray}
P'_E(n_p;N_{pe}) = P_E(n_p) \cdot L_E(N_{pe} - n_p) \label{eq:qprime} \\
L'_E(n_l;N_{pe}) = L_E(n_l) \cdot P_E(N_{pe} - n_l) \label{eq:lprime}
\end{eqnarray}

We call Eq.~(\ref{eq:qprime}) and~(\ref{eq:lprime}) the primed or constrained distributions. Note that these distributions are perfectly anti-correlated. 
The sum of the primed distribution over all possible values of $n_t$ is the free distribution.
\begin{eqnarray}
\sum_{n_t=0}^{N_{pe}} P'_E(n_p;N_{pe}) = P_E(n_{p}) \\
\sum_{n_t=0}^{N_{pe}} L'_E(n_p;N_{pe}) = L_E(n_{p})
\end{eqnarray}

\begin{figure}[htbp] 
     \includegraphics[width=3.5in]{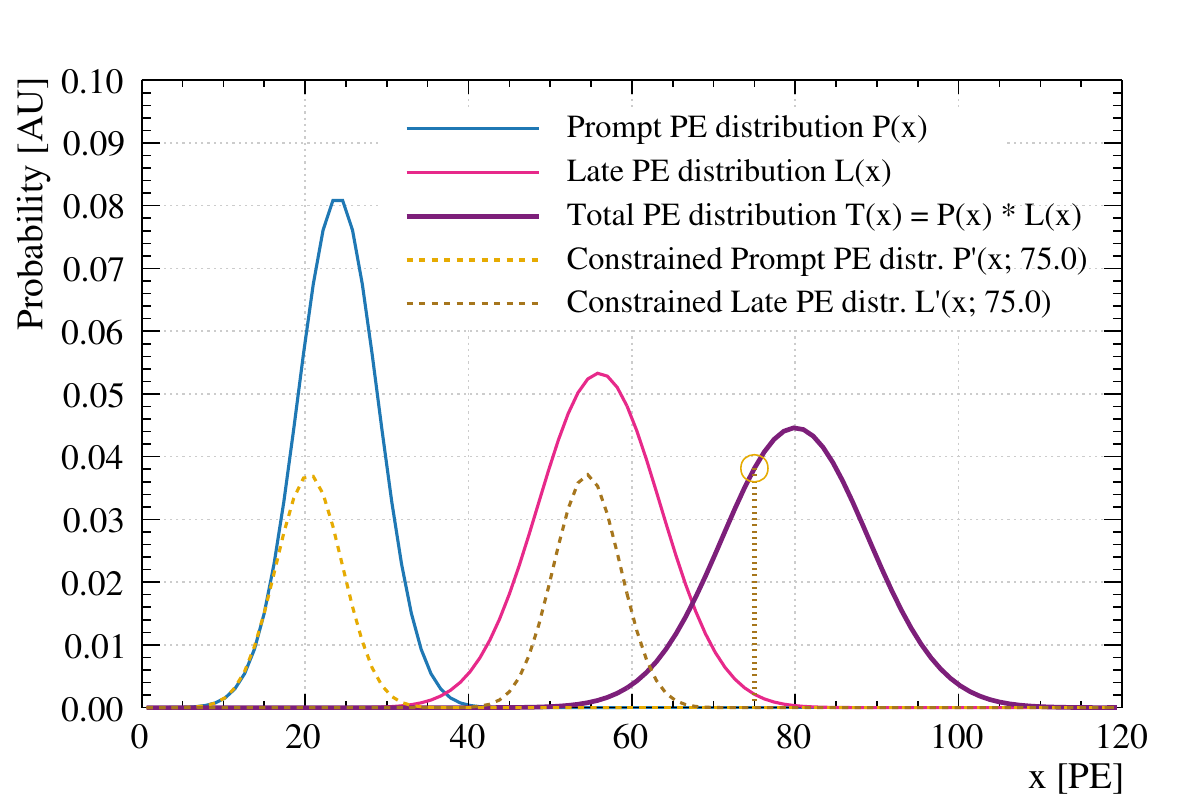}
     \caption{The relationships between the free and constrained distributions, Eqs.~(\ref{eq:qprime}, \ref{eq:lprime}), shown for the case of Gaussian distributions.  
The TotalPE distribution (bold purple) is a Gaussian with mean $\mu_t = 80$~PE, $\bar{f_p} = 0.3$, and has associated Gaussian free parent Prompt (blue) and Late (pink) distributions with means 
$\mu_p = 24$~PE and $\mu_l=56$~PE. 
Only a subset of events within the free parent Prompt and Late distributions contribute to those events of TotalPE=75. These subsets are the constrained distributions shown in orange and brown. 
The constrained distributions in this specific example happen to be also Gaussians.}
     \label{fig:gausparents} 
\end{figure}

The relationship between the free and primed Prompt and Late distributions is illustrated in Fig.~\ref{fig:gausparents} for the case of Gaussian parent distributions (see details in~\ref{sect:gausparents}).

By definition, each event from the population with constant $n_t = N_{pe}$ has a discrimination parameter $f = n_p/N_{pe}$; that is the \fprompt\ distribution for fixed TotalPE is entirely determined by the Prompt distribution.
Hence the \fprompt\ distribution for events of common energy $E$ that are detected with fixed TotalPE $N_{pe}$ is equal to the distribution $P'$ after a simple variable transformation
\footnote{If $n=f \cdot N_{pe}$, then $P(f)df = P'(n)dn$; hence $P(f) = P'(n) dn/df = P'(n) N_{pe}$.}.
\begin{eqnarray}
R_E(f;N_{pe}) = N_{pe} \cdot P'_E(n_p=f\cdot N_{pe};N_{pe}) \label{eq:rebasic}
\end{eqnarray}

Eqs.~(\ref{eq:rebasic}) and (\ref{eq:qprime}) are the basic building blocks of {\bf any } \fprompt\ distribution as will be discussed in the following sections.

\subsection{Mono-energetic events at any TotalPE}

One can obtain the \fprompt\ distribution for all events of the same energy E, regardless of each event's particular $n_t$,
by summing the \fprompt\ distributions at fixed  
TotalPE = $N_{pe}$ over all values of $N_{pe}$:
\begin{eqnarray}
R_E(f) = \sum_{N_{pe}=0}^\infty N_{pe} \cdot P'_E(f N_{pe};N_{pe}) \label{eq:pstar}
\end{eqnarray}
This distribution fully describes the \fprompt\ distribution of data from a mono-energetic calibration source and without TotalPE binning, as long as there are negligible background contributions.

\subsection{Fixed TotalPE and multiple energies}
More generally, the apparatus detects events from a variety of sources emitting an energy spectrum $N(E)$, so that events with a range of energies contribute to the \fprompt\ distribution for any given TotalPE. 
We build the primed Prompt distribution for this case by taking the primed Prompt distributions for each energy in the spectrum (Eq.~(\ref{eq:qprime})) and adding them up,
each weighted by the probability $T_E(N_{pe})$ that an event of energy $E$ is detected at $N_{pe}$:
\begin{equation}
P'(n_p; N_{pe}) = \int_0^\infty P'_E(n_p;N_{pe})\; T_E(N_{pe}) N(E) \; dE  \\ \label{eq:qpe}
\end{equation}
And in \fprompt\ it is
\begin{eqnarray}
R(f;N_{pe}) &=& N_{pe} \cdot P'\left(n_p;N_{pe}\right) \nonumber \\
			&=&  N_{pe} \cdot P'\left(f\cdot N_{pe};N_{pe}\right) \label{eq:rpe}
\end{eqnarray}
For practical calculation, the integral over $E$ in Eq.~(\ref{eq:qpe}) can be turned into a sum over $\mu_t$.

\subsection{Choice of parent distribution} \label{sect:parentdistr}

Choosing the correct functional form for the free parent distributions is crucial to obtaining an accurate description of the tails, and in principle should be largely motivated by an understanding of the microscopic physics. In a realistic detector, this means that the number of PE detected should be drawn from a compound distribution characterizing each step of the photon emission and detection chain, for example in the form of convolutions between binomial, poisson, and Gaussian distributions as appropriate for each step.

Furthermore, a realistic detector has noise, non-uniformities, and ageing effects that cause the parent distribution's parameters to differ slightly between events. This is the case for example when photons emitted from different locations can have very different photon paths in the detector, or when the overall light yield changes with time. These effects modify the parameters that determine the photon survival probability at each step, namely the rate parameter in a poisson distribution or the success probability in a binomial distribution. Noise, such as electronics noise, modifies the final PE count after the number of PE are drawn from the parent distribution and thus can be modelled by convoluting the parent distribution with the noise distribution over the distribution argument (the PE-count). Effects that modify the photon survival probability also modify the final PE count, but they do so by changing the parent distribution parameters that the number of PE are drawn from, so that over all events, the parent distribution parameters are now themselves distributed as some probability distribution. This situation must be modelled by convoluting the parent distribution with the distribution of its parameter over the parent distribution mean. The technical difference between convoluting over the distribution mean or the distribution argument has a small effect near its peak, but a significant effect in the tails.

To include detector effects that modify the `success probability' or photon survival probability $p$, we distribute $p$ as a beta distribution around a mean of $\bar{p}$. We choose a beta distribution for this because of its flexible shape and because the result is an analytic function: the beta-binomial distribution 
\begin{equation} 
\text{Binomial}(n; \mu, \text{Beta}(\bar{p},b)) = \text{BetaBin}(n;\mu,\bar{p},b)\label{eq:bbin}
\end{equation}
where the Beta distribution is explicitly parametrized in terms of its mean $\bar{p}$ and shape parameter $b$, which are related to the often used $\alpha$ and $\beta$ parameters as $ \bar{p} = \alpha/(\alpha + \beta)$ and $b = \alpha$.
An analytic description minimizes computation time and reduces the complexity of the equations in the final \fprompt\ model. The shape parameter $b$ of the beta distribution can be seen as a dispersion parameter that quantifies the beta-binomial's amount of dispersion over the ideal binomial case. 

Since we desire a description of the \fprompt\ distribution for all $\mu_{p/l}$ of interest, care must be taken to parametrize the distribution in such a way that the relative dispersion remains constant at different $\mu$. The nominal beta-binomial distribution is
\begin{multline}
\text{BetaBin}(n;\mu,\bar{p},\alpha) \equiv
\frac{\Gamma(\frac{\mu}{\bar{p}}+1)}{\Gamma(n+1) \Gamma(\frac{\mu}{\bar{p}}-n+1)} \\
\times\frac{ \Gamma(n+\alpha) \Gamma\left(\frac{\mu}{\bar{p}} - n + \alpha(\frac{1}{\bar{p}} - 1)\right) \Gamma(\frac{\alpha}{\bar{p}})  } {\Gamma\left(\frac{\mu}{\bar{p}} + \frac{\alpha}{p}\right)\Gamma(\alpha)\Gamma\left(\alpha(\frac{1}{p} - 1)\right)} \label{eq:betabin}
\end{multline}
and to obtain a proper behaviour of the dispersion with $\mu$, the parameter $\alpha$ is parametrized as $\alpha = b\mu$. For $p \ll b$, the variance of Eq.~(\ref{eq:betabin}) is then
\begin{equation}
\sigma^2_{\text{BetaBin}} = \mu (1 + 1/b) \label{eq:sigmabetabin}
\end{equation}
 
\begin{figure}[htbp]
\begin{center}
     \includegraphics[width=3.5in]{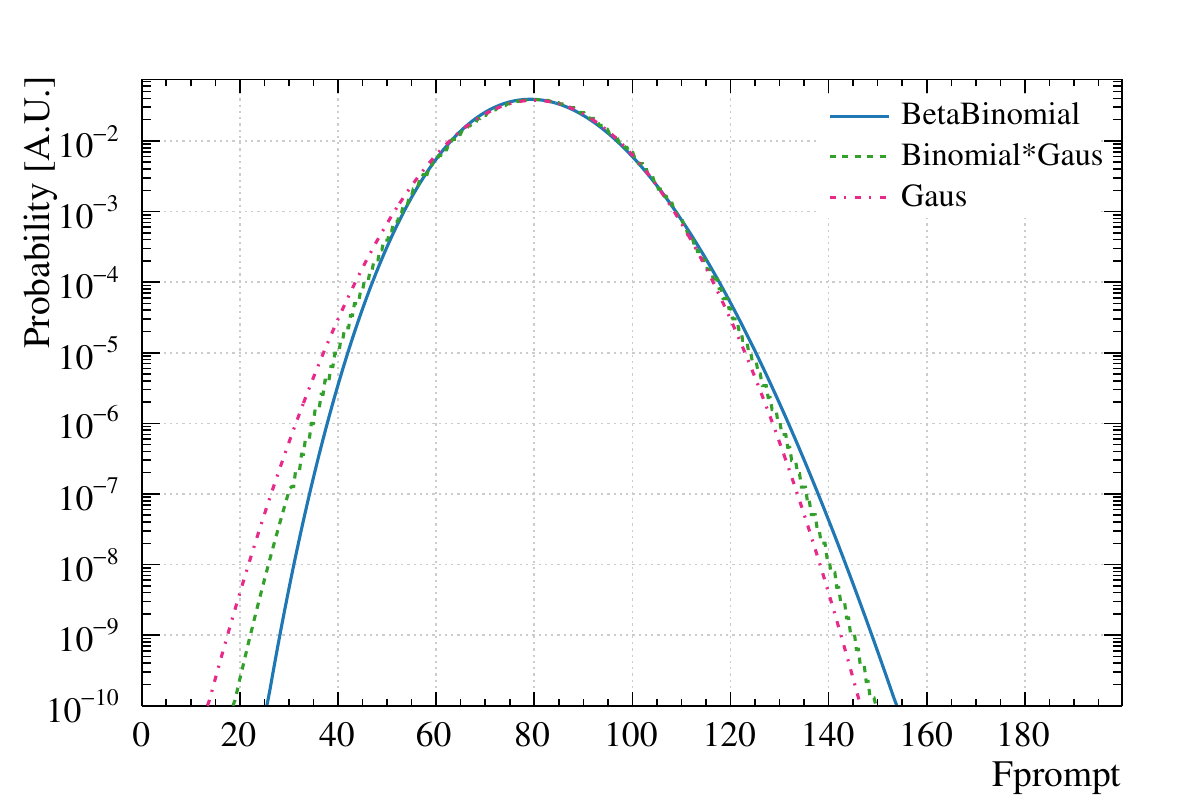}

\caption{Comparison of beta-binomial (blue solid), binomial*Gaussian (green dashed), and Gaussian (red dash-dot) distributions at $\mu = 80$~PE and binomial probability $\bar{p} = 0.06$. All distributions have a variance of $\sigma^2 = 112$~PE$^2$. }
\label{fig:betabin}
\end{center}
\end{figure}

The difference in shape between the beta-binomial (binomial convolved with beta distribution with respect to the binomial probability), binomial convolved with a Gaussian (over the distribution argument), and pure Gaussian distributions is illustrated in Fig.~\ref{fig:betabin}. Note in particular the different behaviour of the tails.

Since the overall photon survival probability is relatively small in DEAP-1 probabilistic effects associated with photon detection dominate over those coming from the emission process, so that taking a single binomial distribution to approximate both was justified in the main part of the paper. This assumption starts to fail in 120--240~PE range for light yields higher than in DEAP-1 and binomial fluctuation between the numbers of emitted prompt and late photons can no longer be neglected. Extending the analytic model to the most general case, i.e. to combine the (anticorrelated) binomial partition between prompt and late scintillation photons with the survival probabilities and noise terms for both populations poses a number of computational challenges, while it can be added to a Monte Carlo model in a rather straightforward way. In this work we have chosen the approach of analytically approximating the parent distributions for a high light yield detector with the detection-related PDFs only and neglecting the emission effects. The additional contribution from emission processes is evaluated with a toy simulation (see Fig.~\ref{fig:d3pleak}). Work on further extensions of the analytical model is ongoing.

\subsubsection{Adding uncorrelated noise} \label{sect:addnoise}
The beta-binomial distribution has an inherent width representing purely statistical fluctuation, or ``statistical noise". 
It also models what we describe as ``detector noise" through a dispersion parameter, quantifying non-uniformity of the detector that leads to $n_p$ and $n_l$ being drawn from a range of 
underlying binomial distributions.

Additional uncorrelated noise has to be added to account for the variation in charge of an SPE pulse and for the signal processing noise from detection electronics. 
We call the standard deviation of these noise components $\sigma_{p,l}$, and model them as having Gaussian shape. 
The two free parent distributions then each consist of the near-ideal beta-binomial probability distribution, convolved with the noise Gaussian over $n_p$ and $n_l$. 

\subsubsection{Adding window noise} \label{sect:windownoise}
Window noise, $\sigma_{\textrm{w}}$, is a detection effect related to the uncertainty in the location of the event peak, and thus in where to stop counting prompt and start counting late light. 
It is 100\% anti-correlated noise that randomly moves counts from the late into the prompt region\footnote{The probability for prompt counts to move into the late region is negligible due to the length of the prompt window and the short singlet decay time.} without affecting the sum count.

If this source of noise were included with the Gaussian noise terms discussed in \ref{sect:addnoise}, multiplying the uncorrelated distributions in Eq.~(\ref{eq:qprime}) would require careful 
consideration of the correlation terms. 
Due to the perfect anti-correlation, it is straight-forward to instead consider this effect separately; 
it randomly adds or subtracts from the $n_p$ in the constrained distribution $P'_\text{E}(n_p;N_{pe})$, which we model by convolving $P'$ with a Gaussian of width $\sigma_{\text{w}}$,
\begin{equation}
P_{wn}'(n_p;N_{pe}) = \sum_{n_p'=0}^\infty P'(n_p';N_{pe}) \cdot \text{Gaus}(n_p',n_p,\sigma_w) \label{eq:windownoise}
\end{equation}
with
\begin{equation}
\text{Gaus}(x,\mu,\sigma) = \frac{1}{\sqrt{2\pi}\sigma}\,\mathrm{e}^{-\frac{1}{2}\left(\frac{x-\mu}{\sigma}\right)^2}
\end{equation}

The distribution in the \fprompt\ variable is found as in Eq.~(\ref{eq:rpe}).

\section{The \fprompt\ distribution for Gaussian parent distributions} \label{sect:gausparents}

For large numbers of PE, the free Prompt and Late distributions can be well described by Gaussians. 
At smaller numbers of PE, and near the tails, the Gaussian approximation is progressively worse, and the measured distributions may only be approximated by Gaussians near their peaks.  
Even though this approach does not describe the \fprompt\ distribution adequately at the TotalPE region of interest in this paper, some general properties or the constrained distributions can be derived using the simple assumption of Gaussian parent distributions.

The Gaussians describing the Prompt and Late distributions are initially of ideal statistical width, given by the root of their mean. These are then convolved with the noise Gaussian
\begin{align}
\begin{split}
P_E(n_p) =& \text{Gaus}\left(n_p; \mu_p(E),\sqrt{\mu_p}\right) \\
         &\ast \; \text{Gaus}\left(0, \mu_p(E),\sigma_{p-}(n_t)
\right) \label{eq:qgaus} \\
	=& \text{Gaus}\left(n_p;\mu_p(E), \sigma_{p}(n_p,E)\right)
\end{split}\\
\begin{split}	
L_E(n_l) =& \text{Gaus}\left(n_l; \mu_l(E),\sqrt{\mu_l}\right) \\
         & \ast \; \text{Gaus}\left(0, \mu_l(E),\sigma_{l-}(n_t)\right) \\
	=&\text{Gaus}\left(n_l; \mu_l(E), \sigma_{l}(n_t,E)\right) \label{eq:lgaus}
	\end{split}
\end{align}
where we used the fact that the convolution of two Gaussians is itself a Gaussian~\cite{Bromiley:ti}, and $\sigma_{p,l-}$ denotes the distribution width without the statistical ($\sqrt{N}$) component.

Using Eq.~(\ref{eq:convtotale}) written as a single Gaussian, we get for the distribution in TotalPE:
\begin{equation}
T_E(n_t) = \text{Gaus}(n_t, \mu_t,\sigma_t)\label{eq:tgaus}
\end{equation}
where 
\begin{align}
\mu_t    & =  \mu_p + \mu_l \\
\sigma_t &= \sqrt{ \sigma_{p}^2 + \sigma_{l}^2 }
\end{align}

Eq.~(\ref{eq:qprime}) is used to obtain the distribution of $n_p$ that contribute to $N_{pe}$ TotalPE:
\begin{align}
\begin{split}
    P_E'(n_p) &=  \frac{1}{2\pi \sigma_{p} \sigma_{l}} \, \mathrm{e}^{-\frac{1}{2}\left(\frac{n-\mu_p}{\sigma_{p}}\right)^2 }
         \,  \mathrm{e}^{-\frac{1}{2}\left(\frac{N_{pe} - n-\mu_l}{\sigma_{l}}\right)^2 }  
\end{split}\\
\begin{split}
		&=  \frac{1}{2\pi \sigma_{p} \sigma_{l}} \, \mathrm{e}^{-\frac{1}{2}\left(\frac{n-\mu_p}{\sigma_{p}}\right)^2}
			\, \mathrm{e}^{-\frac{1}{2}\left(\frac{n - (N_{pe} - \mu_l)}{\sigma_{l}}\right)^2 } 
\end{split}\\
		&= \text{Gaus}(n_p, \mu'_p,\sigma_{pl})\label{eq:qpimegaus}
\end{align}
with~\cite{Bromiley:ti}
\begin{eqnarray}
\sigma_{pl} &=& \sqrt{\frac{\sigma_{p}^2\cdot \sigma_l^2}{\sigma_{p}^2 +\sigma_{l}^2} }\label{eq:sigmapl}\\
\mu'_p &=& \frac{\mu_p \sigma_{l}^2 + (N_{pe} - \mu_l)\sigma_{p}^2}{\sigma_{p}^2 +\sigma_{l}^2} \label{eq:mupprime}
\end{eqnarray}

Note that the equation for $L_E'(n_l)$ follows a parallel argument so that 
\begin{align}
L_E'(n_l) &= \text{Gaus}(n_l, \mu'_l,\sigma_{pl})\label{eq:lpimegaus}
\end{align}
with
\begin{eqnarray}
\mu'_l &=& \frac{\mu_l \sigma_{p}^2 + (N_{pe} - \mu_p)\sigma_{l}^2}{\sigma_{p}^2 +\sigma_{l}^2} \label{eq:mulprime}
\end{eqnarray}

In the case of purely statistical noise, i.e. $\; \sigma_{p} = \sqrt{\mu_p}, \; \sigma_{l} = \sqrt{\mu_l}$ this simplifies to
\begin{eqnarray}
\sigma_{pl}^2  &=& \bar{f_p}\cdot \mu_t(1-\bar{f_p}) \\
\mu'_p &=&  \bar{f_p}\cdot N_{pe} \\
\mu'_l &=& (1-\bar{f_p})\cdot N_{pe}
\end{eqnarray}	

If the noise has terms not proportional to $\sqrt{\mu}$, $\mu'_p$ and $\mu'_l$ will be shifted by some amount. For example, if 
\begin{align}
\sigma_p^2 = a\mu + c_p \\
\sigma_l^2 = a\mu + c_l
\end{align}
then correlated means become:
\begin{align}
\mu'_p = \frac{\mu_p(c_l + a N_{pe}) + c_p(N_{pe} - \mu_l)}{a\mu_t + c_p + c_l} \\
\mu'_l = \frac{\mu_l(c_p + a N_{pe}) + c_l(N_{pe} - \mu_p)}{a\mu_t + c_p + c_l}
\end{align}
The shift is such that the mean \fprompt\ is unaffected.

Eq.~(\ref{eq:pstar}) evaluates to the uncorrelated Hinkley function~\cite{hinkley}, also called the Gaussian ratio distribution (after a variable transformation from $w = n_p/n_l$ to $f=n_p/(n_p+n_l)$ where then $ w = f/(1-f)$).

Eq.~(\ref{eq:qpe}) becomes:

\begin{multline}
P'(n_p;N_{pe}) = \int_0^\infty \text{Gaus}\left(n_p;\mu'_p\text{\small{(E)}}, \sigma_{pl}(n_p,E)\right) \\
\times \text{Gaus}\left(N_{pe}, \mu_t(E),\sigma_t(E)\right) \cdot N(E)\; dE
\end{multline}
Written in \fprompt\ and setting $E\cdot Y = \mu_t$:

\begin{align}
R(f;N_{pe}) &= N_{pe} \cdot P' \left(N_{pe} f;N_{pe}\right) \nonumber \\
\begin{split}\label{eq:gausFinalP}
    &= \int_0^\infty \frac{N_{pe}}{2\pi\; \sigma_t\text{\tiny{(E)}} \;\sigma_{pl}\text{\tiny{(E)}}}
\mathrm{e}^{-\frac{1}{2}\left(\frac{f - \mu'_p\text{\tiny{(E)}}/N_{pe}}{\sigma_{pl}\text{\tiny{(E)}}/N_{pe}}\right)^2}  \\
    &\, \, \times \mathrm{e}^{-\frac{1}{2}\left(\frac{N_{pe} - \mu_t\text{\tiny{(E)}}}{\sigma_t\text{\tiny{(E)}}}\right)^2} N(\mu_t) \; d\mu_t\\ 
\end{split}
\end{align}
and 
\begin{equation}
\sigma_t \;\sigma_{pl}\text{\tiny{(E)}} = \sigma_p\text{\tiny{(E)}} \sigma_l\text{\tiny{(E)}}
\end{equation}
\begin{figure}[htbp] 
     \includegraphics[width=3.5in]{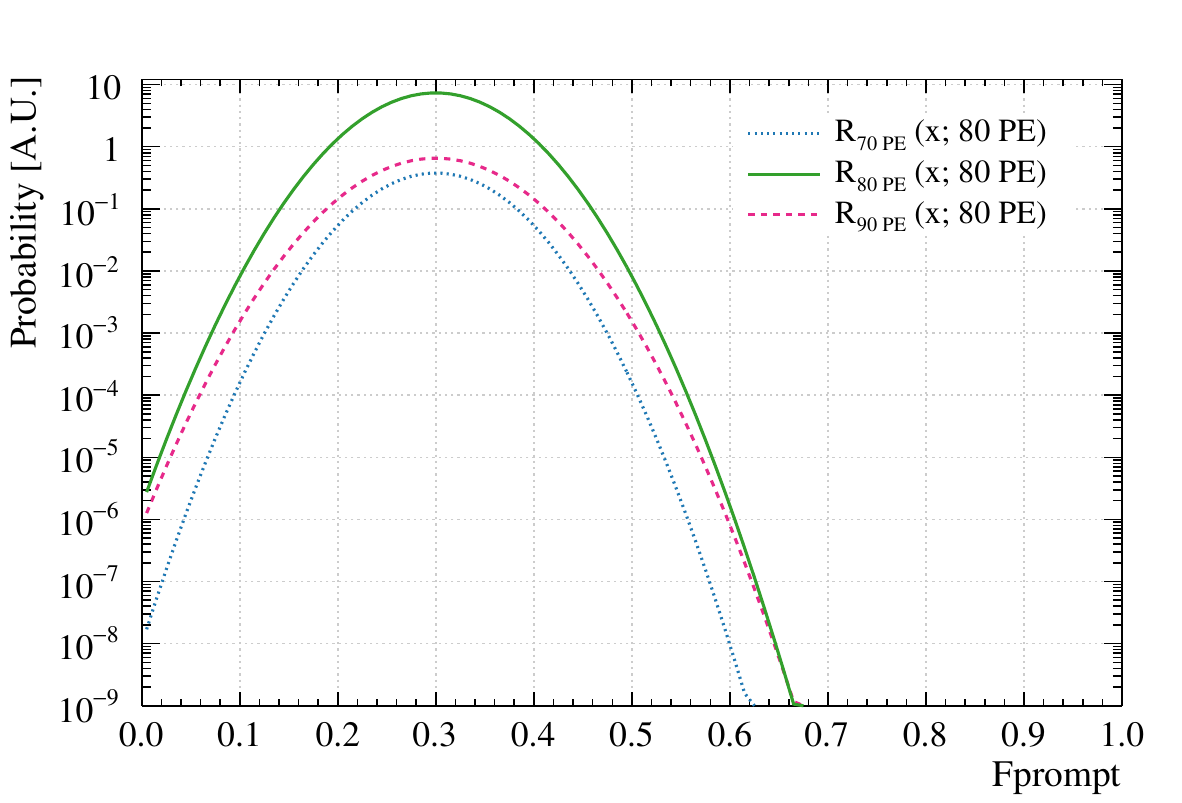}
     \caption{The contributions to the \fprompt\ distribution at a detected 80~PE, for a source that emits events of only three different energies (corresponding to TotalPE distribution means of 70, 80, and 90~PE) each with the same probability. The observed \fprompt\ distribution would be the sum of the three curves.}
     \label{fig:variousPEat80}
\end{figure}
where we explicitly indicated all parameters that are a function of energy.

At any given TotalPE, the \fprompt\ distribution is composed of the sum of the \fprompt\ distributions from events of many different energies, with each distribution weighted by its 
probability of occurring.

The width of each of these component distributions, $\sigma_{pl}(E)$ is monotonously increasing with energy, thus $\sigma_{pl}(E_1) < \sigma_{pl}(E_2)$ if $E_1 < E_2$, which means that events from 
higher energies contribute more weight to the sum \fprompt\ distributions than events from lower energies. 
In particular, this means that the tails of the sum distribution are more strongly influenced by events from higher energies.

This is illustrated in Fig.~\ref{fig:variousPEat80}, where Eq.~(\ref{eq:gausFinalP}) is drawn for $N_{pe}=80$~PE, and events from three different energies whose TotalPE distribution means, $\mu_t$, are 70, 80, and 90~PE. Each energy is taken as equally likely and the energy resolution is $\sigma_{pl} = \sqrt{1.12 \mu}$. Due to its larger standard deviation, the curve for events from the higher energy is both more dominant and relatively wider than the curve from the smaller energy events. Above 0.6~f, the contribution from higher energy events and the otherwise dominant energy for the given TotalPE bin are about equal.

\section{Mis-calibration of the energy scale} \label{sect:energymiscalibration}

The statistical description of the \fprompt\ distribution requires the precise knowledge of the number of PE in an event, at least on average. It is sensitive to systematic mis-calibration of the energy scale where the measured values of the Prompt and Late means, $\hat{\mu}_p$ and $\hat{\mu}_l$, are wrong by $\delta_p$ and $\delta_l$ and a common constant $a$,
\begin{align}
\hat{\mu}_p &= a\mu_p + \delta_p \\
\hat{\mu}_l &= a\mu_l + \delta_l \\
\hat{\mu}_t &=  \hat{\mu}_p + \hat{\mu}_l \\
			&=a\mu_t + \delta_t
\end{align}
where $\mu_p$ and $\mu_l$ are the true values. 

The true mean \fprompt\ value $\bar{f_p}$ is 
\begin{equation}
\bar{f_p} = \frac{\mu_p}{\mu_p+\mu_l} = \frac{\mu_p}{\mu_t}
\end{equation}
and assumed constant for all energies, though the extension to energy dependent $\bar{f_p}$ is straight forward.

We derive the functional form of the observed mean \fprompt\, $\hat{f_p}$, the observed prompt and late distribution means, and the observed standard deviation, $\hat{\sigma}_{pl}$, as a function of observed TotalPE $\hat{\mu}_t$.

The observed distribution means can be expressed as
\begin{align}
\hat{\mu}_p &= a\cdot \bar{f_p}  \mu_t + \delta_p\\
			&= a\cdot \bar{f_p}  \frac{1}{a}( \hat{\mu}_t - \delta_t) + \delta_p \\
			&= \bar{f_p} ( \hat{\mu}_t - \delta_t) + \delta_p  \label{eq:obspromptmean} \\
\hat{\mu}_l &= (1 -\bar{f_p}) (  N_{pe} - \delta_p -\delta_l) + \delta_l \label{eq:obslatemean}
\end{align}

The measured mean \fprompt\ becomes
\begin{align}
\hat{f_p} &= \frac{\hat{\mu}_p}{\hat{\mu}_p + \hat{\mu}_l} \\
		&= \frac{a\mu_p + \delta_p}{a\mu_p + \delta_p + a\mu_l + \delta_l} \\
		&= \frac{a\mu_p + \delta_p}{a\mu_t + \delta_t} 
\end{align}
dividing by $a\mu_t$
\begin{align}
\hat{f_p} &= \frac{f_p + \delta_p/(a\mu_t)}{1 + \delta_t/(a\mu_t)} 
\end{align}
and re-writing in terms of the measured TotalPE
\begin{align}
\hat{f_p} &= \frac{f_p + \delta_p/(\hat{\mu}_t - \delta_t)}{1 + \delta_t/(\hat{\mu}_t - \delta_t)}  \label{eq:obsfp}
\end{align}

Eq.~(\ref{eq:obsfp}) shows that the mean \fprompt\ obtains a TotalPE-dependence even in the case where the true underlying mean \fprompt\ is constant with energy.

In the Gaussian approximation and with $\sigma_p \simeq \sqrt{\mu_p}$ (i.e. no constant terms), the variance of the correlated Prompt distribution is Eq.~(\ref{eq:sigmapl}) to which we add the window noise:

\begin{align}
\sigma_{pl;w}^2 = (1-f_p)\mu_t f_p + \left( \epsilon_\text{win} (1- f_p) \mu_t \right)^2
\end{align}
Re-writing in terms of the observed number of PE, and realizing that the calibration scaling applies to the width as well, so the measured width is $\hat{\sigma_{pl;w}} = a\sigma_{pl;w}$:
\begin{align}
\hat{\sigma}_{pl;w}^2 &= a^2 \left[ (1-f_p)\frac{1}{a}(\hat{\mu}_t - \delta_t ) f_p + \left( \epsilon_\text{win} (1- f_p) \frac{1}{a}(\hat{\mu}_t - \delta_t ) \right)^2 \right] \\
			&= (1-f_p)a(\hat{\mu}_t - \delta_t ) f_p + \left( \epsilon_\text{win} (1- f_p)(\hat{\mu}_t - \delta_t ) \right)^2 \label{eq:latevar}
\end{align}


\begin{thebibliography}{21}
\bibitem{boulay_astro} M. Boulay and A. Hime, Astropart. Phys. {\bf 25}, 179
(2006).
\bibitem{ardm} A. Rubbia (ArDM), J. Phys. Conf. Ser. {\bf 39}, 129 (2006).
\bibitem{miniclean} R. Keith (MiniCLEAN), AIP Conf. Proc. {\bf 1441}, 518 (2012).
\bibitem{darkside} P.~Agnes et al. (DarkSide), Phys. Lett. B {\bf 743}, 456 (2015).
\bibitem{warp} P. Benetti et al. (WArP), Astropart. Phys. {\bf 28}, 495 (2008).
\bibitem{d1arxiv} M. Boulay et al., e-print arXiv:0904.2930v1 (2009).
\bibitem{Miyajima:1974zz} M. Miyajima et al., \newblock Phys. Rev. A {\bf 9}, 1438 (1974).
\bibitem{benetti} P. Benetti et al., \newblock Nucl. Instr. Meth. A {\bf 574}, 83 (2007).
\bibitem{Looslipaper} H.\,H. Loosli,
\newblock {Earth and Planetary Science Letters} {\bf 63}, 51
(1983).
\bibitem{Mul70} R.\,S. Mulliken, \newblock J. Chem. Phys. {\bf 52}, 5170 (1970).
\bibitem{Ged72} A. Gedanken et al., \newblock J. Chem. Phys. {\bf 57}, 3456 (1972).
\bibitem{hitachi} A. Hitachi and T. Takahashi, \newblock Physical Review B {\bf 27}, 5279 (1983).
\bibitem{gastleretal} D. Gastler et al.,
\newblock Phys. Rev. C {\bf 85}, 065811 (2012).
\bibitem{Cao2014} H. Cao et al. (SCENE), Phys. Rev. D {\bf 91}, 092007 (2015).
\bibitem{Creus2015} W. Creus et al., JINST {\bf 10}, P08002 (2015).
\bibitem{doke} T. Doke et al., \newblock Nucl. Instr. Meth. A {\bf 269}, 291 (1988).
\bibitem{suemoto} T. Suemoto and H. Kanzaki, \newblock J. Phys. Soc. Jpn. {\bf 46}, 1554 (1979).
\bibitem{Acciarri:2008kv} R. Acciarri et al. (WArP), \newblock JINST {\bf 5}, P06003 (2010),
\newblock e-print arXiv:0804.1217.
\bibitem{Lippincott:2008ad} W.\,H. Lippincott et al.,
\newblock Phys. Rev. C {\bf 78}, 035801 (2008),
e-print arXiv:0801.1531, see also Phys. Rev. C {\bf 81}, 039901(E) (2010).
\bibitem{burtonpowell} W.\,M. Burton and B.\,A. Powell, Appl. Opt. {\bf 12} 87 (1973).
\bibitem{Lidgard} J.\,J. Lidgard, Pulse shape discrimination studies in liquid argon for the {DEAP-1} detector,
\newblock Master's thesis, Queen{'}s University, 2008.
\bibitem{SAES} SAES PS4-MTR-3-R1 getter specifications,
http://www.saesgetter.com.
\bibitem{Eoin} E. O{'}Dwyer, Radon Background Reduction in DEAP-1 and DEAP-3600,
\newblock Master's thesis, Queen{'}s University (Dec. 2010).
\bibitem{geant4} S. Agostinelli et al.,
\newblock Nucl. Instr. Meth. A {\bf 506}, 250 (2003).
\bibitem{geant4b} J. Allison et al.,
\newblock IEEE Transactions on Nuclear Science {\bf 53},
270 (2006).
\bibitem{snolab} SNOLAB website, \url{http://www.snolab.ca}.
\bibitem{paradorn} P. Pasuthup, Characterization of Pulse-Shape Discrimination for Background Reduction in the DEAP-1 Detector,
\newblock Master's thesis, Queen{'}s University, 2009.
\bibitem{root} R. Brun and F. Rademakers,
\newblock Nucl. Instr. Meth. A {\bf 389}, 81 (1997).
\bibitem{hpcvl} High Performance Computing Virtual Laboratory,
http://www.hpcvl.org.
\bibitem{ford} Section 6.1 in R. J. Ford, Calibration of {SNO} for the Detection of $^8$B Neutrinos,
\newblock PhD thesis, Queen{'}s University, 1998.
\bibitem{sno_spe} C. Jillings, A photomultiplier tube evaluation system for the Sudbury Neutrino Observatory,
Master's thesis, Queen{'}s University, 1992.
\bibitem{Pollmann:2012ad} P.-A. Amaudruz et al. (DEAP), Astropart. Phys. {\bf 62}, 178 (2015).
\bibitem{Acciarri:2008kx} R. Acciarri et al. (WArP), \newblock JINST {\bf 5}, P05003 (2010),
\newblock e-print arXiv:0804.1222.
\bibitem{Cowan} G. Cowan, ``Statistical Data Analysis'', Oxford University Press, 1998. Chapter 9.9.
\bibitem{tcald_thesis} T. Caldwell, Searching for dark matter with single phase liquid argon,
PhD thesis, University of Pennsylvania, 2015.
\bibitem{tcald_paper} M. Akashi-Ronquest et al.,
\newblock Astropart. Phys. {\bf 65}, 40 (2015).
\bibitem{NEST} M. Szydagis et al., JINST {\bf 6} P10002 (2011).
\bibitem{Regenfus} C. Regenfus et al., J. Phys. Conf. Ser. {\bf 375} 012019 (2012).
\bibitem{BoulayTAUP} M.\,G.~Boulay (DEAP), J. Phys. Conf. Ser. {\bf 375}, 012027 (2012).
\bibitem{GalbiatiPaper} D. Acosta-Kane et al.,
\newblock Nucl. Instr. Meth. A {\bf 587}, 46 (2008).
\bibitem{UAr} P.~Agnes et al. (DarkSide),
\newblock e-print arXiv:1510.00702 (2015).
\bibitem{cdms} CDMS collaboration,
\newblock Science {\bf 327}, 1619 (2010).
\bibitem{ds50} P. Agnes et al. (DarkSide), Phys. Rev. D {\bf 93}, 081101 (2016).
\bibitem{pico60} C. Amole et al. (PICO), Phys. Rev. D {\bf 93}, 052014 (2016). 
\bibitem{pandax} X. Xiao et al. (PandaX), Phys. Rev. D {\bf 92}, 052004 (2015).
\bibitem{xenon100} J. Angle et al.,
\newblock Phys. Rev. Lett. {\bf 109}, 181301 (2012).
\bibitem{lux} D.\,S.~Akerib et al. (LUX), Phys. Rev. Lett. {\bf 116}, 161301 (2016).
\bibitem{Lewin:1995rx} J.\,D. Lewin and P.\,F. Smith,
\newblock Astropart. Phys. {\bf 6}, 87 (1996).
\bibitem{Bromiley:ti} P.\,A. Bromiley, ``Products and Convolutions of Gaussian Probability Density Functions'', TINA memo 2003-003, Imaging Sciences Research Group, Institute of Population Health, School of Medicine, University of Manchester
\bibitem{hinkley} D.\,V. Hinkley,
\newblock Biometrika {\bf 56}, 635 (1969).
\end{thebibliography}
\end{document}